\documentclass[12pt,a4paper]{article}
\usepackage{amsmath} 
\usepackage{amsthm} 
\usepackage{amsfonts}
\usepackage{amssymb}

\numberwithin{equation}{section}

\newtheorem{theorem}{Theorem}[section]
\newtheorem{cor}[theorem]{Corollary} 
\newtheorem{lemma}[theorem]{Lemma}
\newtheorem{prop}[theorem]{Proposition}
\theoremstyle{definition} 
\newtheorem{defi}{Definition}[section]
\newtheorem{rem}[defi]{Remark}

\newcommand\pseven{\psi_0}
\newcommand\steven{\psi^*_0}
\newcommand\pshes{\psi^*_{\ov r}}
\newcommand\sthes{\psi^*_{\ov r}}
\newcommand\psodd{\psi_1}
\newcommand\stodd{\psi^*_1}
\newcommand\reven{0}
\newcommand\rodd{1}

\newcommand\stpm{\psi^*_{\ov r}}

\newcommand\Ha{\widehat H}

\newcommand\half{\tfrac{1}{2}}
\newcommand\num{D_\g}
\newcommand\rao{{{|\h_0}}}

\newcommand\algsi{\widehat L (\g, \si)}
\newcommand\alkmusi{\widehat L (\k_\mu, \si_{|\k_\mu})}

\newcommand\ov{\overline} 

\renewcommand\o{\varpi}

\renewcommand\({\left(} 
\renewcommand\){\right)} 
 
\newcommand\rhat{\widehat\rho} 
\newcommand\rkhat{\rhat_\k}
\newcommand\be{\beta}
\newcommand\de{\delta}

\newcommand\ka{\widehat \k}
\newcommand\op{\widehat{so(\p)}}

\newcommand\g{\mathfrak g}
\newcommand\ga{\widehat{\mathfrak g}} 
\newcommand\h{\mathfrak h}
\newcommand\ha{\widehat{\mathfrak h}}

\renewcommand\sp{\text{\it Span\,}} 
 
\newcommand\bb{\mathfrak b}
\newcommand\D{\Delta} 
\renewcommand\l{\lambda} 
\newcommand\Dp{\Delta^+}
 
\newcommand\Da{\widehat\Delta}
\newcommand\Pia{\widehat\Pi} 
\newcommand\Dap{\widehat\Delta^+}
\newcommand\Wa{\widehat{W}} 
\renewcommand\d{\delta}
\renewcommand\r{\mathfrak r}
\renewcommand\t{\otimes} 
\renewcommand\a{\alpha}
\renewcommand\aa{\mathfrak a} 
\renewcommand\b{\bb}
 
\renewcommand\k{\mathfrak k}
\newcommand\lie{\mathfrak l}
\newcommand\alie{\widehat \lie}
\newcommand\arho{\widehat\rho}

\renewcommand\th{\theta}

\newcommand\ganz{\mathbb Z}
 
\newcommand\s{\sigma}
\renewcommand\L{\Lambda}

\renewcommand\aa{\mathfrak a}
\newcommand\la{\langle} 
\newcommand\ra{\rangle}
\renewcommand\u{\mathfrak u} 
\newcommand\real{\mathbb R}
\newcommand\hard{\ha_\real^*} 
\newcommand\hauno{\ha^*_1}
\newcommand\hazero{\ha^*_0} 
\newcommand\huno{\h^*_1}
\newcommand\hzero{\h^*_0}

\newcommand\e{\epsilon} 
\newcommand\C{\mathbb C}
\newcommand\R{\mathbb R}

\newcommand\si{\sigma} 
\newcommand\Si{\Sigma} 

\newcommand\What{\widehat W}

\renewcommand\num{D_\g}

\renewcommand\rao{{{|\h_0}}} 
\renewcommand\ha{\widehat{\mathfrak h}}
\renewcommand\hard{\ha_\real^*} 
\renewcommand\hauno{\ha^*_1}
\renewcommand\hazero{\ha^*_0} 
\renewcommand\huno{\h^*_1}
\renewcommand\hzero{\h^*_0}
\newcommand\Wakmu{\Wa_{\k_\mu}}
\newcommand\Wakmusi{\Wa_{\k_\mu,\si}}

\newcommand{\fg}{\mathfrak{g}}

\newcommand{\fk}{\mathfrak{k}}

\newcommand\p{\mathfrak p}
\newcommand{\fs}{\mathfrak{s}}
\newcommand{\CC}{\mathbb{C}}
\newcommand{\fp}{\mathfrak{p}}
\newcommand{\rank}{{\rm rank}}

\newcommand{\ZZ}{\mathbb{Z}}

\catcode`\@=11
\font\tenln    = line10
\font\tenlnw   = linew10

\newskip\Einheit \Einheit=0.5cm
\newcount\xcoord \newcount\ycoord
\newdimen\xdim \newdimen\ydim \newdimen\PfadD@cke \newdimen\Pfadd@cke

\newcount\@tempcnta
\newcount\@tempcntb

\newdimen\@tempdima
\newdimen\@tempdimb

\newdimen\@wholewidth
\newdimen\@halfwidth

\newcount\@xarg
\newcount\@yarg
\newcount\@yyarg
\newbox\@linechar
\newbox\@tempboxa
\newdimen\@linelen
\newdimen\@clnwd
\newdimen\@clnht

\newif\if@negarg

\def\@whilenoop#1{}
\def\@whiledim#1\do #2{\ifdim #1\relax#2\@iwhiledim{#1\relax#2}\fi}
\def\@iwhiledim#1{\ifdim #1\let\@nextwhile=\@iwhiledim
        \else\let\@nextwhile=\@whilenoop\fi\@nextwhile{#1}}

\def\@whileswnoop#1\fi{}
\def\@whilesw#1\fi#2{#1#2\@iwhilesw{#1#2}\fi\fi}
\def\@iwhilesw#1\fi{#1\let\@nextwhile=\@iwhilesw
         \else\let\@nextwhile=\@whileswnoop\fi\@nextwhile{#1}\fi}

\def\thinlines{\let\@linefnt\tenln \let\@circlefnt\tencirc
  \@wholewidth\fontdimen8\tenln \@halfwidth .5\@wholewidth}
\def\thicklines{\let\@linefnt\tenlnw \let\@circlefnt\tencircw
  \@wholewidth\fontdimen8\tenlnw \@halfwidth .5\@wholewidth}
\thinlines
\def\PfadDicke#1{\PfadD@cke#1 \divide\PfadD@cke by2 \Pfadd@cke\PfadD@cke
\multiply\PfadD@cke by2}
\long\def\LOOP#1\REPEAT{\def\BODY{#1}\ITERATE}
\def\ITERATE{\BODY \let\next\ITERATE \else\let\next\relax\fi \next}
\let\REPEAT=\fi
\def\Punkt{\hbox{\raise-2pt\hbox to0pt{\hss$\ssize\bullet$\hss}}}
\def\DuennPunkt(#1,#2){\unskip
  \raise#2 \Einheit\hbox to0pt{\hskip#1 \Einheit
          \raise-2.5pt\hbox to0pt{\hss$\bullet$\hss}\hss}}
\def\NormalPunkt(#1,#2){\unskip
  \raise#2 \Einheit\hbox to0pt{\hskip#1 \Einheit
          \raise-3pt\hbox to0pt{\hss\twelvepoint$\bullet$\hss}\hss}}
\def\DickPunkt(#1,#2){\unskip
  \raise#2 \Einheit\hbox to0pt{\hskip#1 \Einheit
          \raise-4pt\hbox to0pt{\hss\fourteenpoint$\bullet$\hss}\hss}}
\def\Kreis(#1,#2){\unskip
  \raise#2 \Einheit\hbox to0pt{\hskip#1 \Einheit
          \raise-4pt\hbox to0pt{\hss\fourteenpoint$\circ$\hss}\hss}}

\def\Line@(#1,#2)#3{\@xarg #1\relax \@yarg #2\relax
\@linelen=#3\Einheit
\ifnum\@xarg =0 \@vline
  \else \ifnum\@yarg =0 \@hline \else \@sline\fi
\fi}

\def\@sline{\ifnum\@xarg< 0 \@negargtrue \@xarg -\@xarg \@yyarg -\@yarg
  \else \@negargfalse \@yyarg \@yarg \fi
\ifnum \@yyarg >0 \@tempcnta\@yyarg \else \@tempcnta -\@yyarg \fi
\ifnum\@tempcnta>6 \@badlinearg\@tempcnta0 \fi
\ifnum\@xarg>6 \@badlinearg\@xarg 1 \fi
\setbox\@linechar\hbox{\@linefnt\@getlinechar(\@xarg,\@yyarg)}%
\ifnum \@yarg >0 \let\@upordown\raise \@clnht\z@
   \else\let\@upordown\lower \@clnht \ht\@linechar\fi
\@clnwd=\wd\@linechar
\if@negarg \hskip -\wd\@linechar \def\@tempa{\hskip -2\wd\@linechar}\else
     \let\@tempa\relax \fi
\@whiledim \@clnwd <\@linelen \do
  {\@upordown\@clnht\copy\@linechar
   \@tempa
   \advance\@clnht \ht\@linechar
   \advance\@clnwd \wd\@linechar}%
\advance\@clnht -\ht\@linechar
\advance\@clnwd -\wd\@linechar
\@tempdima\@linelen\advance\@tempdima -\@clnwd
\@tempdimb\@tempdima\advance\@tempdimb -\wd\@linechar
\if@negarg \hskip -\@tempdimb \else \hskip \@tempdimb \fi
\multiply\@tempdima \@m
\@tempcnta \@tempdima \@tempdima \wd\@linechar \divide\@tempcnta \@tempdima
\@tempdima \ht\@linechar \multiply\@tempdima \@tempcnta
\divide\@tempdima \@m
\advance\@clnht \@tempdima
\ifdim \@linelen <\wd\@linechar
   \hskip \wd\@linechar
  \else\@upordown\@clnht\copy\@linechar\fi}

\def\@hline{\ifnum \@xarg <0 \hskip -\@linelen \fi
\vrule height\Pfadd@cke width \@linelen depth\Pfadd@cke
\ifnum \@xarg <0 \hskip -\@linelen \fi}

\def\@getlinechar(#1,#2){\@tempcnta#1\relax\multiply\@tempcnta 8
\advance\@tempcnta -9 \ifnum #2>0 \advance\@tempcnta #2\relax\else
\advance\@tempcnta -#2\relax\advance\@tempcnta 64 \fi
\char\@tempcnta}

\def\Vektor(#1,#2)#3(#4,#5){\unskip\leavevmode
  \xcoord#4\relax \ycoord#5\relax
      \raise\ycoord \Einheit\hbox to0pt{\hskip\xcoord \Einheit
         \Vector@(#1,#2){#3}\hss}}

\def\Vector@(#1,#2)#3{\@xarg #1\relax \@yarg #2\relax
\@tempcnta \ifnum\@xarg<0 -\@xarg\else\@xarg\fi
\ifnum\@tempcnta<5\relax
\@linelen=#3\Einheit
\ifnum\@xarg =0 \@vvector
  \else \ifnum\@yarg =0 \@hvector \else \@svector\fi
\fi
\else\@badlinearg\fi}

\def\@hvector{\@hline\hbox to 0pt{\@linefnt
\ifnum \@xarg <0 \@getlarrow(1,0)\hss\else
    \hss\@getrarrow(1,0)\fi}}

\def\@vvector{\ifnum \@yarg <0 \@downvector \else \@upvector \fi}

\def\@svector{\@sline
\@tempcnta\@yarg \ifnum\@tempcnta <0 \@tempcnta=-\@tempcnta\fi
\ifnum\@tempcnta <5
  \hskip -\wd\@linechar
  \@upordown\@clnht \hbox{\@linefnt  \if@negarg
  \@getlarrow(\@xarg,\@yyarg) \else \@getrarrow(\@xarg,\@yyarg) \fi}%
\else\@badlinearg\fi}

\def\@upline{\hbox to \z@{\hskip -.5\Pfadd@cke \vrule width \Pfadd@cke
   height \@linelen depth \z@\hss}}

\def\@downline{\hbox to \z@{\hskip -.5\Pfadd@cke \vrule width \Pfadd@cke
   height \z@ depth \@linelen \hss}}

\def\@upvector{\@upline\setbox\@tempboxa\hbox{\@linefnt\char'66}\raise
     \@linelen \hbox to\z@{\lower \ht\@tempboxa\box\@tempboxa\hss}}

\def\@downvector{\@downline\lower \@linelen
      \hbox to \z@{\@linefnt\char'77\hss}}

\def\@getlarrow(#1,#2){\ifnum #2 =\z@ \@tempcnta='33\else
\@tempcnta=#1\relax\multiply\@tempcnta \sixt@@n \advance\@tempcnta
-9 \@tempcntb=#2\relax\multiply\@tempcntb \tw@
\ifnum \@tempcntb >0 \advance\@tempcnta \@tempcntb\relax
\else\advance\@tempcnta -\@tempcntb\advance\@tempcnta 64
\fi\fi\char\@tempcnta}

\def\@getrarrow(#1,#2){\@tempcntb=#2\relax
\ifnum\@tempcntb < 0 \@tempcntb=-\@tempcntb\relax\fi
\ifcase \@tempcntb\relax \@tempcnta='55 \or
\ifnum #1<3 \@tempcnta=#1\relax\multiply\@tempcnta
24 \advance\@tempcnta -6 \else \ifnum #1=3 \@tempcnta=49
\else\@tempcnta=58 \fi\fi\or
\ifnum #1<3 \@tempcnta=#1\relax\multiply\@tempcnta
24 \advance\@tempcnta -3 \else \@tempcnta=51\fi\or
\@tempcnta=#1\relax\multiply\@tempcnta
\sixt@@n \advance\@tempcnta -\tw@ \else
\@tempcnta=#1\relax\multiply\@tempcnta
\sixt@@n \advance\@tempcnta 7 \fi\ifnum #2<0 \advance\@tempcnta 64 \fi
\char\@tempcnta}
\def\Diagonale(#1,#2)#3{\unskip\leavevmode
  \xcoord#1\relax \ycoord#2\relax
      \raise\ycoord \Einheit\hbox to0pt{\hskip\xcoord \Einheit
         \Line@(1,1){#3}\hss}}
\def\AntiDiagonale(#1,#2)#3{\unskip\leavevmode
  \xcoord#1\relax \ycoord#2\relax %\advance\xcoord by -0.05\relax
      \raise\ycoord \Einheit\hbox to0pt{\hskip\xcoord \Einheit
         \Line@(1,-1){#3}\hss}}
\def\Pfad(#1,#2),#3\endPfad{\unskip\leavevmode
  \xcoord#1 \ycoord#2 \thicklines\ZeichnePfad#3\endPfad\thinlines}
\def\ZeichnePfad#1{\ifx#1\endPfad\let\next\relax
  \else\let\next\ZeichnePfad
    \ifnum#1=1
      \raise\ycoord \Einheit\hbox to0pt{\hskip\xcoord \Einheit
         \vrule height\Pfadd@cke width1 \Einheit depth\Pfadd@cke\hss}%
      \advance\xcoord by 1
    \else\ifnum#1=2
      \raise\ycoord \Einheit\hbox to0pt{\hskip\xcoord \Einheit
        \hbox{\hskip-\PfadD@cke\vrule height1 \Einheit width\PfadD@cke
depth0pt}\hss}%
      \advance\ycoord by 1
    \else\ifnum#1=3
      \raise\ycoord \Einheit\hbox to0pt{\hskip\xcoord \Einheit
         \Line@(1,1){1}\hss}
      \advance\xcoord by 1
      \advance\ycoord by 1
    \else\ifnum#1=4
      \raise\ycoord \Einheit\hbox to0pt{\hskip\xcoord \Einheit
         \Line@(1,-1){1}\hss}
      \advance\xcoord by 1
      \advance\ycoord by -1
    \fi\fi\fi\fi
  \fi\next}
\def\hSSchritt{\leavevmode\raise-.4pt\hbox to0pt{\hss.\hss}\hskip.2\Einheit
  \raise-.4pt\hbox to0pt{\hss.\hss}\hskip.2\Einheit
  \raise-.4pt\hbox to0pt{\hss.\hss}\hskip.2\Einheit
  \raise-.4pt\hbox to0pt{\hss.\hss}\hskip.2\Einheit
  \raise-.4pt\hbox to0pt{\hss.\hss}\hskip.2\Einheit}
\def\vSSchritt{\vbox{\baselineskip.2\Einheit\lineskiplimit0pt
\hbox{.}\hbox{.}\hbox{.}\hbox{.}\hbox{.}}}
\def\DSSchritt{\leavevmode\raise-.4pt\hbox to0pt{%
  \hbox to0pt{\hss.\hss}\hskip.2\Einheit
  \raise.2\Einheit\hbox to0pt{\hss.\hss}\hskip.2\Einheit
  \raise.4\Einheit\hbox to0pt{\hss.\hss}\hskip.2\Einheit
  \raise.6\Einheit\hbox to0pt{\hss.\hss}\hskip.2\Einheit
  \raise.8\Einheit\hbox to0pt{\hss.\hss}\hss}}
\def\dSSchritt{\leavevmode\raise-.4pt\hbox to0pt{%
  \hbox to0pt{\hss.\hss}\hskip.2\Einheit
  \raise-.2\Einheit\hbox to0pt{\hss.\hss}\hskip.2\Einheit
  \raise-.4\Einheit\hbox to0pt{\hss.\hss}\hskip.2\Einheit
  \raise-.6\Einheit\hbox to0pt{\hss.\hss}\hskip.2\Einheit
  \raise-.8\Einheit\hbox to0pt{\hss.\hss}\hss}}
\def\SPfad(#1,#2),#3\endSPfad{\unskip\leavevmode
  \xcoord#1 \ycoord#2 \ZeichneSPfad#3\endSPfad}
\def\ZeichneSPfad#1{\ifx#1\endSPfad\let\next\relax
  \else\let\next\ZeichneSPfad
    \ifnum#1=1
      \raise\ycoord \Einheit\hbox to0pt{\hskip\xcoord \Einheit
         \hSSchritt\hss}%
      \advance\xcoord by 1
    \else\ifnum#1=2
      \raise\ycoord \Einheit\hbox to0pt{\hskip\xcoord \Einheit
        \hbox{\hskip-2pt \vSSchritt}\hss}%
      \advance\ycoord by 1
    \else\ifnum#1=3
      \raise\ycoord \Einheit\hbox to0pt{\hskip\xcoord \Einheit
         \DSSchritt\hss}
      \advance\xcoord by 1
      \advance\ycoord by 1
    \else\ifnum#1=4
      \raise\ycoord \Einheit\hbox to0pt{\hskip\xcoord \Einheit
         \dSSchritt\hss}
      \advance\xcoord by 1
      \advance\ycoord by -1
    \fi\fi\fi\fi
  \fi\next}
\def\Koordinatenachsen(#1,#2){\unskip
 \hbox to0pt{\hskip-.5pt\vrule height#2 \Einheit width.5pt depth1 \Einheit}%
 \hbox to0pt{\hskip-1 \Einheit \xcoord#1 \advance\xcoord by1
    \vrule height0.25pt width\xcoord \Einheit depth0.25pt\hss}}
\def\Koordinatenachsen(#1,#2)(#3,#4){\unskip
 \hbox to0pt{\hskip-.5pt \ycoord-#4 \advance\ycoord by1
    \vrule height#2 \Einheit width.5pt depth\ycoord \Einheit}%
 \hbox to0pt{\hskip-1 \Einheit \hskip#3\Einheit
    \xcoord#1 \advance\xcoord by1 \advance\xcoord by-#3
    \vrule height0.25pt width\xcoord \Einheit depth0.25pt\hss}}
\def\Gitter(#1,#2){\unskip \xcoord0 \ycoord0 \leavevmode
  \LOOP\ifnum\ycoord<#2
    \loop\ifnum\xcoord<#1
      \raise\ycoord \Einheit\hbox to0pt{\hskip\xcoord \Einheit\Punkt\hss}%
      \advance\xcoord by1
    \repeat
    \xcoord0
    \advance\ycoord by1
  \REPEAT}
\def\Gitter(#1,#2)(#3,#4){\unskip \xcoord#3 \ycoord#4 \leavevmode
  \LOOP\ifnum\ycoord<#2
    \loop\ifnum\xcoord<#1
      \raise\ycoord \Einheit\hbox to0pt{\hskip\xcoord \Einheit\Punkt\hss}%
      \advance\xcoord by1
    \repeat
    \xcoord#3
    \advance\ycoord by1
  \REPEAT}
\def\Label#1#2(#3,#4){\unskip \xdim#3 \Einheit \ydim#4 \Einheit
  \def\lo{\advance\xdim by-.5 \Einheit \advance\ydim by.5 \Einheit}%
  \def\llo{\advance\xdim by-.25cm \advance\ydim by.5 \Einheit}%
  \def\loo{\advance\xdim by-.5 \Einheit \advance\ydim by.25cm}%
  \def\o{\advance\ydim by.25cm}%
  \def\ro{\advance\xdim by.5 \Einheit \advance\ydim by.5 \Einheit}%
  \def\rro{\advance\xdim by.25cm \advance\ydim by.5 \Einheit}%
  \def\roo{\advance\xdim by.5 \Einheit \advance\ydim by.25cm}%
  \def\l{\advance\xdim by-.30cm}%
  \def\r{\advance\xdim by.30cm}%
  \def\lu{\advance\xdim by-.5 \Einheit \advance\ydim by-.6 \Einheit}%
  \def\llu{\advance\xdim by-.25cm \advance\ydim by-.6 \Einheit}%
  \def\luu{\advance\xdim by-.5 \Einheit \advance\ydim by-.30cm}%
  \def\u{\advance\ydim by-.30cm}%
  \def\ru{\advance\xdim by.5 \Einheit \advance\ydim by-.6 \Einheit}%
  \def\rru{\advance\xdim by.25cm \advance\ydim by-.6 \Einheit}%
  \def\ruu{\advance\xdim by.5 \Einheit \advance\ydim by-.30cm}%
  #1\raise\ydim\hbox to0pt{\hskip\xdim
     \vbox to0pt{\vss\hbox to0pt{\hss$#2$\hss}\vss}\hss}%
}
\catcode`\@=12

\begin{document}

\title{Decomposition rules for conformal pairs
associated to symmetric spaces and abelian subalgebras of $\ganz_2$-graded
Lie algebras} \author{Paola Cellini\\ Victor G.
Kac\\Pierluigi M\"oseneder Frajria\\Paolo Papi}
\date{}
\maketitle
\begin{abstract}
We give uniform formulas
for the branching rules of level 1 modules over orthogonal affine Lie algebras
for all conformal pairs associated to symmetric spaces. We also 
provide a combinatorial
intepretation of these formulas in terms of certain abelian subalgebras of     
simple Lie algebras.
\end{abstract}

\section{Introduction}
\label{intro}

A pair $(\fs ,\fk)$, where $\fs$ is a finite-dimensional
semisimple Lie algebra over $\CC$ and $\fk$ is a reductive
subalgebra of $\fs$, such that the restriction of the Killing form of
$\fs$ to $\fk$ is non-degenerate, is called a \emph{conformal
  pair} if there exists an integrable highest weight module $V$
over the affine Kac--Moody algebra $\widehat{\fs}$, faithful on
each simple component of $\fs$, such that the restriction to $\widehat{\fk}$ of
each weight space of the center of $\fk$ in $V$ decomposes into a
finite direct sum of irreducible $\widehat{\fk}$-modules.  In such a
case $\fk$ is called a \emph{conformal subalgebra} of $\fs$.

It is well-known that any integrable highest weight
$\widehat{\fs}$-module, when restricted to $\widehat{\fk}$, decomposes
into a direct sum of irreducible $\widehat{\fk}$-modules \cite{Kac},
but almost always this decomposition is infinite.

The first cases of a finite decomposition, found in \cite{KacPeterson},
are as follows.  Let $\fg =\fk \oplus \fp$ be the eigenspace
decomposition of an inner involution of a simple Lie algebra
$\fg$ such that $\fk$ is semisimple.  This defines an embedding
$\fk \subset so (\fp)$.  It was shown by Kac and Peterson in
\cite{KacPeterson},
by an explicit decomposition formula, that the restriction of the
spinor representations of $\widehat{so(\fp)}$ to $\widehat{\fk}$ is a
finite direct sum of irreducible $\widehat{\fk}$-modules.  Thus, the
pair $(so (\fp), \fk)$ is conformal.

Due to their importance for string compactifications, a series of
papers on conformal pairs appeared in the second half of the
1980s in physics literature.  First of all, a connection to
representation theory of the Virasoro algebra was established.
Namely, it was found that the decomposition in question is finite
if and only if the following numerical criterion holds:  the
central charges of the Sugawara construction of the Virasoro
algebra for $\widehat{\fs}$ and $\widehat{\fk}$ are equal \cite{GNO}.
This immediately has led to a conclusion: the decomposition in
question has a chance to be finite only if the level of the
$\widehat{\fs}$-module $V$ is equal to~$1$, and if it is finite for
one of the $\widehat{\fs}$-modules of level~$1$, it is also finite
for all others.  Furthermore, Goddard, Nahm and Olive show
\cite{GNO} that the observation of Kac and Peterson can be
reversed.  Namely all conformal pairs $(so (\fp), \fk)$ are
obtained from an involution (not necessarily inner) of a semisimple Lie algebra $\fg$,
and all such pairs are conformal.  However, they obtain this
result using the above numerical criterion, and do not find
actual decompositions.

All conformal subalgebras $\fk$ for all simple Lie algebras $\fs$
were classified in \cite{BB} and \cite{SW} by making use of the
numerical criterion.  Also, it was pointed out in \cite{AGO}
that, using the conformal pairs $(so_{2n}, g\ell_n)$ and
$(so_{4n}, sp_{2n} \times s\ell_2)$, one can reduce the study of
conformal subalgebras in all classical Lie algebras to that in
$so_n$.

Around the same time the general problem of restricting
representations of affine Lie algebras to their subalgebras was
treated, using the theory of modular forms.  Namely it was
observed in \cite{KP} that the branching rules are described by
certain modular functions, called branching functions, for which
one can write down explicit transformation formulas.  This idea
was further developed in \cite{KacW}, where the above mentioned
``modular constraints'', along with the ``conformal
constraints'', provided by the Virasoro algebra, allowed to
compute easily branching functions (which are constants in the conformal
pair case) in many interesting cases, and, in principle, in any given case.
The technology, developed
in \cite{KacW} was subsequently used in \cite{KS} to find
all the decompositions of all integrable highest weight modules
of level~$1$ over affine Lie algebras $\widehat{\fs}$, restricted to
affine subalgebras $\widehat{\fk}$, where $\fk$ is a conformal
subalgebra of a simple exceptional Lie algebra $\fs$.  The
branching rules of some other conformal embeddings were
subsequently found in \cite{ABI}, \cite{LL} and a few other papers,
written in the 1990s.

The problem of finding a general conceptual formula for branching
rules for level~$1$ integrable highest weight modules over
~$\widehat{so(\p)}$, when restricted to $\widehat{\fk}$, where the
conformal embedding of $\fk$ in $so(\p)$ is defined by the
eigenspace decomposition of an involution of a semisimple Lie
algebra $\fg = \fk + \fp$, has remained an
open problem.

In the present paper we completely solve this problem.  The
solution turned out to be intimately related to recent
developments in the study of abelian subalgebras of simple Lie
algebras, that began with a paper of Kostant \cite{Kostant1} and
continued in \cite{IMRN}, \cite{CP}, \cite{CP2}, \cite{CP3},
\cite{Kostant2}, \cite{Pan1}, \cite{Pan2}, \cite{Suter}.

Let us explain our main observation on the example of a conformal
embedding $\fk \subset so (\fk)$, where $\fk$ is a simple Lie
algebra, via the adjoint representation of $\fk$.  In this case
the restriction to $\widehat{\fk}$ of the basic $+$~vector
representations of $\widehat{so (\fk)}$ decomposes into a direct
sum of $2^{\rank\,\fk}$ irreducible $\widehat{\fk}$-modules (this
decomposition was found already by Kac and Wakimoto \cite{KacW}), and,
remarkably, these $2^{\rank\,\fk}$ $\widehat{\fk}$-modules are in a
canonical one-to-one correspondence with all abelian ideals of a
Borel subalgebra of $\fk$.

Our main result is that for all conformal pairs associated to
symmetric spaces, the decomposition of level~$1$ modules over
$\widehat{so(\fp)}$ is described in terms of a certain class of
abelian subalgebras of semisimple $\ZZ_2$-graded Lie algebras $\fg=\fk
\oplus \fp$, studied and classified recently by Cellini,
M\"oseneder Frajria and Papi \cite{IMRN}.

We hope that the connection of representation theory of affine
Lie algebras to the theory of abelian subalgebras of simple Lie
algebras will shed a new light on the latter theory as well.  So
far we obtained only partial results in this direction.
\vskip10pt
Now we describe our results in a special case which might give the flavour of the general case. 
First remark that $\widehat{so(\p)}$ is an affine algebra of type
$B^{(1)}$ or $D^{(1)}$ according to whether $\dim(\p)$ is odd or even.
Hence the level $1$ modules are the fundamental representations
associated to the extremal nodes of the Dynkin diagram. They are the 
basic, vector and spin representations, and have been studied since a long time.
\par
We consider in detail  the case of the basic and vector representation of $\ka$
and furthermore we assume that $\k$ is semisimple. Denote  by 
$L(\tilde\L_0)$ the basic representation, by $L(\tilde\L_1)$ the vector representation and set
$L=L(\tilde\L_0)+L(\tilde\L_1)$. The first step in our analysis consists in calculating the
character of the $\ka$-module $L$. This is done using the explicit description of the 
action given in \cite{KacPeterson}. To be more precise we need to fix some notation. Let $\h_0$ be a
 Cartan subalgebra of $\k$; denote
by $\D_\k$ the set of $\h_0$-roots of $\k$ and by $\D(\p)$ the set
of $\h_0$-weights of $\p$. Fix a set of positive roots $\Dp_\k$ and let 
$\b_0$ be the corresponding Borel subalgebra. 
Let $\si$ denote the involution which induces  the decomposition 
$\g=\k\oplus\p$ and denote by $\hat L(\g,\si)$ the extended
 loop algebra associated to the pair $(\g,\si)$ (see \cite{Kac}, Ch.~8) and by $\Wa$ its Weyl
group. The choice of
$\Dp_\k$ induces natural choices $\Dap,\,\Dap_\k$ for the positive
roots of $\hat L(\g,\si),\,\ka$ respectively. 
Here, as above, $\ka$ denotes the untwisted affine algebra associated to $\k$; we also 
set $\d_\k$ to be its fundamental imaginary root and $\Wa_\k$ its  Weyl group. 
Finally let $\Dp$ denote a set of positive roots (w.r.t. the centralizer of $\h_0$ in $\g$) of 
$\g$ compatible with that of $\k$ (see \ref{compatibile}).
This allows us to define
\begin{align*}
&\Dp(\p)=\Dp_{|\h_0}\cap\D(\p),\\
&\Dap(\p)=\{(m+\frac{1}{2})\d_\k+\a\mid\a\in\Dp(\p),\,m\in\ganz\}.
\end{align*}
Then it turns out that, up to an exponential factor, $ch(L)$ equals
\begin{equation*}
\prod_{\a\in\Dap(\p)}\left(1+e^{-\a}\right)^{mult(\a)}.
\end{equation*}
(see \eqref{carattere}). To extract from the previous formula information on the $\ka$-module structure
of
$L$ we generalize an idea used in \cite{Parthasaraty}  in the finite dimensional equal rank case. We
introduce a natural  map $\psi_0:\ka\to \widehat L(\g,\si)$, whose transpose induces a bijection
$\psi^*_0:\Da\leftrightarrow\Dap_\k\cup\Dap(\p)$. Using this map and the Weyl-Kac character formula we get the following decomposition into irreducible
$\ka$-modules
\begin{equation}\label{uno}L(\tilde\L_\epsilon)=\sum_{{u\in W'_{\si,0}}\atop{\ell(u)\equiv
\epsilon\,mod 2}}L(\psi^*_0(u\rhat)-\rhat_\k+\frac{1}{2}\epsilon\d_\k).\end{equation} Here
$W'_{\si,0}$ is the set of minimal right coset representatives of 
$(\psi^*_0)^{-1}\Wa_\k\psi^*_0$ in $\Wa$, $\rhat,\rhat_\k$ are the sum of fundamental weights in $\Dap, \Dap_\k$ and
$\epsilon=0,1$. 
A more accurate statement of formula \eqref{uno} is given in Theorem \ref{decompositioneven}.
\vskip5pt
The combinatorial interpretation of formula \eqref{uno} arises from the fact
that, if $C_1$ denotes the fundamental alcove of $\Wa$, then the set 
$\bigcup_{w\in W'_{\si,0}}w\ov{C_1}$ is the polytope studied in \cite{IMRN}
in connection with the problem of enumerating $\b_0$-stable abelian subalgebras of $\p$.
This coincidence allows us to give a much more explicit rendering of formula \eqref{uno}.
For instance, if $\si$ is an automorphism of type $(0,\ldots,1, \ldots,0;1)$ with $1$
in a  position corresponding to a short root, then we have 
$$
L(\tilde\L_\epsilon)=\bigoplus_{A\in \Sigma\atop |A|\equiv
\epsilon\,\,mod\,2}L\left(\L_{0,\k}+\langle
A\rangle-\frac{1}{2}(|A|-\epsilon)\d_\k\right),$$
where $\Sigma$ is the set of $\b_0$-stable abelian subalgebras of $\p$, and for
$A\in\Sigma$, $\langle A\rangle,\,|A|$ denote the sum (resp. the number) of the roots in
$A$. The general case is completely
described in Theorem \ref{decoeabeliani}.\par The situation in the case of the spin representation
($\k$ semisimple) is  much more complex. For instance, we need to use the factorization of the
involution  $\sigma$  as
$\mu\eta$ with 
$\eta$ inner and $\mu$ diagram automorphism to write down
the auxiliary map 
$\psi_1:\ka\to L'(\g,\sigma)$ which plays the role of $\psi_0$ in the spin case and which is the crucial tool for manipulating the character. Moreover the target algebra
$L'(\g,\sigma)$ is different according to whether we are considering the equal rank
case, the $A_{2n}^{(2)}$ case or the remaining non equal rank cases.
Surprisingly enough, we obtain a decomposition formula which is  quite similar
to  \eqref{uno}:
\begin{equation}\label{due}
ch(X_r)=2^{\lfloor\frac{N-n}{2}\rfloor} \sum\limits_{u\in W'_{\si,\rodd}}
ch(L(a_0\stodd(u\rhat') -
\rhat_\k)).\end{equation}
We refer the reader to Proposition \ref{spindec} for the undefined notation. As far 
as the combinatorial interpretation is concerned, we get again a description of formula \eqref{due}
in terms of  abelian subalgebras, but in the non equal rank the right 
class to consider is that of  {\it noncompact} subalgebras (cf. Definition
\ref{noncomp}) which are stable under the Borel subalgebra
$\b_0\cap\k\cap\k_\mu$ of the subalgebra $\k\cap\k_\mu$ of $\mu$-fixed points in $\k$. The relevant
results in this direction are Theorems
\ref{decoeabelianispin}, \ref{decoeabelner}, \ref{tipoa22n}.\par
A few words on the case in which $\k$ has a non-trivial center. In this case we 
 describe  the finite decomposition of an eigenspace of the
center on the level $1$ modules. Also in this case there is a special subset of 
abelian stable subalgebras of $\p$ which plays an important role in the description of
the decompositions. We describe in detail the finite decomposition of each 
eigenspace of the center (see Theorems \ref{finalehs}, \ref{finale2hs}).
\vskip10pt
The paper is organized as follows.  In Section~\ref{preliminaries} we
recall the
necessary information on the structure and representation theory
of affine Lie algebras, as well as the construction of all
level~$1$ modules over the affinization of orthogonal Lie
algebras $so_n$.
In Section~\ref{basicandvector} we find the
decompositions of the basic $+$~vector representations of
$\widehat{so(\p)}$, restricted to $\widehat{\fk}$, $\k$ semisimple, and in
Section~\ref{spin}
 we solve the same problem for the spinor representations.  In
 Section~\ref{hermitian}  we deal with the case when $\k$ has
a non-trivial center.
Finally in Section~\ref{examples} we consider some concrete examples and discuss
 connections  of the theory of
 abelian subalgebras  to modular invariance.

    \section{Preliminaries}\label{preliminaries}
 \subsection{Lie algebra involutions and
affine algebras} \label{prelims}Let $\g$ be a semisimple Lie algebra over
$\C$, $\si$
an involutive automorphism of $\g$ and let $\g=\k\oplus\p$ be the
corresponding eigenspace decomposition. Let $\h_0$ be a Cartan
subalgebra  of the reductive subalgebra $\k$ and let $\mathfrak z$ be the
centralizer of
$\h_0$ in $\g$. Let
$(\cdot,\cdot)$ denote a non-degenerate invariant symmetric bilinear  form on $\g$.
Then  (see \cite{Helgason1}, Lemma 5.3 or
\cite{Kac} Lemma 8.1),  $\mathfrak
z$ is a Cartan subalgebra of $\g$ and $(\k,\p)=0$. In particular
$(\cdot,\cdot)_{|\k\times\k}$ is a nondegenerate invariant form on $\k$
and  $(\cdot,\cdot)$
is nondegenerate when restricted to $\h_0$, so we can induce a form,
still denoted by $(\cdot,\cdot)$ on $\h_0^*$.

    Consider the root system $\D_\k$ of the pair $(\k,\h_0)$ and fix a
subset of positive roots $\Dp_\k$. Let $\b_0$  denote the corresponding
Borel subalgebra of $\k$.
We denote by  $\D(\p)$ the set
of $\h_0$-weights of $\p$.

    Let $L(\g)$ be the loop algebra of $\g$: $$ L(\g)=\C[t,t^{-1}]\t \g.
$$ Let $L(\g,\si)$ be the subalgebra $$
L(\g,\si)=\(\sum_{n\in\ganz,\ n\text{ even}}t^n\t \k\)\oplus\(
\sum_{n\in\ganz,\ n\text{ odd} }t^n\t\p\) $$ and consider the extended
loop algebra $ \widehat L(\g)=L(\g)\oplus\C K'\oplus\C d'$ with bracket
defined by \begin{align} [t^n\t &X+\l K'+\mu d',t^m\t
Y+\l_1K'+\mu_1d']=\notag\\&=t^{n+m}\t[X,Y]+\d_{n,-m}n(X,Y)K' +\mu_1nt^n\otimes
X+\mu mt^m\t Y.\label{bracket} \end{align} (The construction of the
extended loop
algebra is done in \cite{Kac} only for $\g$ simple, but everything
extends to semisimple $\g$ in a straightforward way). Set $ \widehat
L(\g,\sigma)=L(\g,\si)\oplus\C K'\oplus\C d'$. Clearly $\widehat
L(\g,\si)$ is a subalgebra of $\widehat L(\g)$.

    Set $\ha=(1\t\h_0)\oplus\C K'\oplus\C d'$ and let $\Da$  denote the
set of nonzero $\ha$-weights of $\widehat L(\g,\si)$. Define
$\d'\in\ha^*$ by setting $$ \d'(1\t \h_0)=\d'(K')=0\quad\d'(d')=1.$$

    We identify $\hzero$ and the subset $ \{ \l\in\ha^*\mid
\l(d')=\l(K')=0\}$ of $\ha^*$.\par
\noindent{\bf Notation.} If $\l\in\ha^*$ we denote by
$\overline\l$ its restriction to $1\t\h_0$.

    The set of roots of $\widehat L(\g,\si)$ is \begin{align*}
\Da=&\{k\d'+\a\mid \a\in\D_\k,\ k\text{ even}\}\cup\{k\d'+\a\mid
\a\in\D(\p),\ k\text{ odd}\}\\&\cup\{k\d'\mid k\in 2\ganz,\ k\ne0\}.
\end{align*} We set $\Dap=\Dp_\k\cup\{\a\in\Da\mid \a(d')>0\}$. Let
$\Pia$ be the corresponding set of simple roots. Denote by $\Wa$ the  Weyl group generated by $\Pia$.\par Assume now that
$\si$ is indecomposable (i.e. $\g$ has no non-trivial $\si$-stable ideals).
Then either
$\g$
is simple, or $\g$ is a direct sum of two copies of a simple Lie algebra
$\k$ and
$\s$
permutes the summands. We call the latter the {\it complex} case.\par The
following
two propositions give a summary of Exercises 8.1--8.4 from
\cite{Kac}. The
proof can be found in \S~5 of \cite{Helgason1}, Ch.X.
\begin{prop}\label{isaffine} \begin{enumerate} \item
$|\Pia|=n+1$, where $n$ is the rank of $\k$. \item
If $\Pia=\{\a_0,\dots,\a_n\}$, then 
$\overline\a_0,\dots,\overline\a_n$ span
$\h_0$.\item\label{iscartan}$(\overline\a_i,\overline\a_i)>0$ for all $i$ and
$$
a_{ij}=2\frac{(\overline\a_i,\overline\a_j)}{(\overline\a_i,\overline\a_i)}\in-
\ganz_+,\text{ if $i\ne j$}. $$ \item  $A=(a_{ij})$ is a generalized Cartan
matrix of an affine type. \end{enumerate} \end{prop}

We label the $\a_i$ so that the corresponding Dynkin  diagram is
one of those displayed at  pp. 54--55 of \cite{Kac}.

If $\a\in\Da$, then $\a$ can be written uniquely as $\sum_{i=0}^n
m_i(\a)\a_i$ with $m_i(\a)\in\ganz$.
    Write $\a_i=s_i\d'+\overline\a_i$. By Proposition
\ref{isaffine}.\ref{iscartan}, $\overline\a_i\ne0$. Let
$h_{\overline\a_i}$ be the unique element of $\h_0$ such that
$\overline\a_i(h)=(h_{\overline\a_i},h)$ for all $h\in\h_0$. Set $
h_i=\frac{2}{(\overline\a_i,\overline\a_i)}h_{\overline\a_i}$ and fix
$t^{s_i}\t X_i\in \widehat L(\g,\si)_{\a_i}$, $t^{-s_i}\t Y_i\in
\widehat L(\g,\si)_{-\a_i}$ in such a way that
$(X_i,Y_i)=\frac{2}{(\overline\a_i,\overline\a_i)}$. Then
$[X_i,Y_i]=h_i$. It follows that $$ [t^{s_i}\t X_i,t^{-s_i}\t
Y_i]=\frac{2s_i}{(\overline\a_i,\overline\a_i)}K'+h_i. $$ Set
$\a_i^\vee=\frac{2s_i}{(\overline\a_i,\overline\a_i)}K'+h_i$ and
$\Pia^\vee=\{\a_0^\vee,\ldots,\a_n^\vee\}$. In the following proposition 
we use the notation of \cite{Kac}, Ch.~1.
\begin{prop}\label{iso}
The triple $(\ha,\Pia,\Pia^\vee)$ is a realization of $A$ and the
map $$ e_i\mapsto t^{s_i}\t X_i\quad f_i\mapsto t^{-s_i}\t Y_i $$
extends  to a Lie algebra  isomorphism  of the affine
Kac-Moody
algebra $
\g(A)$ to the Lie algebra $\widehat L(\g,\si).$
\end{prop}

Let $a_0,
\dots, a_n$ (resp. $a^\vee_0, \dots, a^\vee_n$) be
positive integers with $G.C.D(a_0,\ldots,$ $a_n)=1$ that are coefficients of a linear
dependence between the rows (resp. columns) of the matrix $A$:
$\sum\limits_{i=0}^na_i\overline\a_i=0$ and
$\sum\limits_{i=0}^na^\vee_ih_i=0$.
Set $ \d=\sum_{i=0}^na_i\a_i $ and
notice that $\d=(\sum_{i=0}^na_is_i)\d'$. We also let $K=\sum_{i=0}^n
a_i^\vee\a_i^\vee$
 be the canonical central element.

    \begin{prop}\label{parametri} Set $k=\frac{2}{\sum_{i=0}^na_is_i}$.
Then 
$k=1$ if $\sigma$ is inner and $k=2$ otherwise.\end{prop}
\begin{proof}
Since $2\d'=k\d$ is a root (cf. \cite{Kac}, Theorem 5.6~b)) we deduce that
$k$ is an integer, hence $k\in\{1,2\}$. Since $\d'=\frac{k}{2}\d$ is a
root if and only if $\mathfrak z\ne\h_0$, we see that $k=2$ if and only
if $\si$ is not of inner type.
\end{proof}

    \begin{rem} If $\g$ is simple of type $X_N$, then $\widehat
L(\g,\si)$ is an
affine
Kac-Moody algebra of type
$X^{(k)}_N$.
If $\g=\k\oplus \k$, where $\k$  is simple of type $X_N$, then
 $\widehat L(\g,\si)$ is of type
$X^{(1)}_N$.
  \end{rem}

\begin{rem}\label{simple} If $\g$ is simple, using the terminology of
\cite{Kac}, we have that  $\sigma$ is an automorphism of type $(s_0,\ldots,s_n;k)$.
In the complex case we have $k=2,\,s_0=1$.\end{rem}

 If $\g$ is simple, we choose $(\cdot,\cdot)=k^{-1}(\cdot,\cdot)_n$, where
 $(\cdot,\cdot)_n$ is the invariant form on $\g$ 
such  that the square root lenght of a long
root is
$2$. We will call $(\cdot,\cdot)_n$ a {\it normalized} invariant form. If
$\g=\k\times \k$ we define
 $(\cdot,\cdot)$  by
$$((X,Y),(X',Y'))=\half((X,X')_n+(Y,Y')_n),$$ where
 $(\cdot,\cdot)_n$
 is the normalized invariant form on  $\k$.

We define a  standard invariant  form $(\cdot,\cdot)$ on
$\widehat
L(\g,\si)$ by setting 
\begin{align}\label{forma1}
&(\a_i^\vee,h)=\frac{2}{(\overline\a_i,\overline\a_i)}\a_i(h)\quad\text{for
$i=0, \dots,n$ and $h\in\ha$}\\ \label{forma2}
&(d',d')=0.\end{align}

 We want to prove that the previous formulas define a normalized invariant form on $\widehat L(\g,\sigma)$, i.e.
$(\theta,\theta)=2$, where $\theta=\sum_{i=1}^na_i\a_i$.
  Let $\nu:\ha\to\ha^*$ be the induced isomorphism and let
$(\cdot,\cdot)$ be the induced form on $\ha^*$. Since $$
\a_i(h)=\frac{(\overline\a_i,\overline\a_i)}{2}(\a_i^\vee,h) $$ we see
that $\nu^{-1}(\a_i)=\frac{(\overline\a_i,\overline\a_i)}{2}\a_i^\vee$.
It follows that $$ (\a_i,\a_j)=(\overline\a_i,\overline\a_j), $$ hence,
if $\l,\mu\in\sp(\Pia)$, then
\begin{equation}\label{prodottoscalare}
 (\l,\mu)=(\overline\l,\overline\mu).
\end{equation}

  We need to discuss the relationship between the  roots of $\g$ and the
 roots of $\widehat L(\g,\sigma)$. Write $\mathfrak z=\h_0\oplus\aa_\C$, where
$\aa_\C=\p\cap\mathfrak z$.
Let $\D$ be the $\mathfrak z$-root system of $\g$.  There are three types
of roots in $\D$:
those such that $\a_{|\aa_\C}=0$ and whose root vector $X_\a$ is in
$\k$, those  such that $\a_{|\aa_\C}=0$ and whose root vector $X_\a$ is in
$\p$,  and those such
that $\a_{|\aa_\C}\ne0$.  These are usually called respectively
 compact imaginary roots, noncompact imaginary (or singular imaginary) roots,
 and complex roots.
 To avoid confusion with standard Kac-Moody terminology we call them
compact,
 noncompact,
 and complex.
  If $\a$ is a complex root, then the
corresponding root vector decomposes as $$ X_\a=u_\a+v_\a $$ with
$u_\a\in\k$ and $v_\a\in\p$. Then $u_\a$ is a root vector in $\k$ for
the root $\a_{|\h_0}$ and $v_\a$ is a weight vector in  $\p$ for the
weight $\a_{|\h_0}$ in $\D(\p)$. In particular $\a$ is a complex root if
and only if $\a_{|\h_0}\in\D_\k\cap\D(\p)$.
 It follows that $\a\in\D$ is a compact root if and only if $\a_\rao\in\Da$
and
 $\d'+\a_\rao\not\in\Da$,
 $\a\in\D$ is a noncompact root if and only if $\a_\rao\not\in\Da$ and
$\d'+\a_\rao\in\Da$, and
 $\a\in\D$ is a complex root if and only if $\a_\rao\in\Da$ and
$\d'+\a_\rao\in\Da$.
 More precisely if $k=1$, then $\h_0=\mathfrak z$ hence, if $\a\in\Da$, then
$\ov\a=\be_\rao$
 with $\be$
 compact or noncompact. It follows that $(\a,\a)=(\ov\a,\ov\a)=(\be,\be)_n$.
 In particular, since $\a_0$ is a long root, $(\a_0,\a_0)=2$.
 If $k=2$ and $\g$ is simple, then $\d'=\d$ hence, if $\a\in\Da$, then  $\ov\a=\be_\rao$ with
$\be$
 compact or noncompact if and only if $\a$ is a long root. It follows that
 if $\a$ is a long root and $\ov\a=\be_\rao$, then
 $(\a,\a)=(\ov\a,\ov\a)=k(\be,\be)_n=4$. In particular, since $\a_0$ is a
short root,
 $a_0(\a_0,\a_0)=2$.
 In the complex case one checks directly that $(\a_0,\a_0)=2$ in this case
too.
 We have proven
 \begin{lemma}\label{normalized}
 The form $(\cdot,\cdot)$ on $\widehat L(\g,\si)$ defined above
 is a normalized standard invariant form.
 \end{lemma}

  Since $(d',K)=\sum
a^\vee_i\frac{2}{(\a_i,\a_i)}\a_i(d')
 =\sum a^\vee_i\frac{2s_i}{(\a_i,\a_i)}$, and $K=\sum_{i=0}^n\frac{2a_i^\vee s_i}{(\a_i,\a_i)}K'$ we see that
$(d',K')=1$. Also remark that
\begin{equation}\label{diprimo}
(d',h)=0\text{ if $h\in\h_0$}
\end{equation}
which is easily proved by observing that
$$(d',h_i)=(d',\a^\vee_i-2\frac{s_i}{(\a_i,\a_i)}K')=\frac{2}{(\a_i,\a_i)}\a_i(d'
)-2\frac{s_i}{(\a_i,\a_i)}=0.$$

Let $H$ be the unique element of $\h_0$ such that $\ov{\a_i}(H)=s_i$
 for $i=1,\dots,n$.
Then easy calculations show that  $d=\frac{a_0k}{2}(d'-H-\half(H,H)K')$ is a scaling element for
$\widehat L(\g,\si)$ and
$(d,d)=0$. It follows that, if we
define $\L_0\in\ha^*$  by setting $\L_0(\a^\vee_i)=\d_{i0}$ and
$\L_0(d)=0$,
then the form $(\cdot,\cdot)$ on $\widehat L(\g,\si)$ is given by the
formulas of
\cite[\S~6.2]{Kac}. In particular we see that $\sum
2\frac{s_ia_i^\vee}{(\a_i,\a_i)}=\sum
a_is_i=\frac{2}{k}$. Hence $K=\frac{2}{k}K'$.

    \subsection{The Lie algebra $\ka$ and the character formula}
    In general $\k$ is a reductive Lie algebra, hence we can write
    $\k=\k_0\oplus\sum_{S=1}^M\k_S$, where $\k_0$ is the center of $\k$
and $\k_S$
    are the simple ideals of $\k$.  If $S>0$, we denote  by $\Pi_S$ the
set of simple roots of $\k_S$. Let also
$W_S$ be the relative Weyl group: $W_S=\langle s_\a\mid
\a\in\Pi_S\rangle$, $\D_S$ the relative root system,
$\D_S=W_S\Pi_S$, and $\th_S$  the highest root of $\D_S$.
    We recall that the dimension of $\k_0$ is at most one.

    We define the affine Lie algebra $\ka$ as follows.
    Consider
the standard loop algebra $\tilde \k=L(\k)=\oplus_{S} L(\k_S)$. On each
simple ideal $\k_S$ let $(\cdot,\cdot)_S$ be the normalized invariant form.
Set also $(\cdot,\cdot)_0$ to be the normalized invariant form of $\g$
restricted to $\k_0$.
 We
then let $\k'_S=L(\k_S)\oplus\C K_S$ be the central extension of
$L(\k_S)$ with bracket defined as usual as $$ [t^m\otimes X,t^n\otimes
Y]=t^{n+m}\otimes [X,Y]+\d_{m,-n}m(X,Y)_S K_S. $$ Set finally
$\ka=(\oplus_S \k'_S)\oplus \C d_\k$, where $d_\k$ is the derivation
$t\frac{d}{dt}$ on $L(\k)$ extended by setting $[d_\k,K_S]=0$.
Set $\ka_S=\k'_S\oplus\C d_\k$. We can extend the form $(\cdot,\cdot)_S$ on
all of
$\ka_S$ by setting $(K_S,\k_S)_S=(K_S,K_S)_S=(d_\k,\k_S)_S=(d_\k,d_\k)_S=0$
and
$(d_\k,K_S)_S=1$.

    We denote by $\Wa_\k$ the Weyl group of $\ka$. It
is a group of linear transformations on $\ha_\k^*$, where $$ \ha_\k=1\otimes
\h_0\oplus(\oplus_S\C K_S)\oplus\C d_\k. $$
If we define $\d_\k\in \ha_\k^*$  setting $$
\d_\k(d_\k)=1,\quad \d_\k(1\otimes \h_0)=\d_\k(K_S)=0, $$
then the set of roots for $\ka$ is $$
\Da_\k=\{n\d_\k+\a\mid \a\in\D_\k,\ n\in\ganz\}\cup\{n\d_\k\mid n\in\ganz,\, n\ne0\}, 
$$
where, as usual, we regard  $\h_0^*$ as a subset of $\ha_\k^*$ by extending
$\l\in\h_0^*$ setting $\l(d_\k)=\l(K_S)=0$.

    We set $\Pi_\k=\cup_{S}\Pi_{S}$ and  $$
\Pia_\k=\Pi_{\k}\cup\{\delta_\k-\th_S\mid S>0 \}.$$
$\Pia$ is  a set of
simple roots for $\ka$  and we denote by $\Dap_\k$ the
corresponding subset of positive roots.

    If $\lambda\in \ha_\k^*$ is a $\Dap_\k$-dominant integral weight, we
denote by $L(\lambda)$ the irreducible integrable $\ka$
module of highest weight $\lambda$.
    We denote by $\Lambda_{j}^{S}$ the fundamental weights of $\ka_S$ and
we set
    $\rkhat=\sum\limits_{j,\,S>0}\Lambda_j^S$.
Recall the Weyl-Kac character formula for
the character of $L(\lambda),\,\l\in\ha_\k^*$:
\begin{equation}\label{caratterek} ch(L(\lambda))=\frac{\sum_{w\in
\Wa_\k}\epsilon(w)e^{w(\lambda+\rkhat)-\rkhat}}{
\prod_{\a\in\Dap_\k}(1-e^{-\a})^{m_\a}}.\end{equation}
Here $m_\a$ denotes the multiplicity of the root $\a$.

    \subsection{Realization of level $1$ modules of $\widehat{so(\p)}$}\label{compatibile}\label{xerre}
If $X\in\k$ set $ad_\p(X)=ad(X)_{|\p}$. Since the action of $\k$ on $\p$
is orthogonal with respect to $(\cdot,\cdot)$, we have an inclusion
$\k\subset so(\p)$ defined by $X\mapsto ad_\p(X)$. We let $\op$  denote
the affine Lie algebra $L(so(\p))\oplus \C K_\p\oplus\C d_\p$, where
$L(so(\p))\oplus\C K_\p$ is the central extension of $L(so(\p))$
defined by setting $$ [t^m\otimes A,t^n\otimes B]=t^{n+m}\otimes
[A,B]+\d_{m,-n}m<A,B>K_\p $$ and $<A,B> =\frac{1}{2}tr(AB)$. Note that
$\widehat{so(\p)}$ is an affine algebra of type $B^{(1)}$ or $D^{(1)}$
according to whether $\dim(\p)$ is odd or even.\vskip10pt In the
following we recall the realization of the level $1$ irreducible modules of
$\widehat{so(\p)}$ described in \cite{KacPeterson}.\vskip10pt

    Fix $r\in\ganz$, set $r'=\lfloor\frac{r}{2}\rfloor$ and consider 
the loop space $\tilde\p=\C[t,t^{-1}]\t \p$.  Define the bilinear form
$\Phi_r$ on $\tilde \p$ by setting $$\Phi_r(t^{m_1}\t X,t^{m_2}\t
Y)=\d_{r+m_1+m_2,-1}(X,Y).$$ Let $Cl_r(\tilde\p)=Cl^+_r(\tilde\p)\oplus
Cl^-_r(\tilde\p)$ be the corresponding Clifford algebra, decomposed into
the sum
of
the even and odd part.
\par If
$m\in
\ganz$ set $\tilde\p_m=\oplus_{i\ge m}(t^i\t \p)$ and
$\tilde\p_m'=\oplus_{i< m}(t^i\t \p)$.  If $r$ is even set $\tilde
U_r=\tilde\p_{-r'}$. Then $\tilde U_r$ is a maximal isotropic subspace
for $\tilde\p$ with respect to $\Phi_r$. If $r$ is odd we choose a
maximal isotropic subspace of $\tilde \p$ as
follows. Recall (see
\cite[\S~9.3.1]{Wallach}) that a set of positive roots $\Dp$ for $\g$
is {\it compatible with $\Dp_\k$} if it is $\sigma$-stable and $\Dp\cap
\D_\k\supseteq\Dp_\k$. Let $\Dp$ be such a positive system.
Set
$\Dp(\p)=\D(\p)\cap\Dp|_{\h_0}$ and
$$\p^\pm=\sum\limits_{\a\in\pm\Dp(\p)}\p_\a.$$ Thus we can write $$
\p=\aa_\C\oplus\p^+\oplus\p^-, $$ where $\aa_\C=\mathfrak z\cap\p$.
Choose a maximal isotropic subspace $\aa$ of $\aa_\C$. Set
$U=\aa\oplus\p^+$ and $$ \tilde U_r=\tilde \p_{-r'}\oplus (t^{-r'-1}\t
U). $$ Let $\widehat \si_r$  denote the left action of $\op$ on the spin
module (defined in \cite{KacPeterson}) $s_r(\tilde\p,\tilde
U_r)=Cl_r(\tilde \p)/Cl_r(\tilde \p)\tilde U_r$ .

    If $r$ is even we let $X_r$ be the subalgebra of $Cl_r(\tilde
\p)$ generated by $\tilde \p'_{-r'}$. If $r$ is odd, set $L=\dim\aa_\C$
and $l=\lfloor\frac{L}{2}\rfloor=\dim\aa$. Fix a basis $\{v_i\}$ of
$\aa_\C$ such that $\{v_i\mid i\le l\}$ is a basis of $\aa$ and
$(v_i,v_{L-j+1})=\d_{ij}$. Then, if $L$ is even, we let $X_r$ be the
subalgebra of $Cl_r(\tilde \p)$ generated by $$ \tilde
\p'_{-r'-1}\oplus(t^{-r'-1}\t (\sp(v_i\mid i>l)\oplus\p^-)) $$ while, if
$L$ is odd,  we let $X_r=X_r^+$ be the subalgebra of $Cl_r(\tilde
\p)$ generated by $$ \(\tilde \p'_{-r'-1}\oplus(t^{-r'-1}\t (\sp(v_i\mid
i>l+1)\oplus\p^-))\)(t^{-r'-1}\t v_{l+1}). $$

    If $rL$ is even (resp. odd) then $Cl_r(\tilde \p)=Cl_r(\tilde
\p)\tilde U_r\oplus X_r$ (resp. $Cl^+_r(\tilde \p)=Cl^-_r(\tilde
\p)\tilde U_r\oplus X_r$) therefore we can identify $s_r(\tilde\p,\tilde
U_r)$ and $X_r$. Set moreover $X_r^\pm=X_r\cap Cl_r^\pm(\tilde\p)$.
Set $m=\lfloor\frac{\dim(\p)}{2}\rfloor$ and let $\tilde\L_0,\ldots,\tilde\L_m$ denote the fundamental
weights of
$\widehat{so(\p)}$ normalized
by setting $\tilde\L_i(d_\p)=0$. Define an element  $\d_\p$ in the
dual
of the Cartan subalgebra of $\widehat{so(\p)}$ requiring that
$\d_\p(d_\p)=1,\,\d_\p(K_\p)=0,\, \d_\p(x)=0$ for any $x$ in the Cartan
subalgebra of $so(\p)$.
\begin{prop}\label{KP}\cite{KacPeterson}
\begin{enumerate}
\item The action
$\widehat \sigma_r$ of $\op$ on $X_r$ is described explicitly in Theorem
1 of \cite{KacPeterson}.
\item If $rL$ is even we have $X_r^+\cong
L(\tilde\L_0),\,X_r^-\cong L(\tilde\L_1-\frac{1}{2}\d_\p)$ if $r$ is even
and $X_r^+\cong
L(\tilde\L_m),\,X_r^-\cong L(\tilde\L_{m-1})$ if $r$ is odd. \item If both
$L$ and
$r$ are odd we have $X_r^+\cong
L(\tilde\L_m)$.
\end{enumerate}
\end{prop}

    We shall conventionally refer to $X_r$ as the {\it basic and vector
representation} if $r$ is even and as the {\it spin representation} if
$r$ is odd.

\vskip20pt
Let $\eta:\ka\to \op$ be the Lie algebra homomorphism such that
$t^m\otimes X\mapsto t^m\otimes ad_{\p}(X)$ and $d_\k\mapsto d_\p$.
Requiring that $\eta$ is a Lie algebra homomorphism fixes the value of
$\eta(K_S)$:
if $h$ is a nonzero element of $\h_0\cap\k_S$, then $$
\eta([t\otimes h,t^{-1}\otimes h])=\eta((h,h)_SK_S) $$ while $$
[\eta(t\otimes h),\eta(t^{-1}\otimes h)]=[t\otimes ad_\p(h),t^{-1}\otimes
ad_\p(h)]=<ad_\p(h),ad_\p(h)> K_\p $$ so
 $$ \eta(K_S)=\frac{<ad_\p(h),ad_\p(h)>}{(h,h)_S}K_\p. $$
Set 
$$j_{S}=\frac{<ad_\p(h),ad_\p(h)>}{(h,h)_S}.$$
Let $\kappa(\cdot,\cdot)$ be the Killing form of $\g$ and
$\kappa_\k(\cdot,\cdot)$
 the Killing form of $\k$. We have
$$ tr(ad_\p(h)ad_\p(h))=\kappa(h,h)-\kappa_\k(h,h). $$
By Corollary 8.7 of \cite{Kac} we have that 
$\frac{\kappa_\k(h,h)}{2(h,h)_S}=h^\vee_S$ where $h^\vee_S$
is the dual Coxeter
number  of $\k_S$ if $S>0$ while $h^\vee_0=0$.
If $\g$ is simple we can apply Corollary~8.7 of \cite{Kac} obtaining
$\frac{\kappa(h,h)}{(h,h)}=2kh^\vee$ (here $h^\vee$ denotes the dual Coxeter number of $\g$). It
follows that
\begin{equation}\label{ratio}
 \frac{\kappa(h,h)}{(h,h)_S}=
 2kh^\vee\cdot\frac{(h,h) }{(h,h)_S}.
\end{equation}
In the complex case one checks directly that
\eqref{ratio} still holds.
The final outcome is that
$j_{S}$ is independent of the
choice of $h$ and that we
can write  
\begin{equation}\label{js}
j_S=kh^\vee\cdot\frac{(h,h) }{(h,h)_S}-h^\vee_S.
\end{equation}

Notice that, if $S>0$,
$\frac{(h,h)}{(h,h)_S}= \frac{2}{(\theta_S,\theta_S)}$. Setting
$n_S=\frac{2k}{(\theta_S,\theta_S)}=\frac{a_0k(\a_0,\a_0)}{(\theta_S,\theta_S)}$
for $S>0$ and $n_0=1$, we see that,
since $a_0k(\a_0,\a_0)$
is the length of a long root of $\widehat L(\g,\sigma)$, we can rewrite
\eqref{js} as
$$
j_S=n_Sh^\vee-h^\vee_S
$$
where $n_S$ is an integer (=1,2,3 or 4).

\vskip5pt
The map $\eta$ defines a representation $\si_r$ of $\ka$ on $X_r$ by
setting $\si_r=\widehat \si_r\circ\eta$. Using Theorem 1 of
\cite{KacPeterson},
we
can describe explicitly the action $\s_r$ of the Cartan subalgebra
$\ha_\k$ as follows. Fix  weight vectors $X_\a\in\p_\a$ such that
$(X_\a,X_{-\a})=1$ and set $$ \xi_{i,\a}=\begin{cases}t^i\t
X_\a&\text{if $rL$ is even}\\ \sqrt{-2}(t^i\t X_\a)(t^{-r'-1}\t
v_{l+1})&\text{if $rL$ is odd} \end{cases}. $$ Set also $$
v_{i,j}=\begin{cases}t^i\t v_j&\text{if $rL$ is even}\\ \sqrt{-2}(t^i\t
v_j)(t^{-r'-1}\t v_{l+1})&\text{if $rL$ is odd} \end{cases}. $$ Set $$
J_-=\{(i,\a)\mid i\le-r'-1,\ \a\in\D(\p)\}\cup\{(i,j)\mid i\le-r'-1,\
j=1,\dots,L\} $$ if $r$ is even, and \begin{align*} J_-=& \{(i,\a)\mid
i<-r'-1,\ \a\in\D(\p)\}\cup\{(-r'-1,\a)\mid \a\in-\Dp(\p)\}\\
&\cup\{(i,j)\mid i<-r'-1,\ j=1,\dots,L\}\cup\{(-r'-1,j)\mid L-j+1\le l\}
\end{align*} if $r$ is odd. Putting any total order on $J_-$, the
vectors \begin{equation}\label{vector} v_{i_1,j_1}\dots
v_{i_h,j_h}\xi_{m_1,\be_1}\dots\xi_{m_k,\be_k} \end{equation} with
$(i_1,j_1)<\dots<(i_h,j_h)$ and $(m_1,\be_1)<\dots<(m_k,\be_k)$ in $J_-$
form a basis for $X_r$. If $v$
is a vector given by \eqref{vector} and $h\in\h_0$ then
$$\si_r(h)v=(\sum_{s=1}^k\be_s(h)+
\half\d_{r,2r'+1}\!\!\!\sum_{\a\in\Dp(\p)}\a(h))v
$$ while
$$
\si_r(d_\k)v=(\sum_{s=1}^h (i_s+\frac{r+1}{2})+\sum_{s=1}^k
(m_s+\frac{r+1}{2}))v.
$$
Since $K_\p$ acts as the identity on $X_r$ we
find that $$ \si_r(K_S)v=j_Sv. $$

    It follows that $v$ is a weight vector having weight
\begin{equation}\label{peso}
\sum_{S}j_S\L_0^S+\sum_{s=1}^h(i_s+\frac{r+1}{2})\d_\k+\sum_{s=1}^k((m_s+
\frac{r+1}{2})\d_\k+\be_s)+
\d_{r,2r'+1}\rho_n,
\end{equation}
where by definition $\rho_n=\half\sum_{\a\in\Dp(\p)}\a$.\par
We shall
use this formula in the next sections, treating separately the cases $r$
even, $r$ odd to obtain the decomposition of the $\widehat{so(\p)}$-modules
$X_r$ with respect to the subalgebra $\eta(\ka)$.

\section{Decomposition of the basic and vector
representation (semisimple case)}\label{basicandvector}
Here we assume that $r$ is even and $\k$ is semisimple.
Set $c_S=\frac{(h,h)}{(h,h)_S}$, where $h$ is any
nonzero element of $\h_0\cap\k_S$. As already observed $c_S$ does not
depend on $h$.
Define a linear map $\pseven:\ka\to\widehat L(\g,\si)$ by setting
\begin{equation}\label{psi1}
\pseven(t^n\t
X)=t^{2n}\otimes X\quad\pseven(d_\k)=d'/2\quad\pseven(K_S)=2c_SK'=kc_SK.
\end{equation}

Let
$\steven:\ha^*\to (\ha_\k)^*$ denote the transpose of $\pseven$ (restricted to
$\ha^*$).
Clearly
$\d_\k=2\steven(\d')$.  Set
$$
\Dap(\p)=\{(m+1/2)\d_\k+\a\mid\a\in\D(\p),\,m\ge0\}.
$$

Let us
record the following facts.
\begin{lemma}\label{lemmapsi}\ %
\begin{enumerate}
\item The map $\steven$
defines a bijection between  $\Dap$ and  $\Dap_\k\cup\Dap(\p)$.
\item
$\steven(\ha)$ is stable under the action of $\Wa_\k$.
\end{enumerate}
\end{lemma}
\begin{proof} For the first statement, remark
that $\steven(m\d'+\a)=\frac{m}{2}\d_\k+\a$ for all
$\a\in\D_\k\cup\D(\p)\cup\{0\}$. To
prove the second statement, note that if $\a=m\d_\k+\be$ with $\be$ a
nonzero root in $\D_\k$ and $\lambda=\steven(\mu)$, then
$\a=\steven(2m\d'+\be)$, hence $$
s_\a\lambda=\lambda-\lambda(\a^\vee)\a=
\steven(\mu-\lambda(\a^\vee)(2m\d'+\be)).
$$
\end{proof}

Since $\pseven$ is onto, $\steven$ is bijective on its image so, for
each $w\in\Wa_\k$, we can define
$$
\hat w=(\steven)^{-1}w\steven.
$$
Then we set 
$\Wa_{\si,\reven}=\{\hat w\mid w\in\Wa_\k\}$.

\begin{lemma}\label{embed}
Let $\a\in\Da_\k$ be a real root.
If $\a=\steven(\gamma)$ then $\pseven(\a^\vee)=\gamma^\vee$.
In particular $\hat s_\a=s_\gamma$.
\end{lemma}
\begin{proof}
Assume that $\a$ is a real root of $\ka_S$. We define a bilinear form $\{\cdot,\cdot\}$
on
$\ha_\k\cap\ka_S$ by setting
$$
\{h,k\}=(\pseven(h),\pseven(k)).
$$
Obviously, if $h,k\in\h_0$ then $\{h,k\}=(h,k)=c_S(h,k)_S$. We claim that
$$
\{\cdot,\cdot\}=c_S(\cdot,\cdot)_S.
$$
Indeed, if $h\in\h_0\cap\k_S$, then $\{K_S,h\}=kc_S(K,h)=0$ and
$\{K_S,K_S\}=k^2c_S^2(K,K)=0$. By \eqref{diprimo}
$\{d_\k,h\}=\half(d',h)=0$, $\{d_\k,d_\k\}=\tfrac{1}{4}(d',d')=0$. Finally
$$
\{d_\k,K_S\}=c_S(d',K')=c_S.
$$

Let $\nu_S:\ha_\k\cap\ka_S\to(\ha_\k\cap\ka_S)^*$ be the isomorphism
induced by $\{\cdot,\cdot\}$.
 We claim that, if $\a=\steven(\gamma)\in\Da_S$, then
$\pseven(\nu_S^{-1}(\a))=\nu^{-1}(\gamma)$.
 Indeed write $\nu_S^{-1}(\a)=h_S+aK_S$ with $h_S\in\k_S\cap\h_0$.
 If $h\in\ha$, there exists $h'\in \ha_\k$ such that  $h=\pseven(h')$. Write
$h'=\sum_i( h'_i+b_iK_i)+cd_\k$ with $h'_i\in\k_i\cap\h_0$.
 Then
\begin{align*}
 (\pseven(\nu_S^{-1}(\a)),h)&=(h_S+kc_SaK,\sum_i( h'_i+kc_ib_iK)+\half cd')\\
 &=(h_S+kc_SaK,h'_S+kc_Sb_SK+\tfrac{ c}{2}d')\\
 &=(\pseven(\nu_S^{-1}(\a)),\pseven(h'_S+b_SK_S+cd_\k))\\
 &=\{\nu_S^{-1}(\a),h'_S+b_SK_S+cd_\k\}\\
 &=\a(h')=\gamma(\pseven(h'))=\gamma(h).
 \end{align*}

 In particular
 $$
 \pseven(\a^\vee)=\pseven(\frac{2\nu_S^{-1}(\a)}{\{\nu_S^{-1}(\a),\nu_S^{-1}(\a)\}
})=
 \frac{2\nu^{-1}(\gamma)}{(\nu^{-1}(\gamma),\nu^{-1}(\gamma))}=\gamma^\vee.
 $$
\end{proof}

Let $\Pi_0=\{\a_i\in \Pia\mid s_i=0\}$ and  let
$\Pia_{\si, \reven}=\Pi_0\cup\{k\d-\th_S\mid S>0\}$. Set also
\begin{equation}\label{dasigmaerrepari}
\Da_{\si,\reven}=\D_\k+2\ganz\d'
\end{equation}

 An obvious consequence of
Lemma~\ref{embed} is the following fact.
\begin{cor} \label{rootofk}$\Da_{\si,\reven}$ is a root system and
$\Wa_{\si,\reven}$ is the
corresponding reflection group. In particular $\Wa_{\si,\reven}$
 is the subgroup of $\Wa$ generated by the reflections
$\{s_\a\mid \a\in\Pia_{\si,\reven}\}$.
\end{cor}
\begin{proof}For the first statement observe that $\steven(\Da_{\si,\reven})$
is the set of real roots
in $\Da_\k$.
 As for the second assertion, since $k\d=2\d'$, it is clear
that $\steven$ is a bijection between $\Pia_{\si,\reven}$ and
$\Pia_\k$.
\end{proof}

Set $N=rank(\g)$ and for $\a\in\Dap(\p)$ set $m_\a=1$ if
$\a=(m+\frac{1}{2})\d_\k+\be$
with $\be\in\D(\p)\setminus\{0\}$, while we set $m_\a=N-n$ if $\a=(m+\frac{1}{2})\d_\k$.
Observe that, if $\a=\steven(\beta)$, then $m_\a$ equals the multiplicity $m_\beta$ of
$\beta$ as a root of $\widehat L(\g,\sigma)$ (see \cite{Kac}, Corollary 8.3).\par
Let $\rhat$  be the element of $\ha^*$ such that $\rhat(d')=0$ and
$\rhat(\a_i^\vee)=1$ for $i=0,\dots,n$.
Set
$$\num^-=e^{\sum_S j_S\L_0^S+\rhat_\k}
\prod_{\a\in\Dap(\p)}(1-e^{-\a})^{m_\a}
\prod_{\a\in\Dap_\k}(1-e^{-\a})^{m_\a}.
$$
\begin{lemma}\label{denominbasic} We have
$$
\num^-=e^{\steven(\rhat)}\prod_{\a\in\Dap}(1-e^{-\steven(\a)})^{m_{\a}}.
$$
\end{lemma}
\begin{proof}
By Lemma~\ref{lemmapsi}
$$
\num^-=e^{\sum_S j_S\L_0^S+\rhat_\k}
\prod_{\a\in\Dap}(1-e^{-\steven(\a)})^{m_\a}.
$$
 It remains only to check that
\begin{equation}\label{psirho}
\sum_S j_S\L_0^S+\rhat_\k=\steven(\rhat).
\end{equation}
Since $\pseven(\a^\vee)=\a^\vee$ for $\a\in\Pi_\k$ we see that
$\steven(\rhat)(\a^\vee)=1=(\sum_S j_S\L_0^S+\rhat_\k)(\a^\vee)$ for
$\a\in\Pi_\k$.
We defined $\rhat$ so that $\rhat(d')=0$ hence $\steven(\rhat)(d_\k)=0=
(\sum_S j_S\L_0^S+\rhat_\k)(d_\k)$. It remains only to check that
$\steven(\rhat)(K_S)=(\sum_S j_S\L_0^S+\rhat_\k)(K_S)=j_S+h^\vee_S$, but,
since $\steven(\rhat)(K_S)=kc_S\rhat(K)$, this follows immediately from
\eqref{js} and
the fact that $\rhat(K)=h^\vee$.
\end{proof}

  By
formula \eqref{peso}, the character of $X_r$ can be written as
\begin{equation}\label{carattere} ch(X_{r})=e^{\sum_S
j_S\L_0^S}\prod_{\a\in\Dap(\p)}(1+e^{-\a})^{m_\a}
\end{equation}
 hence
 $$
 ch(X_{r}^+)-ch(X_{r}^-)=e^{\sum_S
j_S\L_0^S}\prod_{\a\in\Dap(\p)}(1-e^{-\a})^{m_\a}
 $$
Applying Lemma~\ref{denominbasic}
 and setting
 $$ D_\k=e^{\rhat_\k}
\prod_{\a\in\Dap_\k}(1-e^{-\a})^{m_\a},
$$
we can write
\begin{equation}\label{carbasic}
ch(X_{r}^+)-ch(X_{r}^-)=\frac{\num^-}{D_\k}=\frac{\prod_{\a\in\Dap}(1-e^{-\steven(\a)})^{m_\a}}{D_\k}
\end{equation}
    By the ``denominator identity" (cf. \cite{Kac}, (10.4.4)), \eqref{carbasic} can be rewritten as
\begin{equation}\label{weylbasic}
\frac{\sum_{w\in\Wa}\epsilon(w)e^{\steven(w(\rhat))}}{D_\k}
\end{equation}

Let
$W'_{\si,\reven}$ be the set of minimal right coset representatives of
$\Wa_{\si,\reven}$ in
$\Wa$. We can rewrite \eqref{weylbasic} as
 $$
\frac{\sum_{u\in
W'_{\si,\reven}}\epsilon(u)\sum_{w\in\Wa_\k}\epsilon(w)e^{w\steven(u(\rhat))-\rkhat
}}
{\prod
_{\a\in\Dap_\k}(1-e^{-\a})^{m_{\a}}}
$$

    Using the Weyl-Kac Character formula in
the formulation \eqref{caratterek},
the final outcome is that, if $r$ is even,
\begin{equation}\label{forabelian} ch(X_r^+)-ch(X_r^-)=\sum_{u\in
W'_{\si,\reven}}\epsilon(u)\,ch(L(\steven(u\rhat)-\rkhat))
\end{equation}
(cf. \cite{Parthasaraty}).
Using \eqref{psirho}, we obtain the
following result
\begin{theorem}\label{decompositioneven} If $\k$ is semisimple and $r$ is
even then for $\epsilon=0$ or $1$ one has:
\begin{equation}\label{decobasic}ch(L(\tilde\L_\epsilon))=\sum_{{u\in
W'_{\si,\reven}}\atop{\ell(u)\equiv\epsilon\,mod\,2}}
ch\(L(\steven(u\rhat-\rhat)+\sum_Sj_S\L_0^S+\frac{1}{2}\epsilon\d_\k)\),
\end{equation}
where
$W'_{\si,\reven}$ is the set of minimal right coset representatives of $\Wa_{\si,0}$
in $\Wa$, $\Wa_{\si,0}$ being given by Corollary \eqref{rootofk} and
$\pseven$ is defined by
\eqref{psi1}.
\end{theorem}
\begin{proof}
If $\l$ is a weight of $X^+_r$ then $\l(d_\k)\in\ganz$, while, if $\l$ is a
weight of $X^-_r$ then
$\l(d_\k)\in \half+\ganz$ (cf. Proposition \ref{KP}).
It follows that $X^+_r$ and $X^-_r$ do not have common components.
\end{proof}

%%%%%%%%%%%%%%%%%%%%%%%%%%%%%%%%%%%%%%%%%%%%%%%%%%%%%%%%%%%%%%%%%%%%

\subsection{Decomposition rules and combinatorics of
roots}\label{abelianibasic}

Let $\Sigma$  denote the set of $\b_0$-stable abelian subspaces 
of $\p$.
Each abelian subspace in $\Sigma$ is a sum of $\h_0$-weight spaces. 
We identify $\mathfrak i\in\Sigma$ and the set $A\subseteq\D(\p)$ such 
that $\mathfrak i=\sum_{\a\in A}\p_\a$.
In this section, we describe the connection between the 
subspaces in $\Sigma$ and the decomposition of the basic and vector 
modules $L(\tilde \Lambda_\epsilon)$ stated in Theorem \ref{decompositioneven}.
\par
The set $\Sigma$ has been studied in \cite{IMRN} in the case of
$\g$ simple. The results of that paper can be easily extended to the complex case.
 In the complex case, the subspaces in
$\Sigma$ turn  out to correspond to more familiar objects. Indeed, we shall 
see that we can view $\Si$ as the set of abelian 
ideals of $\b_0$. We deal at once with this special case.   
\par
We recall some general conventions and facts. 
Let $\lie$ be a simple Lie algebra, $\Phi$ its root system, 
$\b_\lie$ a Borel subalgebra, $\Phi^+$ and $S$ the corresponding set of
positive roots and simple roots, respectively.
A subset $A$  of $\Phi^+$ is called {\it abelian} if $\a+\be\notin\Phi$ 
for all $\a,\be\in A$.  An {\it abelian ideal} of
$\Phi^+$ is an abelian set $A$ such that if $\a\in A$ 
and $\gamma, \a+\gamma\in\Phi^+$, then $\a+\gamma\in A$.
If $A$ is an abelian ideal of $\Phi^+$, then $\sum_{\a\in A}\lie_\a$ 
is an abelian ideal of $\b_\lie$, and, conversely, each abelian ideal 
of $\b_\lie$ is (uniquely) obtained in this way. 
Recently, there has been a great deal of work on these ideals
by several authors (Kostant \cite{Kostant1} \cite{Kostant2}, Cellini-Papi
\cite{CP}\cite{CP2}\cite{CP3}, Panyushev \cite{Pan1}\cite{Pan2}, Suter \cite{Suter}).
There are various explicit descriptions of them and, in particular, 
we know that they are exactly $2^{\rank(\lie)}$.   
\par
Now let $\ov\k$ be a simple Lie algebra, $\g=\ov\k\oplus\ov\k$, and
$\sigma:(x, y)\mapsto (y, x)$ be the switch automorphism of $\g$.
Thus $\k$ is the diagonal copy of $\ov\k$ in $\g$, and 
$\p=\{(x,-x)\mid x\in\ov \k\}$. Then $\p$ is naturally isomorphic to $\k$  
as a $\k$-module, so that what we are going to study is the decomposition of the 
basic and vector representations of $\widehat{so(\k)}$ with respect 
to $\ka$, where $\k$ is any simple Lie algebra. 
\par
Clearly, $\D(\p)=\D_\k\cup\{0\}$ and, through the natural isomorphism 
between $\k$ and $\p$, $\p_\a$ corresponds to $\k_\a$ for all $\a\in \D(\p)$, 
where we intend $\k_0=\h_0$. 
In particular, if a subset $A$ of $\D(\p)$ belongs to $\Sigma$, then by definition 
$\sum_{\a\in A} \p_\a$ is a $\b_0$ stable abelian subspace of $\p$ and therefore
$\sum_{\a\in A} \k_\a$ is a $\b_0$ stable abelian subspace of $\k$. It is easily 
seen that this implies $A\subseteq \Dp_\k$, and hence that $A$ is an abelian 
ideal of $\Dp_\k$. 
\par
Thus, in this case, $\Sigma$ is the set of abelian ideals of $\Dp_\k$. 
The following theorem, which is an easy consequence
of Theorem~\ref{decompositioneven} and the results of \cite{CP},
describes the decomposition of the basic
and vector representations $L(\tilde\L_0)$ and $L(\tilde\L_1)$ of
$\widehat{so(\k)}$ with respect to $\widehat{\k}$ in terms of 
$\Sigma$ (cf. with \cite{KacW}, formula (4.2.13)).
It is the nicest special case of Theorem~\ref{decobasic} below.
\par
Let us fix some notation. 
Set 
$$
\ha_\R=\sp_\R(\a_1^\vee,\dots,\a_n^\vee)+\R K'+\R d'
$$
and
$$
\hauno=\{x\in \hard\mid (x,\d)=1\},\qquad \hazero=\{x\in
\hard\mid (x,\d)=0\}.
$$ 
Let $\pi$ be the canonical projection mod $\d$ and set 
$\huno=\pi \hauno$. We identify $\hzero$ with $\pi \hazero$.
\par
For $\a\in\Dap$ set 
$$ 
H_{\a}=\{x\in \h^*_1\mid (\a,x)=0\} 
$$ 
and $H_\a^+=\{x\in\h_1^*\mid (\a,x)\ge0\}$. Also, let  $C_1$
be the fundamental alcove of $\Wa$,
$$ 
C_1=\{x\in \h_1^*\mid (\a,x)\ge0\,\forall\ \a\in\Pia\}.
$$ 
It is well-known that there 
is a faithful action of $\Wa$ on $\h_1^*$ and that $C_1$ is a 
fundamental domain for this action. 
\par
For $w\in\Wa$ we set 
$$
N(w)=\{\a\in\Dap\mid w^{-1}(\a)\in\Da^-\}.
$$
\par
Finally, for $A\in \Sigma$, we denote by $\la A\ra$ (resp. $|A|$) the sum 
(resp. the number) of elements in $A$.

\begin{theorem} 
Let $\epsilon=0\text{ or }1$. Then one has the following decomposition
of the basic and vector $\widehat{so(\k)}$-modules with respect to $\ka$.
$$
L(\tilde\Lambda_\epsilon)=\bigoplus_{A\in\Sigma\atop
|A|\equiv\epsilon\,mod\, 2}L(h_\k^\vee\Lambda_0^\k+\langle
A\rangle-\frac{1}{2}(|A|-\epsilon)\delta_\k)
$$
(where $h_\k^\vee$  and $\Lambda_0^\k$ are respectively
the dual Coxeter number and $0$th fundamental weight of $\ka$).
\par
Moreover, the highest weight vector $v_A$ of the submodule
$L(h_\k^\vee\Lambda_0^\k+\langle
A\rangle-\frac{1}{2}(|A|-\epsilon)\d_\k)$ is, up to a constant factor, the
following pure spinor (of the spin representation of $Cl_0(\tilde\k)$):
$$v_A=\prod_{\a\in A} (t^{-1}e_\a).$$
\end{theorem}

\begin{proof}  
Under our assumptions, the summation $\sum_S j_S \Lambda_0^S$ in 
\eqref{decompositioneven} has a single
summand, which is $j_\k \Lambda_0^\k$. Clearly, $\algsi$ is 
isomorphic to $\ka$, hence $h^\vee=h_\k^\vee$. Since $k=2$, 
using \eqref{js}, we obtain that $j_\k \Lambda_0^\k=h_\k^\vee \Lambda_0^\k$.  
\par
Now we note that $\Da=\ganz^*\d\cup(\D_\k+\ganz \d)$, and 
$\Da_\k=\psi_0^*(2\ganz^*\d\cup (\D_\k+2\ganz\d))$.
Hence, by Lemma \ref{embed}, we obtain that $\Wa_{\si,\reven}$ is the subgroup 
of $\Wa$ generated by the reflections with respect to roots in $\D_\k+2\ganz\d$.
Then  $\Wa_{\si,\reven}$ is isomorphic to $\Wa$ itself and, moreover, it has 
$2 C_1$ as an alcove. More precisely, $2 C_1$ is the fundamental  alcove of 
$W_{\si,\reven}$ if we choose $\D_\k+2\ganz\d\cap \Dap$ 
as  positive system for the real root system $\D_\k+2\ganz\d$. 
By general theory, it follows that 
$$
W'_{\si,\reven}= \{w\in \Wa\mid wC_1\subset 2C_1\}.
$$
Let $A$ be an abelian ideal of $\Dp_\k$ and consider the set
$-A+\d\subset\Dap$. It is easy to prove  that both $-A+\d$ and its complement
in $\Dap$ are closed under root addition, and hence that
there exists a unique element $w_A\in\Wa$ such that $-A+\d=N(w_A)$. 
Moreover, in \cite{CP} it is proved that $A\mapsto w_A$ is a bijection between
the set $\Sigma$ of abelian ideals of $\Dp_\k$ and the subset 
$\{w\in\Wa\mid wC_1\subset 2C_1\}$.
Now the claim follows directly from Theorem \ref{decompositioneven} and the 
following observations:
\begin{enumerate}
\item 
for $w\in \Wa$, $w(\rhat)-\rhat=-\langle N(w)\rangle$
(see e.g. \cite{Kumar}, Corollary 1.3.22);
\item 
for $\a\in\Dap$, $\steven(-\a+\d)=-\a+\frac{\d_\k}{2}$;
\item 
for  $A\in\Sigma$, $\epsilon(w_A)=(-1)^{|A|}$, and 
$|A|=|-A+\d|=|N(w_A)|=\ell(w_A)$.
\end{enumerate}
The statement on highest weight vectors will be proved in Theorem 
\ref{decoeabeliani}.\end{proof}

We now turn to the combinatorial interpretation of the decomposition
\eqref{decobasic} for general $\g=\k+\p$ with $\k$ semisimple.
We need to recall some results and notation from
\cite{IMRN}. 
%%%%%%%%%%%%%%%%%%%%%%%%%%%%%%%%%%%%%%%%%%%%%%%%%%%%%%%%%%%%%%%%%%%%%%%%%%%%%%%%

 If $\g$ is simple, by the 
classification of Lie algebra involutions (see \cite{Kac}, Ch.8), we have
that there exists an index $p$ such that $s_p=1$ and $s_i=0$ if $i\ne p$.
\par 
Set \begin{equation} \label{dsigma} D_\sigma=\bigcup_{w\in\mathcal W^\sigma_{ab}}
wC_1,\end{equation}
where $$\mathcal W^\sigma_{ab}=\left\{ w\in\Wa\mid
N(w)\subseteq\{\a\in\Da\mid m_p(\a)=1\}\right\}. $$ and, as above,
$N(w)=\{\a\in\Dap\mid w^{-1}(\a)\in\Da^-\}$.
If $w\in\mathcal W^\sigma_{ab}$ we shall say that $w$ is
{\it $\sigma$-minuscule}.

\par
\vskip5pt

Given $w\in \Wa$, a root $\be\in\Dap$ belongs to $N(w)$ if and only
if $H_{\be}$ separates $wC_1$ and $C_1$. It follows that $$
D_\sigma=\bigcap_{{\a\in \Dap,}\atop{m_p(\a)\ne 1}}H^+_{\a}. $$ If
$\a\in\Da$ we set $\Ha_\a=\{x\in\widehat\h^*_\R/\R\d\mid (\a,x)=0\}$ and
$\Ha^+_\a=\{x\in\widehat\h_\R^*/\R\d\mid (\a,x)\ge0\}$. Set also $$
C_\sigma=\bigcap_{{\a\in \Dap,}\atop{m_p(\a)\ne 1}}\Ha_\a^+. $$
Obviously $$ D_\sigma=C_\sigma\cap\h_1^*. $$

Set $\Phi_{\si}=\Pia_{\si,\reven}\cup\{k_p\d+\a_{p}\},$ where
$k_p=k,\,1$ according to whether $\a_p$ is long or short. \begin{prop}
\cite{IMRN} We have that $$ C_\sigma=\bigcap_{\a\in\Phi_\sigma}\Ha^+_\a.
$$ \end{prop}

Now let
\begin{equation}\label{psigma}P_\s=\bigcap\limits_{\a\in\Pia_{\si,\reven}} H_\a^+.
\end{equation}
It is a standard fact that  the set of elements $w\in \Wa$
    such that
$wC_1$ cover the polytope $P_\si$ is the set of minimal  right coset
representatives
for the subgroup of $\Wa$
generated by $s_\a$ with $\a\in\Pia_{\si,\reven}$, which, by
Corollary~\ref{rootofk}, happens
to be $\Wa_{\si,\reven}$. Since obviously  $D_\s\subseteq P_\s$, we have that
$\mathcal
W^\si_{ab}\subset W'_{\si,\reven}$.
We elucidate the precise relation between $\mathcal W^\si_{ab}$ and
$W'_{\si,\reven}$
in the next proposition where,
if $\widehat L(\g,\si)$ is simply laced, we regard all roots
as long. Recall that we denote by $\b_0$ the Borel subalgebra of $\k$
associated
to
our initial choice
of positive roots  in $\D_\k$.
\vskip10pt
Recall that we are assuming that $\k$ is
semisimple. If $\g$ is simple, let  $p$ be the index such that $s_p=1$ and $s_i=0$ if $i\ne
p$.
 In the complex case (and only in this
case) we have $p=0$ (see Remark~\ref{simple}). For $\gamma\in Q^\vee$, we denote by
$t_\gamma$ the translation by $\gamma$ (see \cite[(6.5.2)]{Kac}).

\begin{prop}\label{casistica} \cite{CP},\cite{IMRN}
1).\label{lungocorto} $D_\s=P_\s$ if and only if $p=0$
 or
$\a_p$ is short.  If $\a_p$ is a long root, then $P_\s\setminus
D_\s$ consists exactly of the alcove $w_\s\,C_1$, where
$w_\s=t_{-k\a_p^\vee} w_0^*w_0$, $w_0$ is the longest element  of the
parabolic subgroup of $\Wa$ generated by $\Pia\setminus\{\a_p\}$ and $w_0^*$ is
the longest element of the parabolic subgroup of $\Wa$ generated by
$\Pia\cap\a_p^\perp$.
\par\noindent
2). \label{abeliani} There is a bijection between $\b_0$-stable abelian
subspaces in $\p$
and $\sigma$-minuscule  elements, or, equivalently,
alcoves paving
$D_\sigma$. In this correspondence an element $w\in\mathcal W^\s_{ab}$
such that $N(w)=\{\be_1,\ldots,\be_r\}$
maps to $\bigoplus\limits_{i=1}^r\p_{-\ov{\be_i}}$.

\end{prop}

 Set $\L_{0,\k}=\sum j_S\L_0^S$ and
let $\Sigma$  denote the set of $\b_0$-stable abelian subspaces of $\p$.
Identify $\mathfrak i\in\Sigma$ and the set
$A\subset\D(\p)$ such that $\mathfrak i=\sum_{\a\in A}\p_\a$.
We summarize the connection between abelian subspaces and the decomposition of
$X_r$ in the following proposition.

\begin{theorem}\label{decoeabeliani}
(1) Assume that $p=0$ or $\a_p$ is a
short root. Then
$$
L(\tilde\Lambda_\epsilon)=\bigoplus_{A\in \Sigma\atop |A|\equiv
\epsilon\,\,mod\,2}L\left(\L_{0,\k}+\langle
A\rangle-\frac{1}{2}(|A|-\epsilon)\d_\k\right).$$
 \newline (2) Assume that $\a_p$ is a
long  root and $p\ne0$.
 We
have
$$
L(\tilde\Lambda_\epsilon)=\bigoplus_{A\in \Sigma\atop |A|\equiv
\epsilon\,\,mod\,2}L\left(\L_{0,\k}+\langle
A\rangle-\frac{1}{2}(|A|-\epsilon)\d_\k)\right)\bigoplus\nu L\left(\L_{0,\k}
-y+\frac{1}{2}\epsilon\d_\k\right),$$ where
$$
y:=\steven(\langle N(w_\s)\rangle)=(\sum_{\be\in(\a_p+\Dp_\k)\cap \Dap}
\overline\be)+2\overline\a_p+\(\frac{|(\a_p+\Dp_\k)\cap\Dap|}{2}+2\)\d_\k
$$
and $\nu=\d_{\epsilon,\ell(w_\si) \,mod\,2}$.\vskip5pt
Moreover, in
both cases, the highest weight vector of each component is, up to a
constant factor,
the pure spinor
(of the spin representation of $Cl_r(\tilde\p)$):
\begin{equation}\label{maxima}v_A=\prod_{\a\in A} (t^{-r'-2}e_\a)\end{equation}
where $\p_\a=\mathbb \C e_\a$. A highest weight vector in  the  component
indexed by
$w_\s$ is
\begin{equation}\label{maximal}v_\si=\prod_{\be\in (\a_p+\Dp_\k)\cap\Dap}
(t^{-r'-2}e_{-\overline\be})(t^{-r'-2}e_{-\overline\a_p})(t^{-r'-3}e_{-\overline
\a_p}).
\end{equation}
\end{theorem}
\begin{proof}
It follows immediately from \eqref{psirho} that
\begin{equation}\label{ffb}
\steven(u{\rhat})-\rkhat=
\L_{0,\k}-\steven(\langle
N(u)\rangle).
\end{equation}
hence we can rewrite formulas \eqref{decobasic}  as
$$
ch(X_r)=ch(L(\tilde\L_0))+ch(L(\tilde\L_1))=\sum_{w\in
W'_{\si,\reven}}ch(L(\L_{0,\k}-\steven(\langle
N(w)\rangle)).
$$
By Corollary~\ref{rootofk} and Proposition~\ref{casistica}, we can write
$$
ch(X_r)=\sum_{w\in \mathcal W^\si_{ab}}ch(L(\L_{0,\k}-\steven(\langle
N(w)\rangle))
$$
if $p=0$ or $\a_p$ is short, while
$$
ch(X_r)=\sum_{w\in \mathcal W^\si_{ab}\cup\{w_\si\}}ch(L(\L_{0,\k}-\steven(\langle
N(w)\rangle))
$$
if $p\ne 0$ and $\a_p$ is long.
If  $\a\in\Da$ then $\steven(\a)=\half m_p(\a)\d_\k+\ov\a$.
If $w=w_A$ for some $A\in\Sigma$ and $\a\in N(w_A)$ then $m_p(\a)=1$.
Moreover, if $w_A\in \mathcal W^\si_{ab}$ encodes the subspace $A$, we have
$\epsilon(w_A)=(-1)^{\ell(w_A)}=(-1)^{|A|}$. This justifies the
distribution of the
summands in the  basic and vector modules according to the parity of $|A|$.
\par
The calculation of  $N(w_\s)$ follows by a straightforward computation
using standard properties of the sets $N(w)$ (see \cite{CP3}, 2.5). One gets
\begin{equation}\label{Nw}N(w_\s)=\(\a_p+\Dp_\k\)\cap
\Dap\cup\{\a_p\}\cup\{\a_p+k\d\}.\end{equation} Since
$k\d=k\sum_{i=0}^na_is_i\d=2\d'$ if we apply $\steven$ to each element in
the r.h.s. of \eqref{Nw} and take the sum  we obtain the required
expression for
$y$.\par
We now check that $v_A$ is a highest weight vector. Set $\l_A=
\steven(w_A(\rhat))- \rhat_\k$ be the corresponding 
highest weight.
We will show  that, if 
$\a\in\Pia_\k$, then $\l_A+\a$ is not a weight of $X_r$.
Indeed $\l_A+\rhat_\k=\steven(w_A(\rhat))$ and $\a=\steven(\be)$ with $\be\in\Pia_{\si,0}$ 
so we can write
$\l_A+\rhat_\k+\a=\steven(w_A(\rhat)+\be)$. We observe that
$$
(w_A(\rhat)+\be,w_A(\rhat)+\be)=(\rhat,\rhat)+(\be,\be)+2(w_A(\rhat),\be).
$$
Since $w_A(C_1)\subset P_\si$, we have that $2(w_A(\rhat),\be)\ge 0$, hence
$$
(w_A(\rhat)+\be,w_A(\rhat)+\be)>(\rhat,\rhat).
$$
If $\l_A+\a$ is a weight of $X_r$, then, according to \eqref{carattere},
$\l_A+\a+\rhat_\k\in\sum_S j_S\L_0^S+\rhat_\k-\steven(S)$, where $S$ is the set of weights defined in 
Lemma 3.2.3 of \cite{Kumar}, thus we can write that $\steven(w_A(\rhat)+\be)\in\steven(\rhat-S)$.
It follows that $w_A(\rhat)+\be-\rhat\in -S$. Applying Lemma 3.2.4 of \cite{Kumar} (with 
$\mu=w_A(\rhat)+\be-\rhat$), we find a contradiction.
Obviously the same argument applies also to $w_\si$.
\end{proof}

%%%%%%%%%%%%%%%%%%%%%%%%%%%%%%%%%%%%%%%%%%%%%%%%%%%%%%%%%%%%%%%%%%%%%%%%%%%
%%%%%%%%%%%%%%%%%%%%%%%%%%%%%%%%%%%%%%%%%%%%%%%%%%%%%%%%%%%%%%%%%%%%%%%%%%%
 
\section{Decomposition of the spin representation 
(semisimple case)}\label{spin}
We now consider the case when $r$ is odd and $\k$ is semisimple. We
distinguish two cases: $\g$ not simple (the complex case) and  $\g$ simple.
\subsection{Complex case}
We consider here the case when $\g=\k\times\k, \sigma(X,Y)=(Y,X)$,
$\k$ is
simple and embeds in
$\g$ diagonally. We have that 
$\D(\p)=\D_\k\cup\{0\}$ and we can choose $\Dp(\p)=\Dp_\k$.
In this case $\k$ is simple, so, by \eqref{js}, the sum
$\sum_Sj_S\L_{0,S}$ reduces to one summand, which equals $\rkhat$.
By \eqref{peso}  the character of $X_r$ is
\begin{align*}
&ch(X_r)=\\
&=e^{\rkhat}2^{\lfloor\frac{n}{2}\rfloor}\(\prod_{\a\in\Dap_\k}(1+e^{-\a}
)^{m_\a}\)
=e^{\rkhat}2^{\lfloor\frac{n}{2}\rfloor}\frac{\(\prod_{\a\in\Dap_\k}(1
-e^{-2\a})^{m_\a}\)}
{\(\prod_{\a\in\Dap_\k}(1-e^{-\a})^{m_\a}\)}\\
&=e^{\rkhat}2^{\lfloor\frac{n}{2}\rfloor}\frac{\sum_{w\in\What}\epsilon
(w)e^{2w\rkhat-2\rkhat}}
{\(\prod_{\a\in\Dap_\k}(1-e^{-\a})^{m_\a}\)}=2^{\lfloor\frac{n}{2}\rfloor}\frac{
\sum_{w\in\What}\epsilon(w)e^{w(\rkhat+\rkhat
)-\rkhat}} {\(\prod_{\a\in\Dap_\k}(1-e^{-\a})^{m_\a}\)}\\
&=2^{\lfloor\frac{n}{2}\rfloor}L(\rkhat).
\end{align*}
Thus, we obtain
\begin{prop}(\cite{KacW}, 4.2.2).
In the complex case the spin
representation of $\k\times \k$ restricts to
$2^{\lfloor\frac{rk(\k)}{2}\rfloor}$ copies of the $\ka$-module $L(\rkhat)$.
\end{prop}

\subsection{$\g$ simple case}
We assume now that $r$ is odd  and
$\k$ is a semisimple symmetric subalgebra of a simple algebra $\g$.

\paragraph{Structure theory.}  
By the classification of Lie algebra involutions 
(see \cite{Kac}, Ch.8), we have that there exists
$p\in \{0,\dots,n\}$ such that  $ka_p=2$ and $s_{p}=1$  while $s_{i}=0$
for $i\ne p$.
 Set $\o_p$ to be the unique element of $\h_0$ such that
$\ov\a_i(\o_p)=\d_{ip}$ for
$i=1,\dots,n$.
Set
$$
\mu=\si\circ\exp(\pi iad(\o_p)).
$$
Let $\k_\mu$
denote the set
of $\mu$-fixed points in $\g$.

It is easy to show that $\h_0$ is a Cartan subalgebra of $\k_\mu$:
if $\h'$ is a Cartan subalgebra of $\k_\mu$ containing $\h_0$ then
$[\h',\o_p]=0$,
so $\h'\subset\k$. This implies $\h'=\h_0$. If 
 $m$ is a positive integer such that $\si^m=\mu^m=id$, then,
by Proposition 8.5 of \cite{Kac} (with notation therein) the map
$t^{\frac{m}{2}\o_p}$ is
an isomorphism
$L(\g,\mu,m)\to L(\g,\si,m)$.
In particular
the linear map $t_p:\ov\a+i\d'\mapsto \ov\a-\tfrac{m}{2}(i-\ov\a(\o_p))\d'$
defines a bijection between
$\Da$
and the set of
$\ha$-roots of
$\widehat L(\g,\mu,m)$. It follows that $t_p(\Pia)$
is a set of simple roots for
$t_p(\Dap)$.\par
If $z\in\ganz$, set $L(\g,\mu,m)_z=\{x\in L(\g,\mu,m)\mid d'\cdot x=zx\}$.
Since $t_p(\a_i)=\ov\a_i$ if $i>0$, $t_p(\a_0)=
\frac{m}{ka_0}\d'+\ov\a_0$
and
$\k_\mu= L(\g,\mu,m)_0$, we see
that the set of $\h_0$-roots of $\k_\mu$ is
$\D_f:=\{\ov\a\mid \a\in\Da,\, m_0(\a)=0\}$.

Clearly
 $\Pi_f=\{\ov\a_1,\dots,\ov\a_n\}$ is a set of simple roots for $\D_f$ and
 the corresponding set of positive roots is
 $\Dp_f=\{\ov\a\mid \a\in\Dap,\, m_0(\a)=0\}$.

\paragraph{Explicit description of $\D_\k$ and $\D(\p)$.}
Set
$$
\D_f^0=\{\a\in \D_f\mid , \a(\o_p)\equiv
0\mod 2\},\,
\D_f^1=\{\a\in \D_f\mid \a(\o_p)\equiv 1\mod
2\}
$$
and
let $\D_{f, s}$ and $\D_{f, l}$ be, respectively, the set
of short and long roots in $\D_f$. We let $\D_x^\e=\D_x\cap\D_f^\e$
($x=f,l$ or $f,s$; $\e=0,1$).
%It is clear that if $\a\in\D_f$ and $\mathfrak s_\a$ is any
%$\h_0$-stable subspace of $\g$ of weight $\a$, then
%$\eta_{|\mathfrak s_\a}=1$ (resp. $\eta_{|\mathfrak s_\a}=-1$)
%if and only if $\a\in\D_f^0$ (resp. $\a\in\D_f^1$).
\par

Recall from Section 2 our classification of $\mathfrak z$-roots
of $\g$ into complex, compact, and noncompact roots.
Set
\begin{align*} &\D_{cx} =\{\a\in\D(\p)\mid \a=\be_{|\h_0},\text{
$\be$ complex}\},\\ &\D_{ci}=\{\a\in\D_\k\mid \a=\be_{|\h_0},\text{
$\be$ compact}\},\\ &\D_{ni}=\{\a\in\D(\p)\mid
\a=\be_{|\h_0},\text{ $\be$ noncompact}\};\\
&\Da_{cx}=\{i\d_\k+\a\mid i\in\ganz,\,\a\in\D_{cx}\},\\
&\Da_{ci}=\{i\d_\k+\a\mid i\in\ganz,\,\a\in\D_{ci}\},\\
&\Da_{ni}=\{i\d_\k+\a\mid
i\in\ganz,\,\a\in\D_{ni}\}.\end{align*}

If $k=1$ then $\mathfrak z=\h_0$ and $\si$ is of inner type. It follows that
$\si=\exp(\pi i h)$ for some $h\in \h_0$. Since
$\si(X_{j})=e^{\pi i\ov\a_j(h)}X_{j}=e^{\pi is_j}X_{j}$ for
$j=1\dots,n$, we see that $\si=\exp(\pi iad(\o_p))$ and $\mu=id$. Hence,
in this case,
\begin{equation}\label{radici1}
\D_{cx}=\emptyset,\quad\D(\p)=\D^1_f=\D_{ni},\quad\D_\k=\D^0_f=\D_{ci}.
\end{equation}

Suppose now that $k=2$, so that $\d'=\d$. Recall from \ref{prelims} that
$\a\in\D$ is a
noncompact root if and only if  $\d+\a_\rao$ is a long root of $\Da$,
$\a$ is compact if and only if $\a_\rao$ is a long root of $\Da$, and
 $\a$ is complex if and only if
$\a_\rao\in\Da$ and it is not a long root.\par
Assume that  $k=2$ and $\widehat L(\g,\si)$ is not of type
$A^{(2)}_{2n}$. The following relations hold.
$$\D_\rao=(\D(\p)\setminus\{0\})\cup\D_\k=\overline{\Da}\setminus\{0\}=\D_f$$
The first equality is clear, the second depends on the
fact that $\Da$ is the set of roots of $\widehat
L(\g,\sigma)$, whereas the third follows from  the explicit
description of $\Da$ given in Proposition 6.3~a)  of
\cite{Kac}.
From the above discussion it follows that
\begin{equation}\label{radici2}
\D_{cx}=\D_{f,s},\quad\D_{ci}=\D^0_{f,l},\quad\D_{ni}=\D^1_{f,l}.
\end{equation}
Moreover
$$\D(\p)=\D_{ni}\cup\D_{cx}=\D^1_{f,l}\cup\D_{f,s},
\qquad
\D_\k=\D_{ci}\cup\D_{cx}=\D^0_{f,l}\cup\D_{f,s}.$$

 If
$\widehat L(\g,\si)$ is of type
$A^{(2)}_{2n}$, then $\D_\k$ is the subsystem of $\Da$ generated by
$\{\a_0,\dots,\a_{n-1}\}$. It follows that
$\D_\k$ does not contain long roots of $\Da$, hence $\D_\k=\D_{cx}$. Since
$\ov{\Da}=\D_f\cup\half\D_{f,l}\cup\{0\}$ (see again
\cite{Kac}, Prop. 6.3~b)), arguing as above we have
\begin{equation}\label{radici3}
\D_{ni}=\D_{f,l}\quad\D_{cx}=\D_\k=\half\D_{f,l}\cup\D_{f,s}.
\end{equation}
\vskip20pt
As we have seen in Section \ref{compatibile}, the explicit realization of the spin module depends on the
choice of
 a  set of positive roots 
$\D$ for $\g$ that is compatible with
$\Dp_\k$. We make a particular choice that we now explain.\par
Let $u$ be the longest element in the Weyl group of $\k$, $u'$ the
longest element in the parabolic subgroup corresponding to
$\Pi_\k\setminus\{\a_0\}$, and $w_0=uu'$.
Clearly $w_0$ stabilizes both $\D_\k$ and
$\D(\p)$, hence $\D_{|\h_0}\subset w_0(\half\D_f\cup\D_f)$.
 It is easy to see that
$\Dp_\k\subset w_0(\half\D^+_f\cup\D^+_f)$.
It follows that
$$\Dp=\{\a\in\D\mid\a_\rao\in w_0(\half\Dp_f\cup\Dp_f)\}
$$
is a positive set of roots  for $\D$
compatible with $\Dp_\k$.
Recall that we set $\Dp(\p)=\Dp_{|\h_0}\cap\D(\p)$.
We let
\begin{align*}
\Dp_{cx}=\D_{cx}\cap\Dp_\k,\quad\Dp_{ci}=\D_{ci}\cap\Dp_\k,
\quad
\Dp_{ni}=\D_{ni}\cap\Dp(\p)\\
\Dap_a=\Dp_a\cup\{j\d_\k+\a\mid
j>0,\,\a\in\D_{a}\}\quad(a=cx,ci,ni).
\end{align*}

\paragraph{The algebras $L^\prime(\g,\si)$.} Recall that $(\cdot,\cdot)_n$ denotes the normalized invariant form on $\g$.
Since there is
$\ov\a_i\in\D_f$ such that $\a_i$ is long in $\Da$, it follows that
$(\cdot,\cdot)_n{}_{|\k_\mu}$ is the normalized invariant form on $\k_\mu$.
If
$\D_f$ is a root system of type $Y_n$, we can realize the affine Lie algebra of
type $Y_n^{(1)}$ as the subalgebra $\ka_\mu=L(\k_\mu)\oplus \C
K'\oplus  \C d'$ of $\widehat L(\g)$. We set  $(\cdot,\cdot)=(\cdot,\cdot)_n$ in
\eqref{bracket}, so that $K'$ is the canonical
central element of $\ka_\mu$.
We
denote by $\Da_\mu$ the set of roots of $\ka_\mu$ with
respect to $\ha$ and by $\Wa_{\k_\mu}$ its Weyl group.
If $\ov\th_f$ is the highest root of $\D_f$
with respect to
$\Pi_f$, then $\Pia_\mu=\{-\ov\th_f+\d', \ov\a_1,\dots ,
\ov\a_n\}$ is a set of simple roots of $\ka_\mu$ with
respect to $\ha$. With this choice of the simple roots,
the set of positive roots is
$$\Dap_\mu=\D_f^+\cup
((\D_f\cup\{0\})+\ganz^+\d').
$$
 Let $\L_\mu$ be the
linear functional on $\ha$ which maps $K'$ to $1$ and
$\h_0+\C d'$ to $0$.
\par
Let $(\cdot,\cdot)^\mu$ be the
normalized invariant form  of $\ka_\mu$ such that $(\L_\mu,\L_\mu)^\mu=0$ and
let $\nu:\ha\to\ha^*$ be the isomorphism induced by
$(\cdot,\cdot)^\mu$.
For any subset $R$ of real roots in $\Da_\mu$ we set
$R^\vee=\{\nu(\a^\vee)\mid \a\in R\}$.
Then it is clear that, if $A$ is the generalized Cartan matrix of $\ka_\mu$,
then
$(\ha,
\Pia_\mu^\vee, \nu^{-1}(\Pia_\mu))$ is a
realization of the Cartan matrix ${}^tA$. Let
$\ka_\mu^\vee=\g({}^tA)$ be the twisted affine algebra corresponding
to the given realization of ${}^tA$. By general theory of root systems,
 the set of real roots of $\ka_\mu^\vee$ is  $\Da_{\mu,re}^\vee$,  where
 $\Da_{\mu,re}$ is the set of real roots of $\ka_\mu$. Since
 $(\cdot,\cdot)^\mu$ is a normalized form on  $\ka_\mu$, we have
 that the set of imaginary roots for $\ka_\mu^\vee$ is
$\ganz^*\d'$. It follows that the set of roots
of $\ka_\mu^\vee$ is $\Da_\mu^\vee:=\Da_{\mu,re}^\vee\cup\ganz^*\d'$.
Observe that if $\widehat L(\g,\sigma)$ is of type
$X^{(2)}_N=A_{2l-1}^{(2)},\,D_{l+1}^{(2)},\,E_{6}^{(2)}$, then
$\ka_\mu^\vee$ is of type
$X_{N'}^{(2)}=D_{l+1}^{(2)},\,A_{2l-1}^{(2)},\,E_{6}^{(2)}$
respectively (see \cite{Kac}, 13.9). Moreover, since
$\nu(\a^\vee)=\frac{2}{(\a,\a)^\mu}\a$,  the Weyl group of
$\ka_\mu^\vee$ is
$\Wa_{\k_\mu}$.\par
\begin{rem}\label{molteplicitaimmaginarie} If $rk(\g)=N$, then the number of
short roots in
$\Pi_f^\vee$ is $2n-N$, therefore, as a root of
$\ka_\mu^\vee$, $j\d'$ has multiplicity $2n-N$ if $j$ is
odd and $n$ if $j$ is even.
\end{rem}

We define the Lie
algebra
$L^\prime(\g,\si)$ as follows
$$
L^\prime(\g,\si)=\begin{cases}
\ka_\mu\quad&\text{if $k=1$,}\\
\ka_\mu^\vee&\text{if $k=2$ and $a_0=1$,}\\
\widehat L(\g,\si)^\vee&\text{if $k=a_0=2$.}
\end{cases}
$$
In the last case $\widehat L(\g,\si)$ is of type $A^{(2)}_{2n}$ and
$\widehat L(\g,\si)^\vee$ is realized with a construction analogous to
that performed for $\ka_\mu^\vee$, using the normalized invariant form
of $\widehat L(\g,\si)$.  In particular,
the set of roots of $L'(\g,\si)$ is
\begin{equation}\label{radicia2}
(\half\D_{f,l}+\half(2\ganz-1)\d')
\cup(\D_{f,s}+\ganz\d')\cup(\D_{f,l}+(2\ganz)\d')\cup{\ganz^*\d'}.
\end{equation}

We will denote by $\Da'$ the set of roots of $L'(\g,\si)$ in all cases.
We choose
$(\Da')^+=(\half\Dap_\mu\cup\Dap_\mu)\cap\Da'$ as a set of positive
roots and notice that
the corresponding set $\Pia'$ of simple roots is $\Pia_\mu$,
$\Pia_\mu^\vee$, and
$\{\half(\d'-\ov\theta_f),\ov\a_1,\dots,\ov\a_n\}$ if $a_0k=1$, $2$, and $4$
respectively.
 Let $\rhat'$ denote the sum of the fundamental weights of $L'(\g,\si)$.
Observe that the 
  Weyl group of
 $L'(\g,\si)$ is  $\Wa_{\k_\mu}$.

\paragraph{The map  $\psodd:\ha_\k\to\ha$.}
We already observed that 
$(\cdot,\cdot)_n{}_{|\k_\mu}=(\cdot,\cdot)^\mu_{|\k_\mu}$.
Also recall that we let $c_S=\frac{(h,h)}{(h,h)_S}$, where $h$ is any nonzero
element of $\h_0\cap\k_S$.
 It follows
from the discussion preceding Lemma~\ref{normalized} that
$$
(h ,h)^\mu=kc_S (h , h)_S.
$$
Consider the linear map
$\phi:\ha_\k\to \ha$ defined by
$$
\phi_{|\h_0}=id_{\h_0},\quad \phi(d_\k)=d',\quad
\phi(K_S)=k c_SK'.
$$
We define
\begin{equation}\label{psispin}
\psodd=\phi\circ w_0^{-1}
\end{equation}
so that
$$
\stodd=w_0\circ\phi^*.
$$
 It is clear that $\psodd$ is surjective, hence
$\psodd$ is injective. We denote by $\stodd{}^{-1}$
the inverse of $\stodd:\ha^*\to\stodd(\ha^*)$.

It is immediate from the definition of $\psodd$  that
\begin{align}\label{psistaresplicita1}
&\stodd(\L_\mu)=\sum_Skc_S\L^S_0,\\
\label{psistaresplicita2}
&\stodd(\d')=\d_\k,\\
\label{psistaresplicita3}
&\stodd(\lambda)=w_0(\lambda)\text{ for $\lambda\in\h_0^*$}.
\end{align}
Note that, by \eqref{psistaresplicita2}, \eqref{psistaresplicita3} and
relation $w_0(\D_\k)\subseteq\D_\k$ we have that 
$\Da_\k\subset\stodd(\ha^*)$, hence
$\stodd(\ha^*)$ is $\Wa_\k$-stable.

\begin{lemma}
\label{Weylgroups} For $\a\in \Da_\k$, let $\be$ be the unique element of
 $\Da_\mu$ such that $\stodd(\be)$ is a
multiple of $\a$. Let $s_\a:\ha_\k^*\to \ha_\k^*$ be the reflection with
respect
 to $\a$ and $s'_\be:\ha^*\to
\ha^*$ the reflection  with respect to $\be$. Then
$$
{\stodd}^{-1}s_\a\stodd=s'_{\be}.
$$
\end{lemma}
\begin{proof} The proof is the same as for Lemma~\ref{embed}.\end{proof}

\begin{rem}\label{togliqueste}
We set $\Wa_{\si,\rodd}=(\stodd)^{-1}\Wa_\k\stodd$.
Lemma~\ref{Weylgroups} says that $\Wa_{\si,\rodd}$ is a subgroup of $\Wa_{\k_\mu}$.

If $\widehat L(\g,\si)$ is not of type $A^{(2)}_{2n}$ then, by
Lemma~\ref{Weylgroups},
 $\Wa_{\si,\rodd}$ is generated by the
reflection $s_\a$ with $\a$ a real root in $\stodd{}^{-1}(\Da_\k)$. By
\eqref{psistaresplicita2},
and \eqref{radici1}--\eqref{radici2}, we have that the set of real roots in
$\stodd{}^{-1}(\Da_\k)$ is
\begin{equation}\label{toglialtri}
\Da_{\si,\rodd}:=(\D_{cx}\cup\D_{ci})+\ganz\d'.
\end{equation}

If  $\widehat L(\g,\si)$ is of type  $A^{(2)}_{2n}$, then,
by Lemma~\ref{Weylgroups} and \eqref{radici3}, we have that
$\Wa_{\si,\rodd}$ is the subgroup of $\Wa_{\k_\mu}$ generated by all
reflections with
respect to roots in
\begin{equation}\label{toglia2}
\Da_{\si,\rodd}:=(\D_{f,s}+ \ganz\d')\cup(\D_{f,l}+2\ganz\d').
\end{equation}
\end{rem}

\paragraph
{The character.}
We set $$\Dap_{re}(\p)=\Dp(\p)\cup\{\a+j\d_\k\mid \a\in
\D(\p)\setminus
\{0\},\ j\in \ganz^+\}.$$ From
\eqref{peso} we obtain directly
\begin{equation}\label{ch}
ch(X_r)=
2^{\lfloor\frac{N-n}{2}\rfloor}
e^{\sum_S j_S\L_0^S+\rho_n}
\prod_{j>0}(1+e^{-j\d_\k})^{N-n}
\prod_{\a\in\Dap_{re}(\p)}(1+e^{-\a}).
\end{equation}

Recall that
$$
D_\k=e^{\rhat_\k} \prod_{i>0}(1-e^{-i\d_\k})^n
\prod_{\a\in(\Dap_\k)_{re}}(1-e^{-\a}),
$$
and set
$$
\rho^*=\sum_S j_S\L_0^S+\rho_n+\rhat_\k.
$$

Then dividing and multiplying \eqref{ch} by $D_\k$ yields
\begin{equation}\label{parziale}
ch(X_r)=2^{\lfloor\frac{N-n}{2}\rfloor}{\num^+}/{D_\k},
\end{equation}
where
\begin{align*}
\num^+&=
e^{\rho^*}
\prod_{i>0}(1+e^{-i\d_\k})^{N-n}
\prod_{i>0}(1-e^{-i\d_\k})^n
\\
&\times
\prod_{\a\in\Dap_{ni}}(1+e^{-\a})
\prod_{\a\in\Dap_{cx}}(1-e^{-2\a})
\prod_{\a\in\Dap_{ci}}(1-e^{-\a}).
\end{align*}
If $\widehat L(\g,\si)$ is not of type $A^{(2)}_{2n}$ set
\begin{align*}
\num^-&=
e^{\rho^*}
\prod_{i>0}(1+e^{-i\d_\k})^{N-n}
\prod_{i>0}(1-e^{-i\d_\k})^n
\\
&\times
\prod_{\a\in\Dap_{ni}\cup\Dap_{ci}}(1-e^{-\a})
\prod_{\a\in\Dap_{cx}}(1-e^{-2\a}).
\end{align*}
Observe that $\num^-$ differs from $\num^+$ just in
the product over $\Dap_{ni}$.

If $\widehat L(\g,\si)$ is of type $A^{(2)}_{2n}$ then
set
\begin{align*}
&\Da_{ni}^{even}=\Dp_{ni}\cup(\D_{ni}+2\ganz^+\d_\k)\\
&\Da_{ni}^{odd}=\Dap_{ni}\setminus\Da_{ni}^{even}\\
&\Dap_{f,cx}=(\Dp_{cx}\cap\D_f)\cup(\D_{cx}\cap\D_f+\ganz^+\d_\k).
\end{align*}
Recalling that in this case $N=2n$ and that, by \eqref{radici3},
$\D_{cx}=\half\D_{ni}\cup(\D_{cx}\cap\D_f)$,
we can
rewrite $\num^+$ as
\begin{align*}
\num^+&=
e^{\rho^*}
\prod_{i>0}(1+e^{-i\d_\k})^{n}
\prod_{i>0}(1-e^{-i\d_\k})^n
\\
&\times
\prod_{\a\in\Da_{ni}^{odd}}(1+e^{-\a})
\prod_{\a\in\Da_{ni}^{even}}(1+e^{-\a})
\prod_{\a\in\Da_{ni}^{even}}(1-e^{-\a})
\prod_{\a\in\Dap_{f,cx}}(1-e^{-2\a})\\
&=
e^{\rho^*}
\prod_{i>0}(1-e^{-2i\d_\k})^{n}
\prod_{\a\in\Da_{ni}^{odd}}(1+e^{-\a})
\prod_{\a\in\Da_{ni}^{even}}(1-e^{-2\a})
\prod_{\a\in\Dap_{f,cx}}(1-e^{-2\a}).
\end{align*}
In this case we set
$$
\num^-=e^{\rho^*}
\prod_{i>0}(1-e^{-2i\d_\k})^{n}
\prod_{\a\in\Da_{ni}^{odd}}(1-e^{-\a})
\prod_{\a\in\Da_{ni}^{even}}(1-e^{-2\a})
\prod_{\a\in\Dap_{f,cx}}(1-e^{-2\a}),
$$
that differs from $\num^+$ just in
the product over $\Da^{odd}_{ni}$.

First we show how to compute  $\num^-$ and then we shall compute $\num^+$
from $\num^-$.
\begin{lemma}\label{strho}
\begin{equation*}
\rho^*=
\stodd(a_0\rhat')
\end{equation*}
\end{lemma}
\begin{proof}
 We start from the equal rank case. In this case
$\rhat'=h^\vee\L_\mu+\rho$, where $\rho$ is half the sum of the roots in
$\Dp_f$.
It follows that
$\stodd(\rhat')=h^\vee\stodd(\L_\mu)
 +\stodd(\rho)$.
Since
\begin{equation*}
\stodd(\rho)=w_0(\rho)
 =\half\sum_{\a\in\Dp_\k}\a+\half\sum_{\a\in\Dp(\p)}\a=
\rho_\k+\rho_n \end{equation*}
we can write that
$\stodd(\rhat')=h^\vee\stodd(\L_\mu)+\rho_n+\rho_\k$.
By \eqref{psistaresplicita1},
$\stodd(\L_\mu)=\sum c_S\L_0^S$ Hence, by \eqref{js},  we conclude
that $\stodd(\rhat')=\sum j_S\L_0^S+\rho_n+\rkhat $
as desired.

If $k=2$ and $\widehat L(\g,\si)$ is not of type $A^{(2)}_{2n}$, denoting
by $(h')^\vee$
the dual Coxeter number of $L'(\g,\si)$,
 we have
$\rhat'=(h')^\vee\L_\mu+\rho^\vee$, where $\rho^\vee$ is half the sum of the
roots in
$(\Dp_f)^\vee$ and in turn
$\stodd(\rhat')=(h')^\vee\stodd(\L_\mu)
 +\stodd(\rho^\vee)$.
Since
\begin{equation*}
\stodd(\rho^\vee)=w_0(\rho^\vee)
 =\half\sum_{\a\in\Dp_{ni}}\a+\half\sum_{\a\in\Dp_{ci}}\a+
\sum_{\a\in\Dp_{cx}}\a=
\rho_\k+\rho_n \end{equation*}
we need only to check that $\stodd((h')^\vee\L_\mu)=\sum_S
(j_S+h_S^\vee)\L_0^S$. But  $\stodd((h')^\vee
\L_\mu)=\sum_S(h')^\vee 2c_S\L_0^S$ which equals, by \eqref{js}, $(h')^\vee
\sum_S
\frac{j_s+h^\vee_S}{h^\vee}\L_0^S$. The claim follows because
$h^\vee=(h')^\vee$.

Finally, if $\widehat L(\g,\si)$ is of type $A^{(2)}_{2n}$,
we have
$2\rhat'=h^\vee\L_\mu+2\rho$. Now, from \eqref{radici3}, we
obtain that
\begin{align*}\stodd(2\rho)&=\sum_{\a\in\Dp_{f,s}}\a+
\sum_{\a\in\Dp_{f,l}}\a\\&
=\frac{1}{2}(\sum_{\a\in \Dp_{f,l}}\a+\sum_{\a\in
\frac{1}{2}\Dp_{f,l}}\a+\sum_{\a\in \Dp_{f,s}}\a)\\
&+\frac{1}{2}(\sum_{\a\in \Dp_{f,s}}\a+\sum_{\a\in
\frac{1}{2}\Dp_{f,l}}\a)\\
&=\rho_n+\rho_\k.\end{align*}
Finally, by \eqref{js}, $\stodd(h^\vee\L_\mu)=h^\vee\sum_S 2 c_S\L_0^S=
h^\vee\sum_S\frac{j_S+h_S^\vee}{h^\vee}\L_0^S$ and we
conclude as above.
\end{proof}

\begin{prop}\label{denominat}
$$
\num^-=e^{\stodd(a_0\rhat')}\prod_{\a\in(\Da')^+}(1-e^{-\stodd(a_0\a)})^{m_\a},
$$
where $m_\a$ is the multiplicity of $\a$ as a root of $L'(\g,\si)$.
\end{prop}
\begin{proof}
If $\widehat L(\g,\si)$ is not of type $A^{(2)}_{2n}$, then
formulas \eqref{psistaresplicita1}--\eqref{psistaresplicita3}
imply that $\stodd$ is a bijection between the set $(\Da'_{re})^+$
of positive real roots in $\Da'$ and $\Dap_{ni}\cup\Dap_{ci}
    \cup 2\Dap_{cx}$.
Hence
\begin{equation*}
\num^-=
e^{\rho^*}
\prod_{i>0}(1+e^{-i\d_\k})^{N-n}
\prod_{i>0}(1-e^{-i\d_\k})^n
\prod_{\a\in(\Da'_{re})^+}(1-e^{-\stodd(\a)}).
\end{equation*}
Next we observe that
\begin{align}
\label{immaginarie}
\prod_{i>0}(1+e^{-i\d_\k})^{N-n}
\prod_{i>0}(1-e^{-i\d_\k})^n=
&\prod_{i>0}(1-e^{-2i\d_\k})^{N-n}
 \prod_{i>0}(1-e^{-i\d_\k})^{2n-N}=
\notag\\
&\prod_{i>0}(1-e^{-2i\d_\k})^n
 \prod_{i>0}(1-e^{-(2i-1)\d_\k})^{2n-N}.
\end{align}
hence, using Remark \ref{molteplicitaimmaginarie},
$$
\num^-=e^{\rho^*}
\prod_{\a\in(\Da')^+}(1-e^{-\stodd(\a)})^{m_\a}.
$$
Applying Lemma~\ref{strho}, we obtain the result in this case.

If $\widehat L(\g,\si)$ is of type $A^{(2)}_{2n}$, then, by \eqref{radicia2},
 $\stodd$ defines a bijection between
$2(\Da'_{re})^+$ and
$\Da_{ni}^{odd}\cup2\Da_{ni}^{even}\cup2\Dap_{f,cx}$ hence
$$
\num^-=e^{\rho^*}
\prod_{i>0}(1-e^{-2i\d_\k})^{n}
\prod_{\a\in(\Da'_{re})^+}(1-e^{-2\stodd(\a)}).
$$
By Remark \ref{molteplicitaimmaginarie}, $m(j\d')=n$ for all $j$, hence
$$
\num^-=e^{\rho^*}
\prod_{\a\in(\Da')^+}(1-e^{-2\stodd(\a)})^{m_\a}.
$$
Lemma~\ref{strho} implies the result in this case too.
\end{proof}

Applying Weyl-Kac denominator formula we readily obtain
\begin{cor}\label{WeylKac}
$$
\num^-
=e^{\rho^*}
\sum\limits_{w\in\Wa_{\k_\mu}}
\epsilon(w)e^{a_0(\stodd(w(\rhat')-\rhat'))}.
$$
\end{cor}

\paragraph{Decomposition of $X_r$.}
We now show how to compute $\num^+$ from $\num^-$. This will allow us to
compute
the decomposition of $X_r$.

For $\gamma_1, \dots, \gamma_t\in a_0\stodd((\Da')^+)$, we set
\begin{equation}\label{epsilon}
\e_p(\gamma_1, \dots, \gamma_t)=
\begin{cases}(-1)^{|\{\gamma_1,\dots,\gamma_t\}\cap\Da^{odd}_{ni}|}&
\text{ if
$\widehat L(\g,\si)$ is of type $A^{(2)}_{2n}$}\\
(-1)^{|\{\gamma_1,\dots,\gamma_t\}\cap\Dap_{ni}|}&\text{ otherwise}.
\end{cases}
\end{equation}
Set
$$
h_\si=\begin{cases}d_\k\quad&\text{ if $\widehat L(\g,\si)$ is of type
$A^{(2)}_{2n}$}\\
 \o_p&\text{otherwise}.
 \end{cases}
 $$

By the explicit description of $a_0\stodd((\Da')^+)$ given in the proof
of Proposition \ref{denominat} it is clear
from \eqref{radici1}--\eqref{radici3} that
$(-1)^{(\gamma_1+\cdots+\gamma_t)(h_\si)}= \e_p(\gamma_1, \dots,
\gamma_t)$.  In particular, if we define a function $\e_p$ on the
$\ganz$-lattice $L$ generated by
$a_0\stodd((\Da')^+)$ by setting
\begin{equation*}
\e_p(\l)=(-1)^{\l(h_\si)}.
\end{equation*}
 then, if $\l=\gamma_1+\dots+\gamma_t$ with $\gamma_i\in
a_0\stodd((\Da')^+)$, we have
\begin{equation}\label{epsilonp}
\e_p(\l)=\e_p(\gamma_1,\dots,\gamma_t).
\end{equation}

\begin{lemma}\label{dpiudmeno} We have
$$
\num^-=e^{\rho^*}\sum_{\l\in L} a_\l e^\l,
$$
with $a_\l\in
\ganz$. Moreover
$$
\num^+=e^{\rho^*}\sum_{\l\in L} \e_p(\l)a_\l e^\l.
$$
\end{lemma}
\begin{proof}
By Corollary~\ref{WeylKac}
$$
\num^-
=e^{\rho^*}
\sum\limits_{w\in\Wa_{\k_\mu}}
\epsilon(w)e^{-a_0(\stodd(<N'(w)>))},
$$
where $N'(w)=\{\a\in(\Da')^+\mid w^{-1}(\a)<0\}$, hence the first
assertion follows. The second statement follows
directly from \eqref{epsilonp} and the definition of $\num^+$ and $\num^-$.
\end{proof}

We set $\Da'_{\si,\rodd}=\Da_{\si,\rodd}$ (see \eqref{toglialtri}) if $k=1$ or $\widehat L(\g,\si)$ is of
type $A^{(2)}_{2n}$, while we set
$\Da'_{\si,\rodd}=(\Da_{\si,\rodd})^\vee$ in the other cases. We notice that
$\Da'_{\si,\rodd}$ is a root
system contained in
 $\Da'$ and its associated reflection group is $\Wa_{\si,\rodd}$. By
general theory of reflection
groups (see \cite{Dyer}) the set $$W'_{\si,\rodd}=\{u\in\, \Wa_{\k_\mu}\mid
N'(u)\subseteq
\Da'\setminus\Da'_{\si,\rodd}\}$$
 is a set of minimal coset representatives  of
$\Wa_{\si,\rodd}\backslash \Wa_{\k_\mu}$.
\par
For $w\in \Wa_{\k_\mu}$ set $N^*(w)=N'(w)\cap \Da'_{\si,\rodd}$. Set also
$\ell(w)=|N'(w)|$ and, if $v\in\,\Wa_{\si,\rodd}$, $\ell^*(v)=|N^*(v)|$. 
Now assume $v\in\,\Wa_{\si,\rodd}$
and $u\in W'_{\si,\rodd}$. Since $v(\Da'_{\si,\rodd})=\Da'_{\si,\rodd}$,  we have that
$vN'(u)\subseteq
\Da'\setminus\Da'_{\si,\rodd}$. It is a standard fact that
$N'(v)\subset\Da'_{\si,\rodd}$. In particular
$N'(vu)=N'(v)\,\dot\cup\, (v N'(u)\cap(\D'_\mu)^+)$ (disjoint union),
whence, $N'(vu)\cap
\Da'_{\si,\rodd}=N'(v)\cap \Da'_{\si,\rodd}=N^*(v)$.

 If $\e$ and
$\e^*$ denote the sign functions in $\Wa_{\k_\mu}$ and $\Wa_{\si,\rodd}$, respectively,
then $\e(w)=(-1)^{\ell(w)}$
and $\e^*(v)=(-1)^{\ell^*(v)}$. Notice that the set of real roots in
$\Da'\setminus \Da'_{\si,\rodd}$ maps under $a_0\stodd$ bijectively onto
$\Da_{ni}^{odd}$ if $\widehat L(\g,\si)$ is of type $A^{(2)}_{2n}$, and
onto $\Da_{ni}$
in all the other
cases, therefore
\begin{equation}\label{segni}
\e(vu)=\e^*(v)\e_p(a_0\stodd\la N'(vu)\ra).
\end{equation}
It follows from Corollary~\ref{WeylKac} and \eqref{segni} that
$$
\num^- =e^{\rho^*} \sum\limits_{v\in\Wa_{\si,\rodd}} \sum\limits_{u\in
W'_{\si,\rodd}} \e^*(v) \e_p(a_0\stodd\la N'(vu)\ra)
e^{-a_0\stodd(\la N'(vu)\ra)},
$$
and, by Lemma \ref{dpiudmeno},
$$
\num^+ =e^{\rho^*} \sum\limits_{v\in\Wa_{\si,\rodd}} \sum\limits_{u\in
W'_{\si,\rodd}} \e^*(v) e^{-a_0\stodd\la N'(vu)\ra}.
$$
Clearly, if $v\in \Wa_\k$, then $\e^*(\stodd{}^{-1}v\stodd)=det(\stodd{}^{-1}
v\stodd)=det(v)$. Therefore from the above
equation we obtain
$$
\num^+= \sum\limits_{v\in\Wa_{\si,\rodd}} \sum\limits_{u\in W'_{\si,\rodd}} \e^*(v)
e^{a_0\stodd(vu\rhat')} = \sum\limits_{u\in
W'_{\si,\rodd}} \sum\limits_{v\in \Wa_\k} det(v) e^{v(a_0\stodd(u\rhat'))},
$$
and, since $ch(X_r)=2^{\lfloor\frac{N-n}{2}\rfloor} \frac{\num^+}{D_\k}$,
from \eqref{caratterek} we deduce the
following result.
\begin{prop}\label{spindec}If $\k$ is semisimple and $r$ is odd, then
\begin{equation}
\label{decomposizione}
ch(X_r)=2^{\lfloor\frac{N-n}{2}\rfloor} \sum\limits_{u\in W'_{\si,\rodd}}
ch(L(a_0\stodd(u\rhat') -
\rhat_\k)),\end{equation}
where $\psodd$ is defined by \eqref{psispin}.
\end{prop}

\subsection{Combinatorial interpretation of decompositions of spin modules.}
We will use the following general facts.
Let $\ga_1, \ga_2$ be two affine Kac-Moody algebras,
and for $i=1,2$, let $\ha_i$ be a Cartan subalgebra of
$\ga_i$, $\Da_i$ be the corresponding root system, and
$\Wa_i$ be the Weyl group. Endow $\ga_i$, with a fixed
arbitrary invariant form and $\ha_i^*$ with the form induced by this choice.
We say that a linear isomorphism $f:\ha_1^*\to \ha_2^*$ is
an extended isomorphism of root systems if $f$ is an
isometry and $f\Da_1=\Da_2$. For $\a\in \Da_i$ let $s_\a$
be the reflection on $\ha_i$ with respect to $\a$. The
following result is clear.

\begin{lemma}\label{banale}
If $f$ is an extended isomorphism of root systems, then,
for all $\a\in \Da_1$, $f s_\a f^{-1}=s_{f\a}$. In
particular, if $A\subset \Da_1$, and $W_A$ is the subgroup
of $\Wa_1$ generated by the reflections with respect to
elements of $A$, then $f W_Af^{-1}$ is the subgroup of
$\Wa_2$
generated by the reflections with respect to
elements in $fA$.
\end{lemma}

\begin{lemma}\label{estensione}
Let $f:\h_1^*\oplus\C\d_1\to \h_2^*\oplus\C\d_2$
be a linear map such that $f\Da_1=\Da_2$ and, for all
$\a,\be\in\Da_1$, $(\a,\be)_1=(f\a,f\be)_2$. Then there
exists a unique extension of $f$ to an extended isomorphism
of root systems.
\end{lemma}

\begin{proof}
Since $\C\d_2$ is the orthogonal subspace of
$\h_2^*\oplus\C\d_2$ in $\ha^*_2$, and since $\Da_2$
spans $\h_2^*\oplus\C\d_2$, the conditions $(f\L_0^1,f\a)_2
=(\L_0^1,\a)_1$ for all $\a\in\Da_1$ determine  $f\L_0^1$
modulo $\C\d_2$.
By a direct computation we see that the further condition
$(f\L_0^1,f\L_0^1)_2=0$ determines the component in
$\C\d_2$ of $f\L_0^1$.
\end{proof}
\vskip10pt
\begin{defi}\label{noncomp}
Let us say that a $\h_0$-stable subspace $S$ of $\p$ is
{\it noncompact} if all weights of $\h_0$ on $S$ are
in $\D_{ni}$.
\end{defi}

We will describe the decomposition of $X_r$ in terms of certain noncompact
subspaces
of $\p$.
For the sake of a better exposition we discuss various cases separately:
we  consider
the equal rank case,  the case when $\widehat L(\g,\si)$
is of type $A^{(2)}_{2n}$ and 
the remaining non equal rank cases.

\paragraph{Equal rank case.}In the equal rank case all $\h_0$-stable
 subspaces of $\p$ are noncompact,
for $\D(\p)$ is equal
to $\D_{ni}$, henceforth the final outcome will be very similar to
decomposition of
the basic and vector representations.

Recall that in this case $\mu=Id$, so
 $L'(\g,\si)=\widehat L(\g)$ and $\Da'=\Da_\mu$.
The isomorphism $t^{\o_p}:L(\g,\mu,2)\to L(\g,\si)$ induces a linear
isomorphism
$g:\h_0^*+\C\d'\to \h_0^*+\C \d'$ such that $g(\Da_\mu)=\Da$. Explicitly
\begin{equation}\label{gi}
g:\l+j\d'\mapsto \l+(2 j+\l(\o_p)) \d'.
\end{equation}
By \eqref{prodottoscalare} it is clear that $g$
preserves scalar products of roots.
\par
By \eqref{radici1}, $\D_\k=\D^0_{f}$ so $g(\Da_{\si,\rodd})=\D^0_f+2\ganz\d'$.
Comparing this with \eqref{dasigmaerrepari} we see that
\begin{equation}\label{sonouguali}
g(\Da_{\si,\rodd})=\Da_{\si,0}.
\end{equation}
By Lemma \ref{banale}, $\Wa= g\Wa_{\k_\mu} g^{-1}$ and, by \eqref{sonouguali},
 $g\Wa_{\si,\rodd} g^{-1}=\Wa_{\si,0}$.
Recall that in this case  we have that $p>0$ and $a_p=2$, hence
$g(\d'-\ov\theta)=-\ov\theta=\ov\a_0=\a_0$ and $g(\ov\a_i)=\a_i$ for
$i=1,\dots,n$.
It follows that $g(\Dap_\mu)=\Dap$ hence
 $N(gug^{-1})= g(N'(u))$, for
all $u\in \Wa_{\k_\mu}$.
It follows that
$$
W'_{\si,\rodd}=g^{-1}W'_{\si,0}\,g.
$$
and $g(\rhat')=\rhat$.

Recalling that $N=n$ in this case, we can
 rewrite the decomposition of $X_r$ given in \eqref{decomposizione} as
\begin{align*}
 X_r&=\sum\limits_{u\in W'_{\si,\rodd}}
L(\sum_S j_S\L_0^S+\rho_n+\stodd(u(\rhat')-\rhat')
)\\
&=\sum\limits_{u\in W'_{\si,\rodd}}
L(\sum_S j_S\L_0^S+\rho_n-\stodd(N'(u))
)\\
&=\sum\limits_{u\in W'_{\si,0}}
L(\sum_S j_S\L_0^S+\rho_n-\stodd(g^{-1}N(u))
)
\end{align*}

Applying the discussion of \ref{abelianibasic} we deduce the
 analog of Theorem~\ref{decoeabeliani} for the spin representation in the
equal rank case:

\begin{theorem}\label{decoeabelianispin}Set $m=\lfloor\frac{dim(\p)}{2}\rfloor$.\newline
(1) Assume that $\a_p$ is a
short root. Then
$$
L(\tilde\Lambda_{m-\epsilon})=\bigoplus_{A\in \Sigma\atop
|A|\equiv \epsilon\,\,mod\,2}L\left(\L_{0,\k}+\rho_n+w_0\langle
A\rangle\right).$$
 \newline (2) Assume that $\a_p$ is a
long  root.
 We
have
\begin{align*}
L(\tilde\Lambda_{m-\epsilon})&=\bigoplus_{A\in \Sigma\atop |A|\equiv
\epsilon\,\,mod\,2}L\left(\L_{0,\k}+\rho_n+w_0\langle
A\rangle-k_A\d_\k)\right)\\
&\bigoplus
\nu L\left(\L_{0,\k}+\rho_n -w_0(y)-(k_y+1)\d_\k\right),
\end{align*}
 where
$y$ and $\nu$ are as in Theorem~\ref{decoeabeliani} (2)
and $k_A=|w_0(A)\cap\Dp(\p)|$,
$k_y=|(\a_p+\Dp_\k)\cap(\d'-
\Dp_f)|$.
\vskip5pt
Moreover, in
both cases, the highest weight vector of each component indexed by
$A\in\Sigma$ is, up to a
constant factor, the
pure spinor (of the spin representation of $Cl_r(\tilde\p)$):
\begin{equation}\label{maximas}v_A=\prod_{\a\in w_0(A)\cap\Dp(\p)}
(t^{-r'-2}e_\a)
\prod_{\a\in w_0(A)\cap(-\Dp(\p))} t^{-r'-1}e_\a,\end{equation}
where $\p_\a=\C e_\a$. An highest weight vector for the component
indexed by
$w_\s$ is
\begin{align}\label{maximals}(\prod&_{\be\in (\a_p+\Dp_\k)\cap(\d'- \Dp_f)}
t^{-r'-2}e_{-\overline\be})\\&(\prod_{\be\in
(\a_p+\Dp_\k)\cap(\d'+\Dp_f)}t^{-r'-1}e_{-\overline\be})
(t^{-r'-1}e_{-\overline\a_p})(t^{-r'-2}
e_{-\overline\a_p})\notag.
\end{align}
\end{theorem}
\begin{proof} First of all observe that a weight vector $v$ is in $X^+_r$
if and only if its weight
is equal to $\sum_S j_S\L_0^S+\rho_n+\l$, where $\l$ is
a sum of an even number of elements of $\Da_{ni}$, hence
 $L(\sum_S j_S\L_0^S-\stodd(g^{-1}N(u)) )$ occurs in $X^+_r$ if and only
if $\ell(u)$
is even.

The rest of the result now follows as in Theorem~\ref{decoeabeliani}. Only the
coefficient of $\d_\k$ needs checking.
If
$A\in\Sigma$ and
$\a\in N(w_A)$, then $m_p(\a)=1$ hence  $\a=\d'\pm\ov\a$ with
$\ov\a\in\D^1_f\cap \Dp_f$. If
$\a=\ov\a+\d'$ then $g^{-1}(\a)=\ov\a$, while, if $\a=-\ov\a+\d'$
then $g^{-1}(\a)=-\ov\a+\d'$.
Write $N(w_A)
=\{-\ov\gamma_1+\d',\ldots-\ov\gamma_s+\d',\ov\be_1+\d',\ldots,\ov\be_r+\d'\}$
with
$\ov\be_i,\ov\gamma_i\in\D^1_f$.
Hence
$\stodd(g^{-1}(N(w_A)))=w_0(\sum\ov{\be_i}-\sum\ov{\gamma_i})+s\d_\k$. Since
$A=-\ov{N(w_A)}=
\{\ov{\gamma_1},\ldots\ov{\gamma_s},-\ov{\be_1},\ldots,-\ov{\be_r}\}$
 we have that $s=|w_0(A)\cap\Dp(\p)|=k_A$. The coefficient $k_y$ is computed similarly.
\par
 It remains to check that
 $k_A=0$ for all $A\in\Sigma$  if and only if $\a_p$ is short.
 Let $W_f$  denote the Weyl group of $\D_f$. Clearly
  $\{w\in W_f\mid w(\Dp_f)\supset \stodd{}^{-1}(\Dp_\k)\}\subset W_{\si,\rodd}'$. By
  \cite[Theorem~5.12]{IMRN}, $|W_{\si,\rodd}'|=|\{w\in W_f\mid w(\Dp_f)\supset
\stodd{}^{-1}(\Dp_\k)\}|$
  if and only if $\a_p$ is short. The result follows.
\end{proof}

\paragraph{The non equal rank case with $\mathbf{a_0=1}$.}
It is clear that $\k_\mu$ is $\sigma$-stable, hence we can
consider the subalgebra of $\ka_\mu$
$$
\alkmusi=L(\k_\mu, \si_{|\k_\mu})\oplus \C K'\oplus
\C d'.
$$

Clearly,
$$
\k'=\k\cap \k_\mu, \qquad \p'=\p\cap \k_\mu
$$
are, respectively, the $1$ and $-1$ eigenspaces of
$\si_{|\k_\mu}$ on $\k_\mu$.
Let us denote by $\D_{\k'}$ the $\h_0$-roots of $\k'$, by
$\D(\p')$ the set of weights of $\h_0$ on $\p'$.

Since
$\si_{|\k_\mu}=\exp(\pi i ad(\o_p))$, it is
clear that
\begin{equation}\label{k'p'}
\D_{\k'}=\D_f^0, \qquad \D(\p')\setminus \{0\}=\D_f^1.
\end{equation}

For $w\in\Wa_{\k_\mu}$,
let $N_\mu(w)=\{\a\in\Dap_\mu\mid w^{-1}(\a)<0\}$.
Observe that $W'_{\si,\rodd}=\{u\in\Wa_{\k_\mu}\mid
N_\mu(u)\subset\Da_\mu\setminus\Da_{\si,\rodd}\}$.
This is because  both $W'_{\si,\rodd}$ and
$\{u\in\Wa_{\k_\mu}\mid N_\mu(u)\subset\Da_\mu\setminus\Da_{\si,\rodd}\}$ are the
set of
minimal length
coset representatives.
We notice that the set of real roots in $\Da_\mu\setminus \Da_{\si,\rodd}$
equals the set of real roots in
$\Da'\setminus\Da'_{\si,\rodd}$. The above observation implies that
$N_\mu(u)=N'(u)$ for $u\in W'_{\si,\rodd}$.
In particular
 $$(u\rhat'-\rhat')=
u\rhat_\mu-\rhat_\mu,$$
where $\rhat_\mu$ denotes the sum of the fundamental weights of $\ka_\mu$.

This time the isomorphism
$$
t^{\o_p}:L(\k_\mu,id,2)\to\alkmusi
$$
induces a linear isomorphism $g:\h_0\oplus \C \d'\to\h_0\oplus \C\d'$, still given by
\eqref{gi},   such
that
$g(\Da_\mu)$ is the set $\Da_{\mu,\si}$ of roots of $\alkmusi$.
By Lemma
\eqref{estensione}, we can uniquely extend $g$ to an
extended isomorphism of $\Da_\mu$ with $\Da_{\mu,\si}$,
which we still denote by $g$.
\vskip10pt
We choose $g\Pia_\mu$ as a set of simple roots
for $\Da_{\mu,\si}$, and denote by $\Dap_{\mu,\si}$ the
corresponding positive system of roots. Then it is clear
that $g$ maps $\Dap_\mu$ onto  $\Dap_{\mu,\si}$.
We denote by $\Wakmusi$ the Weyl group of $\alkmusi$ and,
for $w\in \Wakmusi$, we denote by $N_\si$ its negative set with respect to
$\Dap_{\mu,\si}$.
By Lemma \ref{banale}, $\Wakmusi= g\Wakmu g^{-1}$.
Moreover, it  is clear that $N_\si(gug^{-1})= gN_\mu(u)$, for
all $u\in \Wakmu$.

Since $W'_{\si,\rodd}=\{u\in\Wa_{\k_\mu}\mid N_\mu(u)\subset\D_{ni}+\ganz\d'\}$,
by \eqref{radici2},
 we have
that
 $gW'_{\si,\rodd}g^{-1}=\{v\in\Wakmusi\mid
N_\si(v)\subset\D^1_{f,l}+(1+2\ganz)\d'\}$.
 Since the set of real roots in
$\Da_{\mu,\si}\backslash(\D(\k')+2\ganz\d')$ is
 $\D^1_f+(1+2\ganz)\d'$ we have
 in particular that $gW'_{\si,\rodd}g^{-1}$ is precisely the set of all
elements $v$
 in $W'_{\si_{|\k_\mu},0}$ such that $N_\si(v)\subset\D_{ni}+(1+2\ganz)\d'$.
  We actually have a stronger result. 
%Recall that a subset $L$ of $\Dap$ is biconvex
%if it both $L$ and $\Dap\setminus  L$ are closed under root addition. It is well-known that if 
%$L$ is finite, then $L$ is biconvex if and only if it is of the form $N(w), w\in\Wa$.

 \begin{lemma}\label{eminuscolo}
 If $v\in\Wakmusi$ is such that $N_\si(v)\subset\D_{ni}+(1+2\ganz)\d'$, then
 $v$ is $\si_{|\k_\mu}$-minuscule.
 In particular
$$gW'_{\si,\rodd}g^{-1}=\{v\in\mathcal{W}^{\si_{|\k_\mu}}_{ab}\mid \ov{N_\si(v)}
 \subset \D_{ni}\}.$$
 \end{lemma}
 \begin{proof}
 We recall (see \cite{IMRN}) that, if $\si_{|\k_\mu}$ is of type $(s_0,\dots,s_n;k)$,
 then $v$ is $\si_{|\k_\mu}$-minuscule if
 $ht_{\si_{|\k_\mu}}(\a)=1$ for all $\a\in N_\si(v),$ where
$ht_{\si_{|\k_\mu}}(\a)=\sum s_im_i(\a)$.

 We use the well known fact that in a finite root system a long root is the sum
 of two short roots. Suppose now that $N_\si(v)\subset\D_{ni}+(1+2\ganz)\d'$
and that
 $(2m+1)\d'+\a$ is in $N_\si(v)$. By \eqref{radici2} $\a\in\D_f^1$, thus we can
write
 $\a=\be+\gamma$ with $\be\in\D^1_{f,s}$ and $\gamma\in\D_{f,s}^0$. It
follows that
 $(2m+1)\d'+\a=(2m\d'+\gamma)+(\d'+\be)$, hence, by the biconvexity property
of $N_\si(v)$,
 we find that
 $\d'+\be\in N_\si(v)$ unless $m=0$ and $\gamma\not\in\Dap_{\mu,\si}$. If
we write
 $\a=\sum_{i=1}^nm_i\ov\a_i$, then $m_p=\pm1$.
 Since $\d'+\a=\sum_{i=1}^nm_i\a_i$ if $m_p=1$
 and $\d'+\a=2\d'+\sum_{i=1}^nm_i\a_i$ if $m_p=-1$, we see that, in any case,
 $ht_{\si_{|\k_\mu}}(\d'+\a)=1$.
 \end{proof}

We identify $\D_{\k'}$ with the roots in $\a\in \Da_{\mu,
\si}$ such that $\a(d')=0$ and choose  $\Dp_{\k'}=
\Dap_{\mu,\si} \cap \D_{\k'}$ as a set of positive roots  for
$\k'$. We denote by $\b'$ the corresponding
Borel subalgebra of $\k'$.
\begin{rem}\label{biprimo}
By the definition of $g$ we see that
the set of simple roots for $\k'$ is given by
$$
\Pi_{\k'}=\begin{cases}
\{\ov\a_i\mid i\ne 0,p\}&\text{if $\ov\theta_f(\o_p)<2$}\\
\{-\ov\theta_f\}\cup\{\ov\a_i\mid i\ne0,p\}&\text{if $\ov\theta_f(\o_p)=2$}.
\end{cases}
$$
It follows that $\b'=\b_0\cap\k'$.
\end{rem}

Combining Lemma~\ref{eminuscolo} and Remark~\ref{biprimo} with the
results of \cite{IMRN} exposed in \S~\ref{abelianibasic}, we find the
analogue
of Theorem~\ref{decoeabeliani} for this case. Let $\Sigma_{ni}'$ be the the set
of $\b'$-stable abelian noncompact subspaces of $\p'$. Recall that in section \ref{xerre} we set
$L=N-n$ and $l=\lfloor\frac{N-n}{2}\rfloor$.

\begin{theorem}\label{decoeabelner}Set $m=\lfloor\frac{ dim \p}{2}\rfloor$.\newline
(1) Assume that $m$ is even. Then
$$
L(\tilde\Lambda_{m-1})=L(\tilde\Lambda_{m})=2^{l-1}\bigoplus_{A\in
\Sigma'_{ni}}
L\left(\L_{0,\k}+\rho_n+w_0\langle
A\rangle-k_A\d_\k\right).$$
 \newline (2) Assume that $m$ is odd.
 We
have
$$
L(\tilde\Lambda_{m})=2^{l}\bigoplus_{A\in
\Sigma'_{ni}}L\left(\L_{0,\k}+\rho_n+w_0\langle
A\rangle-k_A\d_\k)\right).
$$
In both cases $k_A=|w_0(A)\cap\Dp(\p)|$.
 Moreover the highest weight vectors of each component indexed by
$A\in\Sigma'_{ni}$ are, up to a
constant factor, the
pure spinor (of the spin representation of $Cl_r(\tilde\p)$)
\begin{equation}\prod_{s=0}^{l}
\prod_{l+1\le j_1<\dots <j_s\le L}v_{-r'-1,j_k}\prod_{\a\in w_0(A)\cap\Dp(\p)}
(t^{-r'-2}e_\a)
\prod_{\a\in w_0(A)\cap(-\Dp(\p))} (t^{-r'-1}e_\a)
\end{equation}
if $l$ is even, while, if $l$ is odd, they are
\begin{align*}\prod_{s=0}^{l}&
\prod_{l+1< j_1<\dots <j_s\le L}v_{-r'-1,j_k}(\prod_{\a\in w_0(A)\cap\Dp(\p)}
\left((t^{-r'-2}e_\a)(v_{-r'-1,l+1})\right)\\
&\prod_{\a\in w_0(A)\cap(-\Dp(\p))}\left((t^{-r'-1} e_\a)(t^{-r'-1}
v_{-r'-1,l+1})\right).
\end{align*}
\end{theorem}
\begin{proof} By a direct computation, we see that
$g^{-1}(-\a+\d') =-\a+\d'$ if $\a\in \Dp_f$, while
$g^{-1}(-\a+\d') =-\a$ if $\a\in -\Dp_f$. We can therefore apply the proof of
Theorem~\ref{decoeabelianispin}. We need only to check the decomposition of
$X^+_r=L(\tilde\Lambda_{m})$ and $X^-_r=L(\tilde\Lambda_{m-1})$ when $m$ is
even,
but this follows readily from the
description of the highest vectors.
\end{proof}

\paragraph{The $A^{(2)}_{2n}$-case.}
Recall that in this case $L'(\g,\si)=\widehat L(\g,\si)^\vee$,
and that we chose $\Pia'=\{\half(\d'-\ov\th_f), \ov\a_1,
\dots,\ov\a_n\}$ as root basis for $L'(\g,\si)$.
From the explicit description of $\Da_{\si,\rodd}$,
we obtain that $W'_{\si,\rodd}$ is the set of all
elements $v\in\Wa_{\k_\mu}$ such that $N'(v)$ is included
in the set of short roots of $\Da'$.
If we choose $(\Pia')^\vee=\{\d'-\ov\th_f,
\ov\a_1, \dots,\ov\a_{n-1},\half\ov \a_n\}$ as root basis for $\algsi$,
we obtain $(\Da'{}^+)^\vee$ as positive system for $\algsi$.
Observe that $\Wa=\Wa_{\k_\mu}$. It is clear that if we regard $v\in \Wa_{\k_\mu}$ as an element
of $\Wa$ and denote by $N^\vee(v)$ the negative set
of $v$ with respect to this choice of the positive roots, we
obtain that $N^\vee(v)=(N'(v))^\vee$. In particular,
for $v\in W'_{\si,\rodd}$, $N^\vee(v)=2N'(v)$. Therefore,
as a subset on $\Wa$, $W'_{\si,\rodd}$ is the set of all
elements in $v$ such that $N^\vee(v)$ is included
in the set of long roots of $\Da$.
Now we observe that
$\D_\k=\half\D_{f,l}\cup\D_{f,s}=(\D_f)^\vee$, so
$\{\ov\a_1,\dots,\ov\a_{n-1},\half\ov\a_n\}$
is the set of simple roots corresponding to $\D_\k\cap(\half\Dp_f\cup\Dp_f)$.
It follows that
$w_0(\{\ov\a_1,\dots,\ov\a_{n-1},\half\ov\a_n\})=\{\a_0,\dots,\a_{n-1}\}$.
Since $\ov\theta_f=2\ov\a_1+\dots+2\ov\a_{n-1}+\ov\a_n$ we see that
$w_0(\d'-\ov\theta_f)=\d'+\ov\a_n=\a_n$,
hence $w_0((\Pia')^\vee)=\Pia$ and $w_0((\Da'{}^+)^\vee)=\Dap$.
This says that $w_0W'_{\si,\rodd}w_0^{-1}$ is the set of elements of $\Wa$ such
that
$N(v)$ is included in the set of long roots of $\Da$. Since the set
of long roots of $\Da$ is $\D_{ni}+(1+2\ganz)\d'$ we have that
$w_0^{-1}W'_{\si,\rodd}w_0\subset W'_{\si,0}$. Lemma~\ref{eminuscolo}
applies, so we can conclude that
$$
w_0^{-1}W'_{\si,\rodd}w_0=\{v\in\mathcal W^\si_{ab}\mid \ov{N(v)}\subset\D_{ni}\}.
$$

Arguing as in the previous twisted cases, we finally obtain the analogous of
Theorem~\ref{decoeabeliani} for this case.
Set $\Sigma_{ni}$ to be the set of noncompact $\b_0$-stable abelian
subspaces of $\D(\p)$.
\begin{theorem}\label{tipoa22n}Set $m=\lfloor\frac{dim \p}{2}\rfloor$.\newline
(1) Assume that $m$ is even. Then
$$
L(\tilde\Lambda_{m-1})=L(\tilde\Lambda_{m})=2^{\frac{n}{2}-1}\bigoplus_{A\in
\Sigma_{ni}}
L\left(\L_{0,\k}+\rho_n+\langle
A\rangle-|A|\d_\k\right).$$
 \newline (2) Assume that $m$ is odd.
 We
have
$$
L(\tilde\Lambda_{m})=2^{\lfloor\frac{n}{2}\rfloor}\bigoplus_{A\in
\Sigma_{ni}}L\left(\L_{0,\k}+\rho_n+\langle
A\rangle-|A|\d_\k)\right).
$$
 Moreover the highest weight vectors of each component indexed by
$A\in\Sigma_{ni}$ are, up to a
constant factor, the
pure spinor (of the spin representation of $Cl_r(\tilde\p)$)
\begin{equation*}(\prod_{s=0}^{\frac{n}{2}}
\prod_{\frac{n}{2}+1\le j_1<\dots <j_s\le n}v_{-r'-1,j_k})\prod_{\a\in A}
(t^{-r'-2}e_\a)
\end{equation*}
if $n$ is even, while, if $n$ is odd, are
\begin{align*}\prod_{s=0}^{\lfloor\frac{n}{2}\rfloor}&
\prod_{\frac{n+1}{2}< j_1<\dots <j_s\le n}v_{-r'-1,j_k}\prod_{\a\in A}
(t^{-r'-2}e_\a)(t^{-r'-1}v_{-r'-1,l+1}),
\end{align*}
where $l=\lfloor \frac{n}{2}\rfloor$.
\end{theorem}
\begin{proof}
We know that
$$
ch(X_r)=\sum_{u\in W'_{\si,\rodd}}ch(L(\L_{0,\k}+\rho_n-a_0\stodd(\langle N'(u)\rangle)).
$$
By the above discussion $a_0\stodd(\langle N'(u)\rangle)=\stodd(\langle N(w^{-1}_0uw_0)\rangle)$ so we
can write
$$
ch(X_r)=\sum_{A\in \Sigma_{ni}}ch(L(\L_{0,\k}+\rho_n-\stodd(\langle N(w_A)\rangle)).
$$
The coefficient of $\d_\k$ is computed as in \ref{abelianibasic}.
The rest of the proof follows as in the previous cases.
\end{proof}

%%%%%%%%%%%%%%%%%%%%%%%%%%%%%%%%%%%%%%%%%%%%%%%%%%%%%%%%%%%%%%%%%%%%%%%%%
%%%%%%%%%%%%%%%%%%%%%%%%%%%%%%%%%%%%%%%%%%%%%%%%%%%%%%%%%%%%%%%%%%%%%%%%%

\section{The Hermitian symmetric case} \label{hermitian} 
In this section we discuss the decomposition of a conformal pair
$(so(\p),\k)$ when $\g=\k\oplus\p$ is an infinitesimal Hermitian 
symmetric space.
In this case there exists a node $i\ne0$ such that
$a_i=1$, $s_0=s_i=1$, and $s_j=0$ for $j\ne0,i$. 
It turns out that $\k$ is an equal rank subalgebra
of $\g$ and it is not semisimple.
We can write $\k=\sum_{S>0}\k_S\oplus \k_0$, where 
$\k_0=\C \varpi_i$ and $\varpi_i$ is the unique element of 
$\h_0$ such that $\ov\a_j(\varpi_i)=\d_{ij}$ for $j>0$.
Recall that in this case $$ \ka=\widehat{[\k,\k]}\oplus \ka_0,$$
where $\ka_0=\C[t,t^{-1}]\otimes\k_0\oplus \C K_0$ with bracket 
defined by
$$
[t^n\otimes H+aK_0,t^m\otimes H+bK_0]=\d_{n,-m}(H,H)_nK_0.
$$
As before
$(\cdot,\cdot)_n$ is the normalized invariant form of $\g$.
Let $\ov r=0$ if $r$ is even, $\ov r =1$ if $r$ is odd, and   
$\pshes$, $\Wa_{\s,\ov r}$ be defined as in  Section 
\ref{basicandvector} or \ref{spin}, according to the 
parity of $r$ (note that in this case $a_0=k=1$). 
Then let $\sthes:\ha^*\to (\ha_\k)^*$ denote the transpose of 
the map  $\psi_{\ov r}$ restricted to $\ha_\k$.
\par
The same computation performed in the equal rank case
for $\k$ semisimple would give
\begin{equation}\label{dhs}
ch(X_r^{\pm})=\sum_{{u\in
W'_{\si,\ov r}}\atop{\ell(u)\equiv \epsilon\,mod\,2}}
ch(L(\sthes(u\rhat_{\ov r})-\rkhat))),
\end{equation}
where $\epsilon=0,\,1$ according to whether we are considering the $+$ or $-$ case.
Moreover $W'_{\si,\ov r}$
is the set of minimal right coset  representatives for $\Wa/\Wa_{\si, 0}$    if 
$\ov r=0$ and for $\Wa_{\k_\mu}/\Wa_{\si, 1}$ if
$\ov r=1$, and
$\rhat_{0}=\rhat,\,\rhat_1=\rhat'$. 
\par
We first deal with the basic and vector case and then we transfer our results to the spin case
via map $g$ defined in $\eqref{gi}$. So we assume $\ov r=0$.
The starting point
to provide a more explicit form of  \eqref{dhs} is
a remarkable subset of stable subpaces which has been introduced in
\cite{IMRN}, Section 6. Recall from \eqref{dsigma}  the definition of the polytope
encoding $\b_0$-stable
abelian subspaces of  $\p$ and set
$$D'_\si=D_\si\cap\{x\in \h_1^*\mid (x,\a_i)<0\}.$$
$D'_\si$ corresponds exactly to the set of stable abelian subspaces of $\p$
which include $\g_{-\a_i}$.
Let $\omega_i^\vee$ be the unique element in
$Span_{\R}(\a_1,\ldots,\a_n)$ such that $(\a_j,\omega_i^\vee)=\d_{ij}$.
From the proof of Lemma 6.1 of \cite{IMRN} we deduce  the  following fact.
\begin{lemma}\label{decC}
Consider the group of translations $T_{\ganz\omega_i^\vee}=
\{t_{j\omega_i^\vee}\mid j\in \ganz\}$. Then $\ov D'_\si$ is a fundamental
domain for the action of $T_{\ganz\omega_i^\vee}$ on $\bigcup_{w\in
W'_{\si,0}}w \ov C_1$.
\end{lemma}
Therefore there exists a ``special" subset of stable subspaces of $\p$ such
that
the translates of the corresponding alcoves cover the domain
$W'_{\si,0}\ov{C_1}$.
At this point this
fact gives little information on the weights appearing in the decomposition
\eqref{dhs},
since $T_{\ganz\omega_i^\vee}$ is not included in $\Wa$. This requires some more
work,
which we perform in a general setting.\par
\vskip5pt

Let $\alie=\g(A)$, where $A$ is a generalized Cartan matrix of affine tipe
$X_m^{(1)}$, $\ha_\lie$ its Cartan subalgebra, $\Pia_\lie
=\{\be_0, \be_1, \dots \be_m\}$ and $\Pia_\lie^\vee=
\{\be_0^\vee, \be_1^\vee, \dots \be_m^\vee\}$ the sets of simple
roots and coroots. Moreover, let $\Wa_\lie$ be the Weyl group of
$\alie$,  $\L_0^\lie, \L_1^\lie, \dots, \L_m^\lie$ be the
fundamental weights, and $\arho_\lie=\L_0^\lie+\cdots
+\L_m^\lie$.
\par
As usual we assume that $\Pi_\lie=\{\be_1, \dots \be_m\}$ has
Dynkin diagram of finite type $X_m$, and we denote by $\lie$ the
corresponding finite dimensional simple Lie subalgebra of $\alie$. Also, we
denote by $W_\lie$ the Weyl group of $\lie$, by $\Dp_\lie$ its
set of positive roots, by $\th_\lie$ its highest root, and we
set $\de_\lie=\be_0-\th_\lie$.
\par
Identify $\ha_\lie$ and $\ha^*_\lie$ via the normalized invariant form.
Let $\omega^\vee_1,
\dots,\omega^\vee_m$ be the fundamental coweights of
$\lie$ and, for $i\in \{1, \dots, m\}$, let $w_i\in W_\lie$ be
such that $N(w_i)=\{\a\in \Dp_\lie\mid (\a, \omega^\vee_i)\ne 0\}$. It is
well-known that $w_i$ exists (and it is unique). We denote by
$\widetilde W$ the extended affine Weyl group of $\lie$, i.e.
$\widetilde W=T_{P_\lie^\vee}\rtimes W_\lie$, where
$P_\lie^\vee$ is the coweight lattice. We regard $\widetilde W$
as a group of transformations on $\ha_\lie^*$.
Moreover, we set
$$
Z=\{t_{\omega_i^\vee}w_i\mid i\in \{1,\dots ,m\}, \
(\th_\lie,\omega^\vee_i)=1\}\cup\{1\}.
$$
It is well known that $Z$ is exactly the subgroup of all
elements in $\widetilde W$ that map the fundamental alcove of
$\lie$ to itself (see \cite{IM}). We may identify the fundamental
alcove of $\lie$ with $C_{\alie} \cap \h^*_1$, where $C_{\alie}$
is the fundamental chamber of $\alie$, and $\h^*_1= (\L_0^\lie+
\text{Span}_\real\{\be_0, \dots, \be_m\})/ \real\de_\lie$ (see
\cite{Kac}, Section 6.6, or \cite{CP}, Section 1). Since the
restriction to $\h^*_1$ is a faithful representation of
$\widetilde W$, we obtain
$$
Z=\{v\in \widetilde W\mid v\Pia_\lie=\Pia_\lie\}.
$$

\begin{lemma}
\label{fixrhohat}
For all $v\in Z$,
$$
v\arho_\lie=\arho_\lie.
$$
\end{lemma}
\begin{proof}
We fix $v\in Z\setminus\{1\}$ and set $v^{-1}(\be_i)=\be_{j_i}$
for $i\in\{0, 1\dots, m\}$. We denote by $(\ ,\ )$ the form
induced on $\ha_\lie^*$ by the normalized invariant form of
$\alie$ and we recall that $(\ ,\ )$ is invariant under
$\widetilde W$.  Then, for $i=0, \dots, m$,
$$
v \arho_\lie (\be_i^\vee)= \frac{2(v\arho_\lie, \be_i)}{(\be_i,
\be_i)}= \frac{2(\arho_\lie,v^{-1} \be_i)} {(v^{-1}\be_i,
v^{-1}\be_i)}= \arho_\lie(\be_{j_i}^\vee)=1.
$$
It follows that $v\arho_\lie\equiv\arho_\lie\mod \de_\lie$.
\par
It remains to prove that $(v\arho_\lie, \L_0^\lie)=0$.
We assume  that $v=t_{\omega_i^\vee}w_i$. Since $W_\lie$ fixes $\L_0^\lie$,
 by formula (6.5.2) of \cite{Kac} we have
\begin{equation}\label{Lambda0}
v\L_0^\lie=t_{\omega_i^\vee}\L_0^\lie=\L_0^\lie+\omega_i^\vee-\frac 1 2
|\omega_i^\vee|^2\de_\lie.
\end{equation}
Since  $\arho_\lie=\rho_\lie+h^\vee_\lie\L_0^\lie$, where
$\rho_\lie$ is the sum of fundamental weights of $\lie$ and
$h^\vee_\lie$ is its dual Coxeter number, we obtain
\begin{equation}\label{arho}
v\arho_\lie=v\rho_\lie+h^\vee_\lie(\L_0^\lie+\omega_i^\vee-\frac 1
2 |\omega_i^\vee|^2\de_\lie).
\end{equation}
But $v\arho_\lie-(\rho_\lie+h^\vee_\lie\L_0^\lie)\in \real
\de_\lie$, hence
\begin{equation}\label{rho}
v\rho_\lie=\rho_\lie-h_\lie^\vee \omega_i^\vee+x\de_\lie,
\end{equation}
for some $x\in \real$. It follows that
$$
w_i\rho_\lie=t_{-\omega_i^\vee}(\rho_\lie-h_\lie^\vee
\omega_i^\vee+x\de_\lie)=\rho_\lie-h_\lie^\vee \omega_i^\vee+x\de_\lie-
(\rho_\lie-h_\lie^\vee \omega_i^\vee+x\de_\lie, \omega_i^\vee)\de_\lie,
$$
and since the component of $w_i\rho_\lie$ in $\real \de_\lie$ is
zero, we obtain
\begin{equation}\label{x}
x=- (\arho_\lie, \omega_i^\vee)+ h_\lie^\vee |\omega_i^\vee|^2.
\end{equation}
Combining equations \eqref{arho}, \eqref{rho}, and \eqref{x},
we have
\begin{equation}\label{primo}
(v\arho_\lie, \L_0^\lie)= -(\arho_\lie, \omega_i^\vee)+ \frac 1 2
h^\vee |\omega_i^\vee|^2.
\end{equation}

Now, since $Z$ is a group, there exists $i'\in \{1 \dots, m\}$
such that $v^{-1}= t_{\omega_{i'}^\vee}w_{i'}$.  Therefore, as in
\eqref{Lambda0}, we obtain
$$v^{-1}\L_0^\lie=t_{\omega_{i'}^\vee}\L_0^\lie=\L_0^\lie+\omega_{i'}^\vee-\frac
1 2 |\omega_{i'}^\vee|^2\de_\lie.$$
Hence
\begin{equation}\label{secondo}
(v\arho_\lie, \L_0^\lie)=(\arho_\lie, v^{-1}\L_0^\lie)=
(\arho_\lie, \omega_{i'}^\vee)-\frac 1 2 h^\vee |\omega^\vee_{i'}|^2.
\end{equation}

Since $T_{P_\lie^\vee}$ is normal in $\widetilde W_\lie$,
$v^{-1}=w_i^{-1}t_{-\omega_i^\vee}=t_{-w_i^{-1}\omega_i^\vee}w_i^{-1}$
and since
$T_{P_\lie^\vee}W$ is a semidirect product,
\begin{equation}\label{varpi}
-w_i^{-1}\omega_i^\vee=\omega^\vee_{i'}
\end{equation}
and $w_i^{-1}=w_{i'}$. This implies, in particular, that
$|\omega_i^\vee|^2=|\omega^\vee_{i'}|^2$, and therefore from \eqref{primo}
and \eqref{secondo} we obtain that
$$
-(\arho_\lie, \omega^\vee_{i})+ \frac 1 2 h^\vee |\omega^\vee_{i}|^2=
(\arho_\lie, \omega^\vee_{i'})-\frac 1 2 h^\vee |\omega^\vee_{i}|^2.
$$
At this point, in order to conclude, it suffices to prove
that
\begin{equation}\label{terzo}
(\arho_\lie, \omega^\vee_{i})=(\arho_\lie, \omega^\vee_{i'}).
\end{equation}
By equation \eqref{varpi}, we have that
$(\arho_\lie, \omega^\vee_{i'})=(-w_i\arho_\lie, \omega_i^\vee)$,
hence
$$
(\arho_\lie, \omega^\vee_{i})-(\arho_\lie, \omega^\vee_{i'})=
(2\arho_\lie + w_i\arho_\lie-\arho_\lie, \omega^\vee_{i})=
(\langle\Dp_\lie\rangle-\langle N(w_i)\rangle, \omega^\vee_{i}).
$$
By the definition of $w_i$, the last term of the above
equalities is zero. This proves \eqref{terzo} and hence the lemma.
\end{proof}
Denote by $\Sigma'$ the set of abelian $\b_0$-stable  subspaces of $\p$ whose
corresponding alcoves lie in $D'_\s$.
The previous lemma is the key to read the weight of a factor appearing
in \eqref{dhs} in terms of the weight of a subspace in $\Sigma'$.

\begin{prop}\label{dprimo}  If $A=wC_1,\,w\in W'_{\si,0}$, then there exists a unique
$k\in\ganz$  and a unique $I\in\Sigma'$ such that
\begin{align}\label{pesorho}
&\steven(w(\rhat))-\rhat_\k=\notag\\
&\L_{0,\k}+\langle
I\rangle+kh^\vee\varpi_i+(-\frac{1}{2}\dim(I)+k(|I^+|-|I^-|)
-\frac{k^2}{4}\dim(\p))\d_\k,\end{align}
where $I^\pm= I\cap \pm\Dp(\p)$.
\end{prop}
\begin{proof} By Lemma \ref{decC} we have $A=t_{k\omega_i^\vee}(A')$ for a unique
$k\in\ganz$  and a unique alcove $A'\subset D'_\si$.
Suppose that $A'=w'C_1,\,w'\in\Wa$. Then $wC_1=t_{k\omega_i^\vee}w'C_1$, and hence
there exists a unique $z\in Z$ such that
$w=t_{k\omega_i^\vee}wz$. By Lemma \ref{fixrhohat} and formula  (6.5.2) of
\cite{Kac} 
 we thus obtain
\begin{align}\label{conto}
\steven(w(\rhat))-\rhat_\k&=
\steven(t_{k\omega_i^\vee}w'z(\rhat))-\rhat_\k=
\steven(t_{k\omega_i^\vee}w'(\rhat))-\rhat_\k\notag\\
&=\steven(t_{k\omega_i^\vee}(w'(\rhat)-\rhat))+\steven(t_{k\omega_i^\vee}(\rhat))
-\rhat_\k\notag\\
&=\steven(t_{k\omega_i^\vee}(\langle
I\rangle)-\dim(I)\d')+\steven(t_{k\omega_i^\vee}(\rhat))
-\rhat_\k,
\end{align}
where $I$ is the ideal in $\Sigma'$ corresponding to $w'$. Note that
$\steven(\d')=\frac{1}{2}\d_\k$ and that 
$\steven(\omega_i^\vee)=\nu(\varpi_i)+\frac{|\varpi_i|^2}{2}\d_\k$. Also remark that 
\begin{equation}\label{oss}
(\langle I\rangle,\omega_i^\vee)=|I^+|-|I^-|,\qquad (\rhat,\omega_i^\vee)=\frac{dim(\p)}{4}.
\end{equation}
 Combining \eqref{conto}, \eqref{oss}
 and formula \eqref{primo} we get \eqref{pesorho}.
\end{proof}
Denote by $c_{I,k}$ the coefficient of $\d_\k$ in formula \eqref{pesorho}.
For $q\in\ganz$, denote by $L(\tilde\L_\epsilon)_q$ the eigenspace of
eigenvalue
$q$ for the action of $\k_0$ on $L(\tilde\L_\epsilon)$.
\begin{rem} $L(\tilde\L_\epsilon)_q$ is non zero if and only if
$q\equiv\epsilon\,mod\ 2$ . In fact, by \eqref{peso}, the
weights of $L(\tilde\L_\epsilon)$ are of the form
$\L_{0.\k}-\sum_{j=1}^s\gamma_j,\,\gamma_j\in\Dap(\p),\,
s\equiv\epsilon\,mod\ 2$.
Since $\Dp(\p)=\D^1_f$ in this case, we have
$(\L_{0.\k}-\sum_{j=1}^s\gamma_j)(\varpi_i)=s$.
\end{rem}

From the previous Proposition it follows that
\begin{theorem}\label{finalehs}
\begin{equation*}L(\tilde\L_\epsilon)_q=\!\!\!\!\!\!\!\sum_{{I\in\Sigma'}\atop
{|I^+|-|I^-|\equiv
q\,mod\,\frac{\dim(\p)}{2}}}\!\!\!\!\!\!\!L(\L_{0,\k}+\langle
I\rangle+k_Ih^\vee\nu(\varpi_i)
+(c_{I,k_I}+\frac{\epsilon}{2})\d_\k),\end{equation*}
where $k_I=\frac{2(q-|I^+|+|I^-|)}{\dim(\p)}$.
\end{theorem}
\begin{proof}
Consider the sum
$$
\sum_{w\in W'_{\si,0}}ch(L(\steven(w(\rhat))-\rhat_\k)).
$$
This sum makes sense because, given a weight $\mu$, there is only a finite
number of
elements $w\in W'_{\si,0}$ such that $\mu$ is a weight of
$L(\steven(w(\rhat))-\rhat_\k)$.
Indeed, if  $\mu$ occurs in $L(\steven(w(\rhat))-\rhat_\k)$, then
$\mu=\steven(w(\rhat))-\rhat_\k-\sum_{\a\Dap_\k}n_\a\a$, hence
$\mu(\o_i)=(\steven(w(\rhat))-\rhat_\k)(\o_i)$. It follows from
Proposition~\ref{dprimo} that
$$
\mu(\o_i)=\langle
I\rangle(\o_i)+kh^\vee\nu(\varpi_i)(\o_i)
$$
and there is only a finite number of $I\in \Sigma'$ and $k\in\ganz$ that
satisfy this equation.

We can therefore write
\begin{align*}
\sum_{w\in W'_{\si,0}}ch(L(\steven(w(\rhat))-\rhat_\k))&=\sum_{w\in
W'_{\si,0}}\frac{\sum_{u\in\Wa_\k}\epsilon(u)e^{u\steven(w\rhat)}}{D_\k}\\
&=\frac{\sum_{w\in
W'_{\si,0}}\sum_{u\in\Wa_\k}\epsilon(u)e^{u\steven(w\rhat)}}{D_\k}\\
&=\frac{\num^+}{D_\k}=ch(X_r).
\end{align*}
Thus we can write
\begin{equation}\label{riconto}
L(\tilde\L_\epsilon)=\sum_{k\in\ganz}\sum_{{I\in\Sigma'}\atop
{|I|\equiv \epsilon\,mod\,2}}L(\L_{0,\k}+\langle
I\rangle+kh^\vee\nu(\varpi_i)
+(c_{I,k}+\frac{1}{2}\epsilon)\d_\k).\end{equation}
Observe now that
\begin{align*}(\L_{0,\k}+\langle
I\rangle+kh^\vee\nu(\varpi_i)
+c_{I,k}\d_\k)(\varpi_i)&=
|I^+|-|I^-|+kh^\vee|\varpi_i|^2\\&=|I^+|-|I^-|+k\frac{\dim(\p)}{2}.\end{align*}
The result follows by collecting in \eqref{riconto} the terms with
eigenvalue $q$.
\end{proof}
Arguing as in the semisimple equal rank case we obtain, for the spin representations,
the following result.
\begin{theorem}\label{finale2hs} Set $m=\lfloor\frac{\dim(\p)}{2}\rfloor$. The eigenvalues of $\varpi_i$ on
$L(\tilde\L_{m-\epsilon})$ are of the form
$\frac{\dim(\p)}{4}+q,\,q\in\ganz,\,q\equiv\epsilon\,mod\,2$. The corresponding eigenspaces decompose
as
\begin{equation*}L(\tilde\L_{m-\epsilon})_{\frac{\dim(\p)}{4}+q}=\!\!\!\!\!\!\!\sum_{{I\in\Sigma'}\atop
{|I^+|-|I^-|\equiv q\,mod\,\frac{\dim(\p)}{2}}}\!\!\!\!\!\!\!L(\L_{0,\k}+\langle
I\rangle+\rho_n+k_Ih^\vee\nu(\varpi_i)
+c'_{I,k_I}\d_\k),\end{equation*}
where $k_I=\frac{2(q-|I^+|+|I^-|)}{\dim(\p)}$ and
$$c'_{I,k_I}=(k-1)|I^+|-k|I^-|+(k^2-k)\frac{\dim(\p)}{4}.$$
\end{theorem}

%%%%%%%%%%%%%%%%%%%%%%%%%%%%%%%%%%%%%%%%%%%%%%%%%%%%%%%%%%%%%%%%%%%%%%%%%%%%
%%%%%%%%%%%%%%%%%%%%%%%%%%%%%%%%%%%%%%%%%%%%%%%%%%%%%%%%%%%%%%%%%%%%%%%%%%%%%

\section{Examples and applications}\label{examples}
\subsection{Combinatorial interpretation of decompositions in type $C$}
We want to give a combinatorial interpretation of Theorem
\ref{decompositioneven} and Theorem
\ref{decoeabelianispin} for the pair $\g=sp(V_1\oplus V_2)\supset sp(V_1)\oplus
sp(V_2)=\k$, where
$V_1,V_2$ are complex vector spaces of dimension $2m,2n$ respectively.
It turns out that in this specific case (and indeed only in this) the decomposition formulas
afford bijections between
level $m$ representations of $\widehat{sp(2n)}$ and level $n$
representations of $\widehat{sp(2m)}$.
This result, in the case of the spin representation, appears as Proposition
2 in
\cite{KacPeterson}. In our general setting we are considering the case of a
Lie algebra $\g$ of type
$C_{n+m}$ endowed with an involution $\sigma$ of type $(0,...0,1,0....0;1)$,
where $1$ appears in
position $m$.\par Let $P_{n,m}$ denote the set of $(m+1)$-weak compositions
of $n$, i.e. ordered
$(m+1)$-tuples
$(k_0,\ldots,k_m)$ of non  negative integers such that $\sum_{i=0}^mk_i=n$.
Let also
$S_{h,k}$ denote the set of $h$ elements subsets of $\{1,\ldots,k\}$.
The map
$(k_0,\ldots,k_m)\mapsto\{k_0+1,k_0+k_1+2,\ldots,k_0+\cdots +k_{m-1}+m\}$
is a bijection
$\zeta_{n,m}:P_{n,m}\to S_{m,m+n}$. If $c:S_{m,m+n}\to S_{n,m+n}$ is the
map which associates to an
$m$-element subset of $\{1,\ldots,m+n\}$ its complement, the map
$\zeta_{m,n}^{-1}\circ\, c\,\circ \zeta_{n,m}:P_{n,m}\to P_{m,n}$ is a
bijection, which we denote by
$(k_0,\ldots,k_m)\mapsto (k'_0,\ldots,k'_n)$. Set also
$k''_i=k'_{n-i},\,0\leq i\leq n$.\par
Let $\dot\L_0,\ldots,\dot\L_m,\,\ddot\L_0,\ldots,\ddot\L_n$ be the
fundamental weights of
the simple ideals of $\widehat\k$, assuming that both components have the
Dynkin diagram displayed as
in \cite{Kac}, \S4, Table $Aff\ I$.

\begin{prop}\label{typec} Let $\g,\,\k$ be as above. The following
decomposition formulas for the
level $1$ modules of $\widehat{so(\p)}$
into irreducible
$[\widehat\k,\widehat\k]$-modules hold ($\epsilon=0,1$):
\begin{align*}
&L(\tilde\Lambda_{l-\epsilon}) &&= &&\bigoplus_{{(k_0,\ldots,k_m)\in
P_{n,m}}
\atop{\sum_{i=0}^m i\,k_i\equiv\epsilon\,\,mod\,2}}L(k_0\dot\L_0+\ldots+k_m\dot
\L_m)\otimes L(k'_0\ddot\L_0+\ldots+k'_n\ddot \L_n),\\
&L(\tilde\Lambda_\epsilon) &&= &&\bigoplus_{{(k_0,\ldots,k_m)\in P_{n,m}}
\atop{\sum_{i=0}^m i\,k_i\equiv\epsilon\,\,mod\,2}}L(k_0\dot\L_0+\ldots+k_m\dot
\L_m)\otimes L(k''_0\ddot\L_0+\ldots+k''_n\ddot \L_n).\end{align*}
\end{prop}
The key remark to deduce \ref{typec} from \ref{decompositioneven} and
\ref{decoeabelianispin} is the following combinatorial interpretation of
the sets
$N(w),\,w\in W'_{\si,\ov r}$.
Consider the following rectangle $R_{n,m}$ filled with roots (of $\widehat
L(\g,\s)$) as
displayed in the following figure for $m=2,n=3$:
$${\tiny
\Einheit2.35cm
\Pfad(0,-1),22111\endPfad
\Pfad(0,-1),112\endPfad
\Pfad(0,-1),1112\endPfad
\Pfad(0,-1),122\endPfad
\Pfad(0,0),112\endPfad
\Pfad(0,0),1112\endPfad
\Label\lo{\a_2+\a_3+\a_4}(1,-1)
\Label\lo{\a_1+\a_2+\a_3+\a_4}(1,0)
\Label\lo{\a_1+\a_2+\a_3}(2,0)
\Label\lo{\a_1+\a_2}(3,0)
\Label\lo{\a_2+\a_3}(2,-1)
\Label\lo{\a_2}(3,-1)
\hskip7.6cm}$$
%\par\newpage
Then the  sets $N(w),\,w\in W'_{\si,\ov r}$ can be described as the sets roots lying in
the boxes
under any lattice path from the South-West corner of the rectangle to the
North-East corner.
This is readily checked observing that these sets are biconvex (hence are
of the form
$N(w)$, for some $w\in\widehat W$), that they are either void or intersect
$\Pi$ exactly in  $\a_m$
(hence are of the form
$N(w)$, for some $w\in W'_{\si,\ov r}$), and finally that they are as many as
the above
lattice paths,
hence $\binom{n+m}{n}=|W'_{\si,\ov r}|$ in number (see \cite{IMRN}, Table 5.1).
Now the
proposition follows by
direct computation taking into account that $\L_{0,\k}=n\dot\L_0+m\ddot\L_0,\,
\rho_n+\L_{0,\k}=n\dot\L_m+m\ddot\L_0$, $\stpm(\a_m)=\dot\L_1
+\ddot\L_1$ ($r$ even) and $w_0=s_ms_{m-1}s_m\cdots s_1s_{2}\cdots s_m$,
 $\stpm(\a_m)=w_0(\dot\L_1)
+\ddot\L_1$ ($r$ odd)
. More explicitely,
it is not difficult to  see that if $p_w$ is the lattice path associated to
$w\in W'_{\si,\ov r}$ and
$p_w\leftrightarrow (a_1,\ldots,a_m)\leftrightarrow (b_1,\ldots,b_n)$, where
$0\leq a_1\leq
a_2\leq\ldots\leq n
$ (resp. $m\geq b_1\geq
b_2\geq\ldots\geq 0$) are the lengths of the rows and (resp. columns) of
the subdiagram
of $R_{n,m}$ whose bottom border is  $p_w$, counted from bottom to top (resp. from
left to right), then
$$\L_{0,\k}-\langle\stpm (N(w))\rangle=\sum_{i=1}^n\dot\L_{m-b_i}+
\sum_{i=1}^m\ddot\L_{n-a_i}$$
for $r$ even and
$$\rho_n+\L_{0,\k}-\langle\stpm (N(w))\rangle=\sum_{i=1}^n\dot\L_{b_{i}}+
\sum_{i=1}^m\ddot\L_{n-a_i}.$$
for $r$ odd.
\subsection{A special case}
Suppose that $\si$ is an automorphism of type $(0,\ldots,1,\dots,0;1)$ with
$1$ in a position corresponding to a long simple root (say $\a_p$). We show below how
to calculate the $\ka$-decomposition of the basic and vector representations in terms of a special
class of representatives and how to get information on asymptotic dimension.  We set for shortness
$\Wa=\Wa_{\si,0},\, W'=W'_{\si,0}$. Let $W_f$ the Weyl group generated by $s_1,\ldots,s_n$.
Let  $a(\L)$ denote the  asymptotic dimension of a module $L(\L)$ (see \cite[(2.1.5)]{KacW} for the 
definition).
\begin{prop} 1. The map $w\mapsto w_\si w$ is an involution $i$ on $W'$.\newline  Moreover
we have that 
$i(W'\cap W_f)= W'\setminus(W'\cap W_f).$
\par\noindent
2. Denote by $\Lambda_w=\sum_{i=0}^nb_i\L_i$ the weight of the 
factor indexed by $w\in W'$ in formula \eqref{decobasic}. If $w\in W'\cap W_f$, then   
$\Lambda_{i(w)}=\sum_{i=0}^nb_i\L_{\pi(i)}$,  where
$\pi$ is a suitable permutation of $\{1,\ldots,n\}$. In particular
$a(\Lambda_w)=a(\Lambda_{i(w)})$.\end{prop}
\begin{proof} Consider the set $P_\si$ defined in \eqref{psigma}. By \cite[Lemma 5.9]{IMRN}, , we have
$w_\si(P_\si)\subseteq P_\si$, hence left multiplication by $w_\si$ gives a map $i:W'\to W'$.
By \cite[Lemma 5.11]{IMRN},  we deduce that $0$ does not belong
to $w_\si \ov C_1$; in particular $w_\si\notin W'\cap W_f$. This easily implies 
that 
$i(W'\cap W_f)\subseteq W'\setminus(W'\cap W_f)$.  It is clear that $i$ is injective.  Proposition 5.8
and Theorem 5.12 of 
\cite{IMRN}  give $|W'|=2|W'\cap W_f|$, hence $i(W'\cap W_f)= W'\setminus(W'\cap W_f)$.
Finally $i$ is an involution since
$w_\si$ is an involution. Indeed $w_\s$ is defined in \cite{IMRN} as the product of
certain elements of the extended Weyl groups of the irreducible components
of the extended Dynkin diagram of $\g$ minus the $p$th node; in  \cite{IM} the action of these elements
is completely worked out. This explicit description proves both that $w_\si$ is an involution and that
it acts on each simple component $\ka_S$ by permuting the  fundamental weights.
The assertion on asymptotic dimension follows from the fact that this quantity is invariant under
the action of certain elements in the extended affine Weyl group. More precisely
the invariance  follows from 
\cite[(2.2.15-16)]{KacW} taking into account that $w_\sigma$ is a product of elements in 
$W_0^+$ (in the notation of \cite{KacW}).
\end{proof}
\subsection{More examples}
The following examples should make clear how to use our decomposition formulas in
explicit cases. To avoid cumbersome notation we describe the decomposition as
$[\ka,\ka]$ modules. In other words we consider the weights of the $\ka$-modules
appearing in the decompositions modulo $\d_\k$.

\vskip10pt \noindent{\bf 1.} We describe the
decomposition of $X_{-1}$ when  $\g$ is of type $G_2$ and $\s$ of type
$(0,1,0;1)$. In this case $\ka$ is of type $A_1^{(1)}\times A_1^{(1)}$.
$W_{\si,1}$ is generated inside $\Wa$ by $s_0,s_2, s_1s_2s_1s_2s_1,
s_1s_2s_1s_0s_1s_2s_1s_0s_1s_2s_1$ and
$$
W'_{\si,1}=\{id, s_1, s_1s_0, s_1s_2,
s_1s_2s_0, s_1s_2s_0s_1\}.
$$
According to formula \ref{decoeabelianispin}, the highest weights
of the irreducible components are of the form
$2\dot\L_0+10\ddot\L_0+\rho_n-\stodd(\langle N(u)\rangle)$,  where $u$ ranges
over
$W'_{\si,-1}$. Here and in the following $\dot\L_i$ denotes the $i$-th
fundamental
weight for the
first copy of $A_1^{(1)}$ whereas $\ddot\L_i$ denotes the $i$-th
fundamental weight for the other  copy.  Since $\rho_n=w_0(2\a_1+3\a_2)$ and $\overline
\a_1=-\frac{1}{2}(\overline \a_0+3\overline \a_2)$ we have
\begin{align*}
X_{-1}=&L(2\dot\L_1)\otimes L(10\ddot\L_0)\oplus\\
&L(\dot\L_0+\dot\L_1)\otimes
L(7\ddot\L_0+3\ddot\L_1)\oplus\\
&L(2\dot\L_1)\otimes L(4\ddot\L_0+6\ddot\L_1)\oplus\\
&L(2\dot\L_0)\otimes L(6\ddot\L_0+4\ddot\L_1)\oplus\\
&L(\dot\L_0+\dot\L_1)\otimes
L(3\ddot\L_0+7\ddot\L_1)\oplus\\
&L(2\dot\L_0)\otimes L(10\ddot\L_1).
\end{align*}
\vskip20pt
\noindent{\bf 2.}
We describe the
decomposition of $X_{0}$ when  $\g$ is of type $D_4$ and $\s$ of type
$(0,1,0,0;2)$. In this case $\ka$ is of type $A_1^{(1)}\times
C_2^{(1)}$. $W_{\s,0}$ is generated inside $\Wa$ by
$s_0,s_2,s_3,s_1s_0s_1s_2s_1s_0s_1,$ $s_1s_2s_3s_2s_1s_0s_1s_2s_3s_2s_1$
and we have
\begin{align*}W'_{\si,0}=\{Id,s_1,s_1s_0,s_1s_2, s_1s_0s_1,s_1s_0s_2,s_1s_2s_3,
s_1s_0s_2s_3,s_1s_2s_3s_2,\\s_1s_0s_2s_3s_2,s_1s_2s_3s_2s_
1,s_1s_0s_2s_3s_2s_1\}.\end{align*} According to
formula \ref{decoeabelner}, the highest weights of the irreducible
components are
of the form
$10\dot\L_0+3\ddot\L_0-\steven(\langle N(u)\rangle)$, where $u$ ranges over
$W'_{\si,0}$. Taking into account that
$\overline\a_1=-(\overline \a_0+\overline \a_2+\overline \a_3)$,we get
\begin{align*}X_{0}=&L(10\dot\L_0)\otimes
L(3\ddot\L_0)\oplus\\&
L(8\dot\L_0+2\dot\L_1)\otimes
L(2\ddot\L_0+\ddot\L_2)\oplus\\&
L(6\dot\L_0+4\dot\L_1)\otimes L(\ddot\L_0+2\ddot\L_1)\oplus\\&
L(4\dot\L_0+6\dot\L_1)\otimes
L(\ddot\L_0+2\ddot\L_1)\oplus\\&
L(2\dot\L_0+8\dot\L_1)\otimes L(2\ddot\L_0+\ddot\L_2)\oplus\\&
L(10\dot\L_1)\otimes L(3\ddot\L_0)\oplus\\
&L(10\dot\L_0)\otimes L(3\ddot\L_2)\oplus\\&
L(8\dot\L_0+2\dot\L_1)\otimes
L(\ddot\L_0+2\ddot\L_2)\oplus\\&
L(6\dot\L_0+4\dot\L_1)\otimes L(2\ddot\L_1+\ddot\L_2)\oplus\\&
L(4\dot\L_0+6\dot\L_1)\otimes
L(2\ddot\L_1+\ddot\L_2)\oplus\\&
L(2\dot\L_0+8\dot\L_1)\otimes L(\ddot\L_0+2\ddot\L_2)\oplus\\&
L(10\dot\L_1)\otimes L(3\ddot\L_2).\end{align*}\vskip20pt\noindent{\bf 3.}
It is
easy to see from our formulas that if $\g$ is of type $D_{l+1}$ and $\s$ is of
type$(1,0,\dots,0;2)$ then both the spin  and the basic and vector
representations restrict to the spin  and  basic and vector representations for
$B_l^{(1)}$.\vskip20pt
\noindent{\bf 4.}  Finally we consider the decomposition of
the spin representation  $X_{-1}$ for $\g$  of type $D_4$ and $\s$ of type
$(0,1,0,0;2)$. As in example 2,  $\ka$ is of type $A_1^{(1)}\times C_2^{(1)}$.
$\Wa_{\k_\mu}$ is an affine Weyl group of type $B_3$.
Recall that we chose $\Pia_\mu=\{-\th_f+\d',\ov\a_1,\ov\a_2,\ov\a_3\}$ as a set of
positive roots for $\Da_\mu$. Set $\be_0=-\th_f+\d',\,\be_i=\ov\a_i,i=1,2,3,\,
s_i=s_{\beta_i},\,i=0,1,2,3.$ Then
$${\stodd}^{-1}(\Pia_\k)=\{\be_2,\be_3,\be_0+\be_2+\be_3,
\be_0+\be_1+\be_2,\be_1+\be_2+\be_3\},$$
hence
$\Wa_{\s,1}$  is generated by
$s_2,s_3,s_0s_2s_3s_2s_0,s_0s_1s_2s_1s_0,s_1s_2s_3s_2s_1.$ A
set of minimal right coset representatives is
$$W_{\s,1}'=\{Id,s_0,s_1,s_1s_0,s_1s_2,s_0s_2\}.$$
Taking into account
that
$\overline
\a_1=-(\overline \a_0+\overline \a_2+\overline \a_3)$, and that
$\rho_n=5\dot\L_0+\ddot\L_1$, we
get\newline\begin{align*}X_{-1}=&L(5\dot\L_0+5\dot\L_1)\otimes
L(2\ddot\L_0+\ddot\L_1)\oplus\\&L(3\dot\L_0+7\dot\L_1)\otimes
L(\ddot\L_0+\ddot\L_1+\ddot\L_2)\oplus\\&
L(7\dot\L_0+3\dot\L_1)\otimes L(\ddot\L_0+\ddot\L_1+\ddot\L_2)\oplus\\&
L(5\dot\L_0+5\dot\L_1)\otimes
L(\ddot\L_1+2\ddot\L_2)\oplus\\&L(\dot\L_0+9\dot\L_1)\otimes
L(3\ddot\L_1)\oplus\\&L(9\dot\L_0+\dot\L_1)\otimes
L(3\ddot\L_1).\end{align*}

\subsection{Connections with  modular invariance.}\label{modular}
We now try to use the formulas developed
in the previous sections to obtain information
on the action of $SL(2,\ganz)$ on modified  characters described in
\cite{KacW}.
Here we  consider the very special case when $\sigma$ comes from an
automorphism of
the diagram of $\g$.
This implies that $\g$ is either simple of type $A,D,E$ or of complex type.
Furthermore $\k$ is simple. We shall also assume that $\g$ is not of type
$A_{2n}$. These are precisely the cases in which $W'_{\si,1}=\{1\}$.\par
 Let $h^\vee_\k$ denote the dual Coxeter number of $\k$ and set
$j=h^\vee-h_\k^\vee$.
We denote by
$\dot\L_i$ the $i$-th fundamental weight of $\ka$ and by $P^j_+$ the
set of dominant weights for $\widehat \k$ of level $j$.   Recall that
$N=rk\,
\g$ while
$n=rk\,\k$.  By \eqref{forabelian}  we have that
\begin{equation}\label{1}
ch(L(\tilde\L_0))-ch(L(\tilde\L_1))=\sum_{w\in
W'_{\si,0}}\epsilon(w)ch(L(\steven(w\rhat)-\rkhat)).
\end{equation}
Formula \eqref{decomposizione} becomes in our case
\begin{equation}\label{2}
ch(L(\tilde\L_m))=2^{\lfloor\frac{N-n}{2}\rfloor}ch(L(j\dot\L_0+\rho_n))
\end{equation}
if $N-n$ is odd, and 
\begin{equation}\label{3}
ch(L(\tilde\L_{m-1}))+ch(L(\tilde\L_m))
=2^{\lfloor\frac{N-n}{2}\rfloor}ch(L(j\dot\L_0+\rho_n)),
\end{equation}
if $N-n$ is even. Here $m=\lfloor\frac{\dim(\p)}{2}\rfloor$.
\par
Denote by  $\chi_\L$ is the modified character of $L(\L)$  (see \cite[(1.5.11)]{KacW}),
 and set
$Y=\{h\in\ha^*\mid Re\,\d_\k(h)>0\}$. Moreover we write
$\L_w$ for
$\steven(w\rhat)-\rkhat$. Since the pair $(so(\p),\k)$ is conformal, relation
\eqref{1} translates into
\begin{equation}\label{4}
(\chi_{\tilde\L_0}-\chi_{\tilde\L_1})_{|Y}=\sum_{w\in
W'_{\si,0}}\epsilon(w)\chi_{\L_w},
\end{equation}
whereas \eqref{2} gives
\begin{equation}\label{5}
(\chi_{\tilde\L_m})_{|Y}=2^{\lfloor\frac{N-n}{2}\rfloor}\chi_{j\dot\L_0+\rho_
n}
\end{equation}
($N-n$ odd), and \eqref{3} gives
\begin{equation}\label{6}
(\chi_{\tilde\L_{m-1}}+\chi_{\tilde\L_m})_{|Y}
=2^{\lfloor\frac{N-n}{2}\rfloor}\chi_{j\dot\L_0+\rho_n}
\end{equation}
($N-n$  even). Recall from \cite[Remark 4.2.2]{KacW} that if $N-n$ is odd,
\begin{equation}\label{7}
\chi_{\tilde\L_m}(-\frac{1}{\tau})=\frac{1}{\sqrt{2}}(\chi_{\tilde\L_0}-\chi_
{\tilde\L_1})(\tau)
\end{equation}
and, if $N-n$ is even,
\begin{equation}\label{8}
(\chi_{\tilde\L_{m-1}}+\chi_{\tilde\L_m})(-\frac{1}{\tau})=
(\chi_{\tilde\L_0}-\chi_{\tilde\L_1})(\tau).
\end{equation}
By modular invariance of modified characters,
\begin{equation}\label{9}
\chi_{j\dot\L_0+\rho_n}(-\frac{1}{\tau})=\sum_{\L\in
P^{j}_+}a(\L,j\dot\L_0+\rho_n)\chi_{\L}.
\end{equation}
(here $a(\cdot,\cdot)$ is the function $P_+^j\times P_+^j\to\mathbb C$
defined in
\cite[(2.1.7)]{KacW}). 
Assume $N-n$ even and use
\eqref{4},\eqref{8},\eqref{6},\eqref{9} obtaining
\begin{align*}
\sum_{w\in W'_{\si,0}}\epsilon(w)\chi_{\L_w}(\tau)&=
(\chi_{\tilde\L_0}-\chi_{\tilde\L_1})(\tau)=(\chi_{\tilde\L_{m-1}}+\chi_{\tilde\L_m})(-\frac{1}{\tau})\\
&=2^{\frac{N-n}{2}}\chi_{j\dot\L_0+\rho_n}(-\frac{1}{\tau})=2^{\frac{N-n}{2}}\sum_{\L\in
P^{j}_+}a(\L,j\dot\L_0+\rho_n)\chi_{\L}(\tau).
\end{align*}
The case $N-n$ odd is analogous. We can deduce the following
\begin{prop}\label{result} We have
$a(\L,j\dot\L_0+\rho_n)=0$ unless there exists $w\in W_{\si,0}'$ such that
$\L+\rkhat=\steven(w\rhat)$. In such a case
$a(\L,j\dot\L_0+\rho_n)=2^{-\frac{N-n}{2}}(-1)^{\ell(w)}$.
\end{prop}
\begin{rem}
In the complex case this result was obtained in the same way in \cite{KacW}, (4.2.14).
\end{rem}
\begin{rem}
Recall that, if $\Sigma$ is the set of $\b_0$-stable abelian  subspaces of
$\p$,
then, according to Theorem~\ref{decoeabeliani},
the set $\Sigma$ parametrizes the irreducible components of $X_0$. 
By \cite[(2.2.3)]{KacW}, we know that
$$
\sum_{\L\in P^j_+}|a(\L,j\dot\L_0+\rho_n)|^2=1.
$$
We can therefore deduce that $|\Sigma|=2^{N-n}$ in these cases.
This fact was first proved  in \cite{Pan2} by a different method.
In the complex case we have yet another proof of Peterson's $2^\rank$ abelian ideals Theorem
(see again \cite{KacW}). 
\end{rem}

\begin{rem}
If $\g$ is of type $D_N$, then $\k$ is of type $B_{N-1}$ and one only
obtains again that
$a(\dot\L_{N-1},\dot\L_0)=-a(\dot\L_{N-1},\dot\L_1)=\frac{1}{\sqrt{2}}$ and
$a(\dot\L_{N-1},\dot\L_{N-1})=0$.
\end{rem}
%\bibliographystyle{amsplain}
%\bibliography{abelian}

    \providecommand{\bysame}{\leavevmode\hbox to3em{\hrulefill}\thinspace}

\footnotesize{

\noindent{\bf P.C.}: Dipartimento di Scienze, Universit\`a di Chieti-Pescara, Viale Pindaro 42, 65127 Pescara,
ITALY;\\ {\tt cellini@sci.unich.it}

\noindent{\bf P.MF.}: Politecnico di Milano, Polo regionale di Como, Via
Valleggio 11,
22100 Como, ITALY;\\ {\tt frajria@mate.polimi.it}

\noindent{\bf V.K.}: Department of Mathematics, Rm 2-165, MIT,
77 Mass. Ave, Cambridge, MA 02139;\\
{\tt kac@math.mit.edu}

\noindent{\bf P.P.}: Dipartimento di Matematica, Universit\`a di Roma ``La
Sapienza",
P.le A. Moro 2, 00185, Roma , ITALY;\\
{\tt papi@mat.uniroma1.it}
}

\end{document}